\documentclass{sn-jnl}
\pdfoutput=1
\usepackage[english]{babel} 
\usepackage[utf8]{inputenc} 
\usepackage[title]{appendix} 

\usepackage{graphicx}
\usepackage{caption}
\usepackage{subcaption}

\usepackage{amssymb}
\usepackage{amsmath}
\usepackage{amsthm}
\usepackage{stmaryrd}
\usepackage{amsfonts}
\usepackage{physics}
\usepackage{stackrel}
\usepackage{mathtools}
\usepackage{dsfont}
\usepackage{quiver}
\usepackage{tikz-cd}
\usepackage{enumitem}
\usepackage{fixme}
\fxsetup{status=draft}
\fxusetheme{color}
\FXRegisterAuthor{erg}{aerg}{\textbf{Elias}}
\FXRegisterAuthor{mda}{amda}{\textbf{Matteo}}

\usepackage{xurl}
\usepackage[style=numeric, sorting=none, style=numeric-comp]{biblatex}
\usepackage{csquotes}
\usepackage{hyperref}

\usepackage[noabbrev]{cleveref}

\theoremstyle{definition}
\newtheorem{definition}{Definition}

\theoremstyle{definition}
\newtheorem{example}{Example}

\theoremstyle{remark}
\newtheorem{remark}{Remark}

\theoremstyle{plain}
\newtheorem{theorem}{Theorem}

\theoremstyle{plain}
\newtheorem{proposition}{Prop}

\theoremstyle{plain}

\theoremstyle{plain}
\newtheorem{lemma}{Lemma}

\newcommand{\tT}{\mathcal{T}}
\newcommand{\dD}{\mathcal{D}}

\newcommand{\cC}{\mathcal{C}}

\newcommand{\iI}{\mathcal{i}}
\newcommand{\AAA}{\mathbb{A}}
\newcommand{\BBB}{\mathbb{B}}

\newcommand{\ii}{\mathrm{i}}

\newcommand{\Z}{\mathbb{Z}}

\DeclareMathOperator{\lcm}{lcm}
\usepackage{IEEEtrantools}

\newcommand{\bea}{\begin{equation} \begin{aligned}} 
\newcommand{\eea}{\end{aligned} \end{equation}}

\newcommand{\be}{\begin{equation}} \newcommand{\ee}{\end{equation}}

\usepackage{IEEEtrantools}

\newcommand{\HH}[4][sing]{H^{#2}_{\text{#1}}(#3;#4)}

\newenvironment{tz}[1][]{%
			\begin{tikzpicture}[baseline={([yshift=-.8ex]current bounding 					box.center)},#1] %
				}{%
			\end{tikzpicture} %
			}
			
\pgfkeys{%
  /tikz/on layer/.code={
    \pgfonlayer{#1}\begingroup
    \aftergroup\endpgfonlayer
    \aftergroup\endgroup
  },
  /tikz/node on layer/.code={
     \gdef\node@@on@layer{%
      \setbox\tikz@tempbox=\hbox\bgroup\pgfonlayer{#1}\unhbox\tikz@tempbox\endpgfonlayer\egroup}
    \aftergroup\node@on@layer
  },
  /tikz/end node on layer/.code={
    \endpgfonlayer\endgroup\endgroup
  }
}
\tikzset{slice/.style={draw = gray!50, line width = 0.7pt}}
\tikzset{dot/.style={circle, scale=0.15, fill=black, thick, draw}}

\newcommand{\Zpre}[5]{
\tikzset{line/.style={line width=0.25mm},
curve/.style={line,smooth,tension=1}}
\draw [line] (0,0) -- (2,0) -- (2,2) -- (0,2) -- (0,0);
\ifthenelse{\equal{#4}{}}{
	\ifthenelse{\equal{#2}{1}}{\draw [line,dashed] (0,1) -- (2,1)}{
		\ifthenelse{\equal{#2}{2}}{\draw [line,middlearrow={>},line width=0.5mm] (0,1) -- (2,1)}{}
	};
	\ifthenelse{\equal{#3}{1}}{\draw [line,dashed] (1,0) -- (1,2)}{
		\ifthenelse{\equal{#3}{2}}{\draw [line,middlearrow={>},line width=0.5mm] (1,0) -- (1,2)}{}
	};
}{
	\ifthenelse{\equal{#3}{1}}{
			\draw [line,dashed] (1,0) -- (1,0.7);
			\draw [line,dashed] (1,2) -- (1,1.3);
			\draw [line] (1,0.7) -- (1,1.3);
			
		}{
		\ifthenelse{\equal{#3}{2} \AND \not\equal{#2}{2}}{\draw [line] (1,0) -- (1,2)}{
			\ifthenelse{\equal{#3}{2} \AND \equal{#2}{2}}{
				\draw [line] (1,0) -- (1,0.7);
				\draw [line] (1,2) -- (1,1.3);
				\ifthenelse{\equal{#5}{1}}{
					\draw [line,dashed] (1,0.7) -- (1,1.3);
				}{};
			}{};
		};
	};
	\ifthenelse{\equal{#4}{1}}{
		\ifthenelse{\equal{#2}{1}}{
			\draw [curve,dashed] plot coordinates {(0,1) (0.7,1) (1,1.3)};
			\draw [curve,dashed] plot coordinates {(2,1) (1.3,1) (1,0.7)};
			}{
			\ifthenelse{\equal{#2}{2}}{
			\draw [curve] plot coordinates {(0,1) (0.7,1) (1,1.3)};
			\draw [curve] plot coordinates {(2,1) (1.3,1) (1,0.7)};
			}{};
		};
	}{
		\ifthenelse{\equal{#4}{0}}{
			\ifthenelse{\equal{#2}{1}}{
				\draw [curve,dashed] plot coordinates {(0,1) (0.7,1) (1,0.7)};
				\draw [curve,dashed] plot coordinates {(2,1) (1.3,1) (1,1.3)};
				}{
				\ifthenelse{\equal{#2}{2}}{
				\draw [curve] plot coordinates {(0,1) (0.7,1) (1,0.7)};
				\draw [curve] plot coordinates {(2,1) (1.3,1) (1,1.3)};
				}{};
			};
		};
	};
};
}

\usetikzlibrary{3d}
\def\dl{-135}
\def\dr{-45}
\def\ul{135}
\def\ur{45}

\title{The Higher Structure of Symmetries of Axion-Maxwell Theory}

\author[1,2,3]{\fnm{Michele} \sur{Del Zotto}}
\author[4]{\fnm{Matteo} \sur{Dell'Acqua}}
\author[1]{\fnm{Elias} \sur{Riedel Gårding}}

\affil[1]{\orgdiv{Department of Mathematics}, \orgname{Uppsala University}, \orgaddress{Box 480, \city{Uppsala}, \postcode{SE-75106}, \country{Sweden}}}
\affil[2]{\orgdiv{Center for Geometry and Physics}, \orgname{Uppsala University}, \orgaddress{Box 480, \city{Uppsala}, \postcode{SE-75106}, \country{Sweden}}}
\affil[3]{\orgdiv{Department of Physics and Astronomy}, \orgname{Uppsala University}, \orgaddress{Box 516, \city{Uppsala}, \postcode{SE-75120}, \country{Sweden}}}
\affil[4]{\orgdiv{Center for Cosmology and Particle Physics}, \orgname{New York University}, \orgaddress{\street{726 Broadway}, \city{New York}, \postcode{NY 10003}, \country{United States}}}

\abstract{Generalized symmetries of quantum field theories can be characterized by topological defects/operators organized into a higher category. In this paper we consider the  Axion-Maxwell field theory in four dimensions and,  building on the construction of its topological defects by  Choi, Lam, Shao, Hidaka, Nitta and Yokokura, we discuss field  theoretical methods to compute some aspects of the higher  structure of such category. In particular, we determine explicitly the generalized $F$-symbols for the non-invertible electric 1-form symmetry of the theory. Along the way, we clarify various aspects of the bottom-up worldvolume approach towards the calculus of defects.}

\begin{document}

\maketitle
\tableofcontents

\section{Introduction}
Global symmetries play a central role in the study of physical theories. In recent years a generalization of the concept of symmetry has been introduced \cite{Gaiotto_2015} (see \cite{Cordova:2022ruw,McGreevy:2022oyu,Freed:2022iao,Gomes:2023ahz,Schafer-Nameki:2023jdn,Brennan:2023mmt,Bhardwaj:2023kri,Shao:2023gho,Carqueville:2023jhb} for reviews), establishing the equivalence of symmetries with topological operators and defects of general (co)dimension. 
In particular, these new symmetries need not follow a group-like composition law, but may be non-invertible \cite{Bhardwaj:2017xup,Chang:2018iay,Komargodski:2020mxz,Nguyen:2021naa,Heidenreich:2021xpr,Thorngren:2021yso,Sharpe:2021srf, Huang:2021zvu, Choi:2021kmx, Kaidi:2021xfk, Roumpedakis:2022aik, Bhardwaj:2022yxj, Arias-Tamargo:2022nlf, Choi:2022zal, Kaidi:2022uux, Choi:2022jqy, Cordova:2022ieu, Antinucci:2022eat, Damia:2022bcd, Chang:2022hud, Choi:2022rfe, Bhardwaj:2022lsg, Bartsch:2022mpm, Niro:2022ctq, Chen:2022cyw, Karasik:2022kkq, Decoppet:2022dnz, GarciaEtxebarria:2022jky, Bhardwaj:2022kot, Bhardwaj:2022maz, Bartsch:2022ytj, Hsin:2022heo, Zhang:2023wlu, Choi:2023xjw, Anber:2023pny, Bhardwaj:2023ayw, Bartsch:2023wvv, Copetti:2023mcq, Decoppet:2023bay, Chen:2023czk, Sun:2023xxv, Cordova:2023bja, Antinucci:2023ezl, Choi:2023vgk, Diatlyk:2023fwf, Nagoya:2023zky, Perez-Lona:2023djo, Sela:2024okz, DZGH}. Over the last decade, generalized symmetries have found diverse applications in a wide range of quantum field theories
\cite{Gaiotto:2017yup, Hsin:2018vcg,Garcia-Etxebarria:2018ajm, Buican:2021uyp, Hayashi:2022oxp, Kaya:2022edp, Cordova:2022rer, DelZotto:2022ras, Hayashi:2022fkw, Bhardwaj:2022dyt, Lin:2022xod, Cordova:2022fhg, Choi:2022fgx, Yokokura:2022alv, Giaccari:2022xgs, Cordova:2022qtz,  Apte:2022xtu, Lin:2023uvm, Putrov:2023jqi, Damia:2023ses, Argurio:2023lwl, vanBeest:2023dbu, vanBeest:2023mbs, Cordova:2023ent, Pace:2023kyi, Choi:2023pdp, Cordova:2023her, Damia:2023gtc, Bhardwaj:2023idu, Bhardwaj:2023fca, Brennan:2023kpo, Brennan:2023vsa, Cordova:2023qei, Bhardwaj:2023bbf, Brennan:2023ynm, Dumitrescu:2023hbe, Cordova:2023jip, Cordova:2024ypu, Honda:2024yte, Heckman:2024oot, Antinucci:2024zjp, Seiberg:2024gek, Aloni:2024jpb, Heckman:2024obe, Giacomelli:2024sex, Benedetti:2024kfp, Bonetti:2024cjk, Cordova:2024ypu, Apruzzi:2024htg, DelZotto:2024tae, Armas:2024caa, Santilli:2024dyz, Furlan:2024pvy, Copetti:2024rqj, Cordova:2024vsq, Hsin:2024eyg, Nardoni:2024sos, Honda:2024sdz, Seifnashri:2024dsd, Argurio:2024oym, Ambrosino:2024ggh, Kan:2024fuu, Putrov:2024uor, Bhardwaj:2024wlr, Bhardwaj:2024kvy, Arbalestrier:2024oqg, Honda:2024xmk, Bhardwaj:2024ydc, Choi:2024rjm, Platschorre:2024xxp, Hasan:2024aow, Cappelli:2024elj, Seiberg:2024wgj, Heckman:2024zdo, Reece:2024wrn, Cui:2024cav, Anber:2024gis, Cordova:2024jlk, Benini:2024xjv, Garcia-Valdecasas:2024cqn, Antinucci:2024bcm, Copetti:2024onh, Antinucci:2024ltv, Delcamp:2024cfp, DelZotto:2024arv, Cordova:2024iti, Copetti:2024dcz, Bharadwaj:2024gpj, Choi:2024tri, Dierigl:2024cxm, GarciaEtxebarria:2024jfv, Choi:2024wfm, Argurio:2024ewp, Bolognesi:2024bnm, Dumitrescu:2024jko, Bottini:2024eyv, Bartsch:2024ech, Hidaka:2024kfx, Li:2024nuo}. A new feature of these generalized symmetries is that topological defects can form junctions and networks of various codimension that give the symmetry a \textit{higher structure}. A central question is how to compute the latter for a given field theory. The main aim of this paper is to perform a first step in computing the higher structure of Axion-Maxwell theory in four-dimensions, building on seminal work by Choi, Lam, Shao and Hidaka, Nitta, Yokokura \cite{Choi2022noninvertible,Yokokura:2022alv,Hidaka:2019jtv,Hidaka:2020iaz,Hidaka:2020izy,Hidaka:2021kkf,Hidaka:2021mml,Hidaka:2024kfx} and generalizing previous results about massless QED in four-dimensions \cite{Copetti:2023mcq}.

\medskip

What is the mathematical structure encoding all the physical information present in the topological sector of the (extended) operators of a given QFT $\tT$? It is believed the latter is encoded in a higher tensor category $\cC$ (often called the \textit{symmetry category}).\footnote{\ The latter is natural extension of the concept of a group, viewed as category with one object and morphisms labeled by group elements that are all invertible. From this perspective, the objects of $\cC$ as given by $D$-dimensional field theories $\tT$ and the morphisms are topological interfaces between them (while higher morphisms are interfaces of interfaces of interfaces... and so on). In this paper, we will instead focus on the symmetry category of a given field theory $\tT$, which is $\cC_\tT := \text{End}_\cC(\tT)$, a higher tensor $(D-1)$-category (generically the finiteness requirement typical of a fusion category may fail, e.g. in the symmetry category considered in this paper, which includes $\mathbb{Q}/\Z$ symmetries \cite{Putrov:2022pua}). In particular, the objects of the latter are the codimension $1$ topological operators of $\tT$. In the general setting, these defect can support TQFT on their own, with a spectrum of topological defects and topological interfaces, which we denote with $\operatorname{End}_\cC(\dD)$ and $\operatorname{Hom}_\cC(\dD,\dD')$ respectively. } 
A crucial aspect of the symmetry category is that topological defects can have junctions and intricate topological interfaces corresponding to higher morphisms. For a review we refer our readers to Section 2 of \cite{Copetti:2023mcq}. Not only the topological defects, but also these topological interfaces can have a non-trivial action on the non-topological defects of $\tT$. Many known properties of symmetries, such as for instance 't Hooft anomalies or fractionalization of various kinds are encoded in the higher structure of the symmetry category.\footnote{\ Higher associators and 't~Hooft anomalies are deeply interconnected; see e.g.~\cite{Copetti:2023mcq}.} In particular, the whole higher structure is conserved under duality and symmetry preserving renormalization group (RG) flows. 
This is a key motivation to develop techniques to compute the higher structure of symmetries.

\medskip

 Currently, there are two main approaches to generalized symmetries, which we call the quiche (or SymTFT) approach \cite{Ji:2019jhk, Gaiotto:2020iye, Apruzzi:2021nmk, Freed:2022iao, Kaidi:2022cpf, Freed:2022qnc, Bhardwaj:2023wzd, Baume:2023kkf, Kaidi:2023maf, Bhardwaj:2023fca, Cvetic:2024dzu, Brennan:2024fgj, Antinucci:2024zjp, Bonetti:2024cjk, Nardoni:2024sos} and the worldvolume approach \cite{Aharony:2013hda, Kapustin:2013uxa, Kapustin:2014gua, Kapustin:2014zva, Yoshida:2015cia, Thorngren:2015gtw, Tachikawa:2017gyf, Cordova:2018cvg, Benini:2018reh, Rudelius:2020orz, Hidaka:2020iaz, Hsin:2020nts, Cordova:2020tij, BenettiGenolini:2020doj, Hidaka:2020izy, Brennan:2020ehu, Heidenreich:2020pkc, Hidaka_2020, Hidaka_2021, Hidaka:2021kkf, Kaidi:2021xfk, Apruzzi:2021mlh, Bhardwaj:2021wif, Genolini:2022mpi, Damia:2022seq, Delmastro:2022pfo, Choi2022noninvertible}. The quiche approach relies on the existence of an isomorphism of the field theory of interest with a bulk/boundary system consisting of a TQFT in one higher dimension on a slab with two boundaries, one topological the other a relative theory. The pair of the bulk TQFT and the topological boundary is the quiche, and it encodes all the features of the topological defects of the theory of interest in the well-developed calculus of topological defects in TQFTs. However, this construction relies on the existence of such an isomorphism, which for the most general symmetries is not yet completely understood.\footnote{\ A particularly intriguing open problem is given by continuous global symmetries. For some recent progress in this direction see e.g.~\cite{Antinucci:2024zjp,Brennan:2024fgj,Bonetti:2024cjk,Arbalestrier:2024oqg,Hasan:2024aow}.} Oftentimes, such an isomorphism is not known and one needs to work directly with the QFT of interest to compute the higher structure of its symmetries. This is the context of the worldvolume approach, where the topological defects are constructed using field theoretical techniques without reference to an isomorphism with a higher dimensional system. This paper is in the context of the worldvolume approach to the higher structure. Our focus is to study the higher structure of the symmetries of the 4d Axion-Maxwell theory, whose rich symmetry structure makes it a fruitful setting for case studies of generalized symmetries \cite{Hidaka_2020,Hidaka_2021,brennan2021axions,yokokura2022noninvertible,Choi2023noninvertible,Choi:2023pdp,Hidaka:2024kfx}.

\medskip

Of particular interest are the higher associators, topological defects encoding the associativity property of the fusion product. For 2d theories, the topological codimension one defects are lines, the fusion interfaces are points and the associators (of three topological lines) are numbers (the so called $F$-symbols) satisfying a pentagon identity (which follows from requesting the compatibility of associativity with the fusion of four objects). For higher dimensional field theories, topological codimension $(p+1)$ defects, have fusion interfaces of codimension $(p+2)$, and corresponding associators of $n$ objects of codimenision $p+n$. Higher associators and obstruction to gauging ('t~Hooft anomalies) are deeply interconnected \cite{Cordova:2019bsd,Apte:2022xtu,Antinucci:2023ezl,Sun:2023xxv,Cordova:2023bja}. 




\medskip

A first step towards determining the higher structure of chiral symmetry of $(3+1)d$ massless QED have been obtained in \cite{Copetti:2023mcq}. Among other results, that paper determined the fusion rules, fusion interfaces and F-symbols of the symmetry category, which are 2d TQFTs. The main result of this paper is an analogous analysis for the symmetry of $(3+1)d$ Axion-Maxwell theory. In particular we determine the fusion rules and interfaces as well as the F-symbol 1d TQFTs controlling the associativity of the non-invertible electric 1-form symmetry of this model. We show that these have a non-trivial action on extended operators of the theory and that these satisfy stringent consistency conditions. The symmetry category of Axion-Maxwell theory is richer than the one of massless QED. For the 1-form symmetry, the resulting TQFT coefficients in particular have decomposition (see e.g. \cite{sharpe2022introduction}). Moreover, the 1d F-symbol TQFTs have more symmetries than the ones enjoyed by the 2d F-symbol TQFTs of the 0-form symmetries, which leads to a collection of identities that they have to satisfy.\footnote{\, Neglecting, for simplicity, a possible interplay with the tangential structure.}

\medskip

This work is organized as follows. In \cref{sec: Genueine-top,sec: Top-Junctions,sec: non-top-junction,sec: Noninvertible-actions} we review the previously known data of the symmetry category and charges of $(3+1)d$ Axion Electrodynamics.
In \cref{sec: fusion} we derive the fusion rules and interfaces of the noninvertible electric 1-form symmetry.  Their associator is described in \cref{sec: associator}, where we also check the consistency and analyze the consequences of our results with other previously known structures of the theory.

\medskip

In order to make the work self-contained, we review some mathematical results on the anomalies of invertible symmetries and on generalized gauging in $2$ and $3$ dimensions respectively in \cref{Appendix: cohomology,Appendix: generalized gauging}. Finally, in \cref{Appendix: minimal theories} we review the notion of \textit{minimal theories} and we presented an alternative derivation of the fusion rules of chiral defects in QED, starting from a Lagrangian description.


\medskip

\paragraph{Conventions} For uniformity reasons, we use the following conventions:
\begin{itemize}[nosep]
    \item Dynamical fields are $2\pi$ periodic and their field strengths have periods in $2\pi \Z$.
    \item Background fields are $1$ periodic (i.e. valued in $\mathbb{R}/\Z=U(1)$) and their field strengths have periods in $\Z$.
    \item Symmetry defects' labels are defined mod $\mathbb Z$ 
\end{itemize}

\hfill

\section{Genuine topological defects}\label{sec: Genueine-top}

To fix notations and conventions let us recall that the Euclidean action of the Axion-Maxwell theory is given by:
\begin{equation}
    S = \frac{f^2}{2}\int \dd\theta\wedge *\dd\theta +\frac{1}{2e^2}\int F\wedge *F -i\frac{K}{8\pi^2}\int\theta F\wedge F,
\end{equation}
where $F = \dd{A}$ is the field strength for a $1$-form $U(1)$ gauge field and $\theta$ a $0$-form gauge field, i.e. a compact scalar\footnote{For a more formal discussion on $p$-form gauge fields see \cite{hsieh2022anomaly, Apruzzi_2023}} $\theta \sim \theta + 2\pi$. The two constants $f$ and $e$ are respectively the axion decay constant electric charge and they are unconstrained, while the axion--photon coupling $K$ is an integer (on spin manifolds) in order for the interaction term to be gauge invariant. Neither $A$ nor $\theta$ are gauge invariant, however we can define gauge invariant operators either considering their field strengths, i.e.~$F$ and $d\theta$ together with their Hodge duals, or by considering their holonomies. Hence, the gauge-invariant operators of the theory are given by Wilson lines $W_q = e^{2\pi iq \int_\gamma A}$ and 't Hooft lines $H_m$, as well as the axion $\Theta_n = e^{in\theta}$ and the string worldsheet $S_w$, analogue of the 't Hooft line for $\theta$. 
Due to the axion-photon coupling, the definition of 't Hooft lines and string worldsheets differs slightly from the naive one in terms of boundary conditions for the gauge fields $A$ and $\phi$. We will expand on this remark in \cref{sec: non-top-junction}.

\medskip

Let us recall what happens at trivial coupling $K = 0$. With this choice of coupling, the theory describes two non-interacting free gauge theories: a free Maxwell subtheory with a $U(1)^{(1)}_e \times U(1)^{(1)}_m$ 1-form symmetry (with mixed anomaly) computing the electric and magnetic charge of dyonic lines, and the theory of a free compact boson with a $U(1)^{(0)}_s$ symmetry which acts by shifting the axion field and a $U(1)^{(2)}_w$ measuring the winding number:
\begin{equation}
  \theta\to \theta + 2\pi \Lambda \quad \quad w = \oint_\gamma \frac{\dd\theta}{2\pi}
\end{equation}
again with a mixed anomaly. Let us note that both these symmetries are non linearly realized and thus they are spontaneously broken, we can thus interpret the massless degrees of freedom, i.e. the photon and the axion, as Goldstone bosons \cite{Gaiotto_2015}.

\medskip

We start the analysis of the symmetry category by looking at the genuine defects, i.e.~(higher) endomorphisms of the identity.

\subsection{Invertible defects}
As explained in \cite{Gaiotto_2015}, a sufficient condition for the existence of a global $U(1)^{(p)}$ symmetry is the presence of a conserved $(D-p-1)$-form current. From the previous analysis, we know that the case with $K=0$ has the following conserved currents:
\begin{equation}\label{eq: currents}
  j_s^{(3)} = if^2*\dd{\theta}, \quad
  j_w^{(1)} = \frac{\dd{\theta}}{2\pi}, \quad
  j_e^{(2)} = \frac{-i}{e^2}*F, \quad
  j_m^{(2)} = \frac{F}{2\pi}.
\end{equation}
When $K\neq0$, however, not all of them are conserved anymore: from the equations of motions we have
\begin{equation}\label{eq: conservation}
  \dd{j_s^{(3)}} = \frac{K}{2} j_m^{(2)} \wedge j_m^{(2)}, \quad
  \dd{j_w^{(1)}} = 0, \quad
  \dd{j_e^{(2)}} = K j_w^{(1)} \wedge j_m^{(2)}, \quad
  \dd{j_m^{(2)}} = 0.
\end{equation}
Because of the nontrivial coupling $K$, the electric 1-form and the shift 0-form symmetry are broken to a $\Z_K$ subgroup. However, the $U(1)^{(2)}_w$ winding symmetry and an $U(1)^{(1)}_m$ magnetic one are still conserved:
\begin{equation}
    \eta_{\alpha}^{(\mathrm{m})}(\Sigma^{(2)})\equiv\exp\left(2\pi i\alpha\oint_{\Sigma^{(2)}}j_m^{(2)}\right), \quad\eta_{\alpha}^{(\mathrm{w})}(\Sigma^{(1)})\equiv\exp\left(2\pi i\alpha\oint_{\Sigma^{(1)}}j_w^{(1)}\right).
\end{equation}
Their action on charged objects is given by
\begin{equation*}
  \eta_\alpha^{(m)}:H_m\mapsto\exp(2\pi i\alpha m)H_m, \qquad
  \eta_\alpha^{(w)}:S_w\mapsto\exp(2\pi i\alpha m)S_w.
\end{equation*}

Concerning the unbroken $\Z_K$ shift and electric symmetries, we can define the following operators:
\begin{equation}
  \label{eq: tilde-invertible-definition}
  \begin{aligned}
    \tilde{\eta}_{j}^{(\mathrm{s})}(\Sigma^{(3)})
    &\equiv \exp[2\pi i\frac{j}{K}
      \oint_{\Sigma^{(3)}} \qty(j^{(3)}_w - \frac{K}{8\pi^2} A\wedge F)], \\
    \tilde{\eta}_{j}^{(\mathrm{e})}(\Sigma^{(2)})
    &\equiv \exp[2\pi i\frac{j}{K}
      \oint_{\Sigma^{(2)}} \qty(j^{(2)}_e - \frac{K}{4\pi^2}\theta F)],
  \end{aligned}
\end{equation}
where the equation of motion \cref{eq: conservation} implies that the integrand is formally closed; the integrality condition on the coefficient $j\in\Z$ is needed to ensure gauge invariance of the nonlinear term and the periodicity $j\sim j+K$ follows from the action on charged objects:
\begin{equation}
    \begin{aligned}
      \tilde{\eta}_{j}^{(s)}:
      &\quad \Theta_n \mapsto \exp(\frac{2\pi inj}{K}) \Theta_n,\\
        \tilde{\eta}_{j}^{(e)}:&\quad W_q\mapsto\exp(\frac{2\pi iqj}{K})W_q.
    \end{aligned}
\end{equation}
The extra terms, besides the currents, used to define the operators $\tilde{\eta}_{K}^{(s)}, \tilde{\eta}_{K}^{(e)}$ leads to non trivial action of these ones on other operators. For example, there is a non-trivial interplay between the winding symmetry and 't Hooft lines, since the term $A\wedge F$ term is sensible to them. This kind of structure is known as an higher-group structure as we will show explicitly by looking at the relation among the background fields of the above symmetries.
 
To see the higher group structure of the theory we can (minimally) couple it to a background for all the previously described symmetries \cite{brennan2021axions,Hidaka_2021}:
\begin{equation}\label{Coupled}
  \begin{aligned}
    S &= \frac{f^2}{2}\int_X
        (\dd\theta - 2\pi B_s^{(1)}) \wedge *(\dd\theta-2\pi B_s^{(1)})\\
      &{} + \frac{1}{2e^2}\int_X(F - 2\pi B_e^{(2)})\wedge*(F - 2\pi B_e^{(2)}) \\
      &{} + i\int_X\theta G_w^{(4)} - i\int_X H_m^{(3)}\wedge A \\
      &{} - \dfrac{iK}{8\pi^2}\int_Y
        (\dd\theta - 2\pi B_s^{(1)}) \wedge (F - 2\pi B_e^{(2)})
        \wedge(F - 2\pi B_e^{(2)}),
  \end{aligned}
\end{equation}
where $Y$ is a $5$-dimensional open manifold that bounds the spacetime $X$ and $B_s^{(1)}$ and $B_e^{(2)}$ are background gauge fields quantized so that $\oint B_s^{(1)}, \oint B_e^{(2)} \in \frac{1}{K}\Z$. The field strengths $G^{(4)}_w$ and $H^{(3)}_m$ are defined as
\begin{equation}\label{eq:3group background}
  G_w^{(4)} = \dd B_w^{(3)} + \frac{K}{2} B_e^{(2)} \wedge B_e^{(2)}, \qquad
  H_m^{(3)} = \dd B_m^{(2)} + K B_s^{(1)} \wedge B_e^{(2)}
\end{equation}
to ensure that the 4d path integrand $e^{-S}$ is independent of the choice of $Y$, up to an anomaly term $\exp(2\pi i \frac{K}{2} \int_Y B_s^{(1)} \wedge B_e^{(2)} \wedge B_e^{(2)})$ which depends only on background fields and $Y$. As anticipated, the nonlinear terms in the definition are the hallmark of a nontrivial higher-group structure of the symmetry category. We will explore the consequences of \cref{eq:3group background} in \cref{sec: non-top-junction}.

One may now check that \cref{Coupled} is invariant under modified gauge transformations of the form:
\begin{equation}
  \begin{aligned}
    &\theta \longmapsto \theta + 2\pi \Lambda_s^{(0)}, \quad
      B_s^{(1)} \longmapsto B_s^{(1)} + \dd\Lambda_s^{(0)}, \\
    &A \longmapsto A + 2\pi \Lambda_e^{(0)} \quad
      B_e^{(2)} \longmapsto B_e^{(2)} + \dd\Lambda_e^{(1)}, \\
    &B_w^{(3)} \longmapsto B_w^{(3)} + \dd\Lambda_w^{(2)} - K \qty(B_{e}^{(2)} \wedge \Lambda_{e}^{(1)} + \frac{1}{2}\Lambda_{e}^{(1)} \wedge \dd\Lambda_{e}^{(1)}) , \\
    &B_{m}^{(2)} \longmapsto B_{m}^{(2)} + \dd\Lambda_{m}^{(1)} - K \qty(\Lambda_s^{(0)} B_{e}^{(2)} - B_s^{(1)} \wedge \Lambda_{e}^{(1)} - \dd\Lambda_s^{(0)} \wedge \Lambda_{e}^{(1)}).
    \end{aligned}
\end{equation}
The mixed gauge transformations of $B_{m}^{(2)}$ and $B_w^{(3)}$ are another sign of the higher group structure.

The inflow action can be derived to be \cite{brennan2021axions}:
\begin{equation}
  S_{\text{inflow}} = 2\pi i \int_Y \qty(
  G_w^{(4)} \wedge B_s^{(1)} + H_m^{(3)} \wedge B_e^{(2)}),
\end{equation}
which can be interpreted as the improved version of the mixed $U(1)^{(0)}_s \times U(1)^{(2)}_w$ anomaly of the free compact boson and mixed $U(1)_e^{(1)}\times U(1)^{(1)}_m$ anomaly of the free photon at $K = 0$.

Finally, one could also consider a further conserved current given by the Chern--Weil term
\begin{equation}
  \label{eq: chern-weil}
  j_{CW}^{(3)} = j_w^{(1)} \wedge j_m^{(2)} = \frac{d\theta}{2\pi} \wedge \frac{F}{2\pi};
  \quad \dd{j_{CW}^{(3)}} = 0,
\end{equation}
and define the topological operators:
\begin{equation}
  \eta_{\alpha}^{(CW)}(\Sigma^{(3)}) \equiv
  \exp\left(2\pi i\alpha\oint_{\Sigma^{(3)}} j_{CW}^{(3)} \right).
\end{equation}
However, these defects are trivial.
Indeed, the equation of motion for the electric current \eqref{eq: conservation} implies that, given a $3$-dimensional closed manifold we have
\begin{equation}
  \exp\left(2\pi i\alpha
    \int_{\Sigma^{(3)}}\frac{\dd{\theta}}{2\pi}\wedge \frac{F}{2\pi}\right)
  = \exp\left(\frac{2\pi }{e^2} \frac{\alpha}{K}\oint_{\Sigma^{(3)}} \dd *F\right)= 1,
\end{equation}
using Stokes' theorem. Another, more formal, way to rephrase this fact is that there exists an invertible morphism\footnote{Given an $n$-category $\mathcal{C}$ we denote with $\Omega\mathcal{C}$ its loop category, i.e. the $n-1$-category of the endomorphism of the identty $\Omega\mathcal{C}=\operatorname{End}_{\mathcal{C}}(\mathds{1}_{\mathcal{C}})$}:
\begin{equation}\label{eq:nonfaithful CW}
    \eta_{\alpha}^{(e)}\in \operatorname{Hom}_{\Omega \mathcal{C}}(\mathds{1}^{(0)}; \eta^{(CW)}_\alpha).
\end{equation}

\subsection{Condensation defects}\label{sec: condensation}
Given the invertible symmetry operators, a further class of topological operators is given by condensation defects, obtained by higher gauging of discrete subgroups \cite{Roumpedakis_2023,Choi_2023}; in this case $\Z_N$:
\begin{equation}
  \begin{aligned}
    &\mathcal{C}_{N,\alpha}^{(0,m)}(\Sigma^{(3)})=\sum_{a\in H^{1}(\Sigma^{(3)},\mathbb{Z}_{N})}\exp\left(\frac{2\pi i}{N}\int_{\Sigma^{(3)}} a^{*}\alpha+\frac{2\pi i}{N}\int_{\Sigma^{(3)}} a\cup[ j^{(2)}_m]\right) , \\
    &\mathcal{C}_{N}^{(1,m)}(\Sigma^{(2)})=\sum_{a\in H^{0}(\Sigma^{(2)},\mathbb{Z}_{N})}\exp\left(\frac{2\pi i}{N}\int_{\Sigma^{(2)}}a\cup[ j^{(2)}_m]\right), \\
    &\mathcal{C}_{N}^{(0, w)}(\Sigma^{(3)})=\sum_{\phi\in H^{2}(\Sigma^{(3)},\mathbb{Z}_{N})}\exp\left(\frac{2\pi i}{N}\int_{\Sigma^{(3)}}\phi\cup[ j^{(1)}_w]\right) , \\
    &\mathcal{C}_{N}^{(1, w)}(\Sigma^{(2)})=\sum_{\phi\in H^{1}(\Sigma^{(2)},\mathbb{Z}_{N})}\exp\left(\frac{2\pi i}{N}\int_{\Sigma^{(2)}}\phi\cup[ j^{(1)}_w]\right) ,\\
    &\mathcal{C}_{N}^{(2, w)}(\Sigma^{(1)})=\sum_{\phi\in H^{0}(\Sigma^{(1)},\mathbb{Z}_{N})}\exp\left(\frac{2\pi i}{N}\int_{\Sigma^{(1)}}\phi\cup[ j^{(1)}_w]\right) .
  \end{aligned}
\end{equation}
possibly labelled by\footnote{In \cref{Appendix: cohomology} we discuss the relevant cohomology groups.} $\alpha\in H^3(B\Z_N; U(1))=\Z_N$.
Moreover, one should include simultaneous condensations:
\begin{equation}
  \begin{aligned}
    \mathcal{C}_{N,M,\alpha,\beta}^{(0,m,w)}(\Sigma^{(3)})
    = \sum_{\substack{a\in H^{1}(\Sigma^{(3)},\Z_N) \\ \phi\in H^{2}(\Sigma^{(3)},\mathbb{Z}_{M})}}
    & \exp\left(\frac{2\pi i}{N}\int_{\Sigma^{(3)}} a^{*}\alpha+\frac{2\pi i}{\gcd(N,M)} \int_{\Sigma^{(3)}} (a \times \phi)^{*}\beta\right) \\
    &\times\exp\left(\frac{2\pi i}{N}\int_{\Sigma^{(3)}} a\cup[ j^{(2)}_m]+\frac{2\pi i}{M}\int_{\Sigma^{(3)}}\phi\cup[ j^{(1)}_w]\right) , \\
    \mathcal{C}_{N,M}^{(1,m,w)}(\Sigma^{(2)})
    = \sum_{\substack{a\in H^{0}(\Sigma^{(2)},\Z_N)\\ \phi\in H^{1}(\Sigma^{(2)},\mathbb{Z}_{M})}}
    &\exp\left(\frac{2\pi i}{N}\int_{\Sigma^{(2)}}a\cup[ j^{(2)}_m]+\frac{2\pi i}{M}\int_{\Sigma^{(2)}}\phi\cup[ j^{(1)}_w]\right),
  \end{aligned}
\end{equation}
with a possible choice of discrete torsion $\beta \in \operatorname{Tor}^{\Z}(H^2(B\Z_N; \Z);H^3(B^2\Z_M; \Z))=\Z_{\operatorname{gcd}(N,M)}\subset H^3(B\Z_N\times B^2\Z_M; U(1))$\footnote{In the first line, we regarded $a$ and $\psi$ respectively as maps: $$a:\Sigma^{(3)}\to B\Z_N, \quad \phi:\Sigma^{(3)}\to B^2\Z_2$$ and we denoted with $a\times \phi$ the cartesian product of these maps (with target the cartesian product of the targets):
$$a\times \phi:\Sigma^{(3)}\to B\Z_N\times B^2\Z_M.$$}.
However, a simultaneous condensation of a subgroup of both symmetries (without mixing discrete torsion) factorizes as two independent condensations of the two projections:
\begin{equation}
    \mathcal{C}_{N, M, \alpha,0}^{(0,m, w)}=\mathcal{C}_{N,\alpha}^{(0,m)}\mathcal{C}_{M}^{(0,w)},\quad\quad \mathcal{C}_{N,M}^{(1,m,w)}=\mathcal{C}_{N}^{(1,m)}\mathcal{C}_{M}^{(1,w)},
\end{equation}
since the winding and magnetic symmetry don't interact with each other (i.e.~their subcategory is that of a trivial $2$-group) even at a nontrivial coupling. Meanwhile, the other condensates do not factorize in the same way, because the electric and shift symmetries interact in a nontrivial way, as we will explain in more detail later.

On top of that, in \cite{Choi2023noninvertible} it was shown that the axion--Maxwell theory is self-dual under 1-gauging of the $\Z_N^{(1)} \times \Z_N^{(2)} \subset U(1)^{(1)}_m \times U(1)^{(2)}_w$ subgroup with any choice of $\beta$, in the sense that
\begin{equation}
  \label{eq:self-duality-mixed}
  \mathcal{C}_{N,N,0,\beta}^{(0,m,w)} \sim \mathds{1}.
\end{equation}
This means that there exists an
invertible topological interface between it and the transparent defect $\mathds{1}$, or that $\mathcal{C}_{N,N,0,\beta}^{(0,m,w)}(\Sigma^{(3)})$ is trivial when $\Sigma^{(3)}$ is closed.


\subsection{Non-invertible defects}
We exhausted the clearly conserved currents of the theory. However, there are cases in which it is still possible to define topological operators starting from non conserved currents. From now on, we will work with $K=1$ (unless explicitly specified otherwise).

The conservation \eqref{eq: conservation} of the shift current is spoilt by a term analogous to that of the chiral current in massless QED: $\dd{j_s} = \frac{1}{2} j_m^{(2)} \wedge j_m^{(2)}$. We can thus recover a non-invertible $(\mathbb{Q}/\Z)^{(s)}$ 0-form symmetry by defining the following topological defects \cite{Choi2022noninvertible,cordova2022noninvertible}:
\begin{equation}\label{eq: shift-definition}
  \mathcal{D}^{(s)}_{p/N} \equiv \eta^{(s)}_{p/N} \otimes \mathcal{A}^{N,p} \qty\big[j_m^{(2)}]
  = \exp(2\pi i \frac{p}{N}\oint j_s^{(3)}) \mathcal{A}^{N,p} \qty\big[j_m^{(2)}].
\end{equation}
Here $\mathcal{A}^{N,p}$ is the minimal 3d theory \cite{Hsin_2019} supporting a $\mathbb{Z}_N$ 1-form symmetry with anomaly given in terms of the background\footnote{Here, again, we normalize $B^{(2)}$ according to $\oint B^{(2)} \in \frac{1}{N} \Z$. Then $b^{(2)} \equiv NB^{(2)}$ obeys the other common normalization: $\oint b^{(2)} \in \Z$. This is the difference between expressing $\Z_N$ as $\qty(\frac{1}{N} \Z) / \Z \subset \mathbb{R}/\Z$ or as $\Z / N\Z$. We will usually prefer the latter.} $b^{(2)} \equiv N B^{(2)}$ by the inflow action
\begin{equation}
  S_{\text{inflow}}[b^{(2)}] =
  -\frac{2\pi i p}{N} \int_{X_4} \frac{1}{2} b^{(2)} \wedge b^{(2)},
\end{equation}
and the coupling to the 4d bulk is achieved by setting $b^{(2)} = j_m^{(2)}$; this ensures that $\mathcal{D}^{(s)}_{p/N}$ is topological More details on the $\mathcal{A}^{N,p}$ theory are found in \cref{app:3d}.

With a similar process, one can also recover a noninvertible $(\mathbb{Q}/\Z)^{(e)}$ 1-form symmetry.\footnote{It seems likely that one can relate the constructions of the 0-form and 1-form noninvertible symmetries by dimensional reduction \cite{Nardoni:2024sos}, but we do not pursue this direction here.} In fact, while the electric current is not conserved, \eqref{eq: conservation} states that its exterior derivative is a function of conserved currents: $\dd{j_e^{(2)}} = j_w^{(1)} \wedge j_m^{(2)}$.
One can thus stack the naive electric defect $\eta_\alpha^{(e)}$ with a two-dimensional anomalous TQFT
whose inflow action formally matches $\alpha \dd j^{(2)}_e$:
\begin{equation}
  \label{eq:A2-inflow}
  S_{\text{inflow}}[b^{(1)}, b^{(2)}] =
  -\frac{2\pi i p}{N} \int_{X_3} b^{(1)} \wedge b^{(2)},
\end{equation}
and couple it to the bulk magnetic and winding currents by setting $b^{(1)} = j_w^{(1)}$ and $b^{(2)} = j_m^{(2)}$. For a rational angle $\alpha = \frac{p}{N}$ and backgrounds for $\Z_N^{(i-1)}$ symmetries (with mixed anomaly) $b^{(i)}$, there is a canonical choice of such 2d a TQFT, which we call $\mathcal{A}_2^{N,p}$. It admits a Lagrangian description \cite{Choi2023noninvertible} $\mathcal{A}_2^{N,p}[X_2, b^{(1)}, b^{(2)}] = \int [D\phi\,Dc] e^{-S_{\mathcal{A}_2^{N,p}}[X_2, b^{(1)}, b^{(2)}]}$
 in terms of a $2\pi$-periodic scalar $\phi$ and a 1-form gauge field $c$:
\begin{equation}
  \label{eq:A2-lagrangian}
  S_{\mathcal{A}_2^{N,p}}[X_2, b^{(1)}, b^{(2)}]
  = -i \int_{X_2} \qty(
  N\phi \frac{\dd{c}}{2\pi}
  + x \phi b^{(2)} + y c \wedge b^{(1)})
\end{equation}
where $x, y \in \Z_N$ are any integers such that $xy = p \bmod{N}$ (this ensures that integrating out the fields leaves $\mathcal{A}_2^{N,p}[\partial X_3, b^{(1)}, b^{(2)}] = e^{+S_\text{inflow}[b^{(1)}, b^{(2)}]}$). For concreteness, we may choose $x = 1, y = p$ as in \cite{Choi2023noninvertible}.

We can therefore define the noninvertible electric defect as:
\begin{equation}\label{eq: electric-definition}
  \mathcal{D}^{(e)}_{p/N} \equiv \eta^{(e)}_{p/N} \otimes \mathcal{A}_2^{N,p} \qty\big[j_w^{(1)}; j_m^{(2)}] =
  \exp(\frac{2\pi}{e^2} \frac{p}{N} \oint*F) \mathcal{A}_2^{N,p} \qty\big[j_w^{(1)},j_m^{(2)}];
\end{equation}
it is topological because $2\pi i \frac{p}{N} \int_{X_3} \dd{j_e^{(2)}} = -S_\text{inflow}\qty\big[j_w^{(1)}, j_m^{(2)}]$; the electric symmetry defect $\eta^{(e)}_{p/N}$ cancels the anomaly of $\mathcal{A}_2^{N,p}$.

Finally, there are other arguments to arrive at the same definitions of noninvertible shift and electric defects. For example, both can be seen as half (higher) duality defects \cite{Choi2023noninvertible}; we will make use of this in \cref{sec: Noninvertible-actions}.

All the defects defined so far are built from the corresponding naive defect of the trivially coupled theory up to factors involving currents for \textit{other} symmetries. Thus, their original action is still present:
\begin{equation}
  \begin{aligned}
    \eta_\alpha^{(m)}\colon H_m \mapsto \exp(2\pi i\alpha m)H_m, \quad &\eta_\alpha^{(w)}\colon S_w \mapsto \exp(2\pi i\alpha m)S_w,\\
    \mathcal{D}_{p/N}^{(s)}\colon \exp(i\theta) \mapsto \exp(\frac{2\pi ip}{N})\exp(i\theta), \quad &\mathcal{D}_{p/N}^{(e)}\colon W_q \mapsto \exp(\frac{2\pi ipq}{N})W_q.
  \end{aligned}
\end{equation}
Obviously, the non-invertible defects have additional non-invertible action. In order to derive them, we first need to explore their topological junctions, as we will do in the following paragraph.

\section{Topological junctions}\label{sec: Top-Junctions}
\subsection{Invertible symmetries}
The invertible magnetic and winding symmetries enjoy no nontrivial higher morphisms. More formally:
\begin{equation}
  \begin{aligned}
    \mathrm{Hom}_{\Omega^2\mathcal{C}}(\eta_\alpha^{(m)} \otimes \eta_\beta^{(m)}; \eta_\gamma^{(m)}) =
    \boldsymbol{\delta}_{\alpha+\beta}^{\gamma} L_{\alpha,\beta}^{\gamma} ,\quad &L_{\alpha,\beta}^{\gamma} \simeq \mathrm{Vec}, \\
    \mathrm{Hom}_{\Omega^3\mathcal{C}}(\eta_\alpha^{(w)} \otimes \eta_\beta^{(w)}; \eta_\gamma^{(w)}) =
    \boldsymbol{\delta}_{\alpha+\beta}^{\gamma} m_{\alpha,\beta}^{\gamma} ,\quad &m_{\alpha,\beta}^{\gamma} \simeq \mathbb{C}.
  \end{aligned}
\end{equation}
This means that there is no interface between defects labelled by different group elements; meanwhile, there is always the freedom to dress a submanifold of the defect with a decoupled TQFT of the right dimension.


\subsection{Noninvertible symmetries}
The definitions of both classes of noninvertible defects involve a minimal theory coupled to a current for a bulk symmetry. As we discussed in the previous chapter, this allows for symmetry defects of the bulk symmetry to end topologically on the defects of the minimal theory. More precisely, the defects of the coupled minimal theory will be topological \textit{only if} attached to a bulk defect of one higher dimension supported on a bounding hypersurface.
\begin{figure}[ht!]
  \centering
  \includegraphics[width=0.8\linewidth]{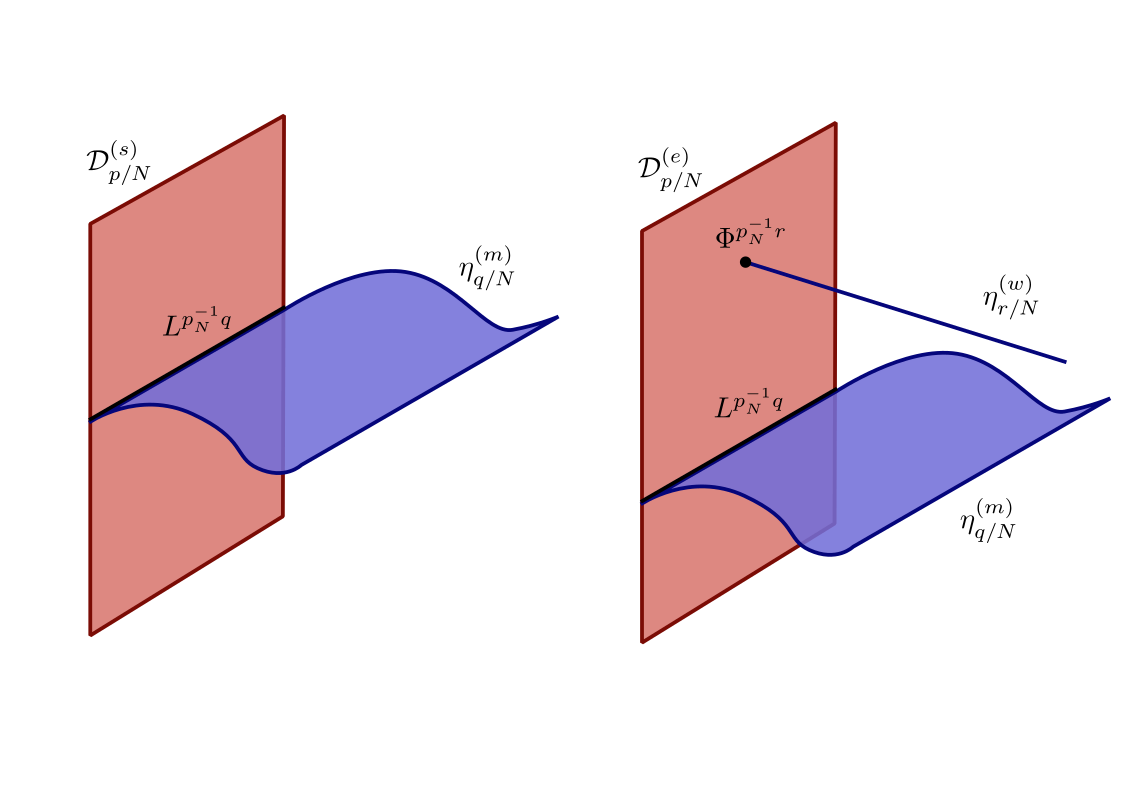}
  \caption{Topological junctions between invertible and noninvertible defects.}
  \label{fig:Topological junctions}
\end{figure}

\paragraph{Shift symmetry}
Concretely, the definition \eqref{eq: shift-definition} of the non-invertible shift symmetry defects in terms of the minimal $\mathcal{A}^{N,p}\left[j_m^{(2)}\right]$ implies that magnetic defects with rational angle $\eta^{(m)}_{\frac{q}{M}}$ can end on shift defects $\mathcal{D}^{(s)}_{\frac{p}{N}}$ for $M|N$: $\eta^{(m)}_{\frac{1}{N}}$ ends on a line $L^{p^{-1}}$ and these lines fuse according to an abelian $\Z_N$ group and braid according to the matrix:
\begin{equation}
    B_{L^{k}, L^{k'}}=\exp\left(\frac{2\pi ip}{N}k k'\right) .
\end{equation}
Formally, we write:
\begin{equation}
    \operatorname{Hom}_{\mathcal{D}^{(s)}_{p/N}}(\eta^{(m)}_{\frac{kp}{N}};\mathds{1}_{\mathcal{D}^{(s)}_{p/N}})=L^k_{\mathcal{D}^{(s)}_{p/N}}.
\end{equation}
We would like to present a more abstract approach, showing how both the existence of the junction and the braiding of the lines are a consequence of the inflow action of the minimal $\mathcal{A}^{N,p}$ theory:
\begin{equation}
    \exp\left(-\frac{2\pi\mathrm{i}p}{2N}\int\mathfrak{P}(B^{(2)})\right).
\end{equation}
One should, in fact, see the defect $\mathcal{D}^{(s)}_{\frac{p}{N}}$ as the result of shrinking a thin (topological) slab with $\eta^{(s)}_{\frac{p}{N}}$ on the left, the classical system (or SPT) with action $\exp\left(-\frac{2\pi\mathrm{i}p}{2N}\int\mathcal{P}(F)\right)$ in the middle and $\mathcal{A}^{N,p}\left[j_m^{(2)}\right]$ on the right. Now, by definition, the topological theory on the right hosts a $\Z_N$ group of lines which are coupled to a magnetic defect in the slab. Again, by definition, the partition function of the slab-boundary system is invariant under topological moves of the aforementioned defects, in the sense that the lines $L^i$ of the minimal $\mathcal{A}^{N,p}\left[j_m^{(2)}\right]$ theories are not topological on their own, but only the following combination is:
\begin{equation}
    L^k(\partial\Sigma^{(2)}) \eta^{(m)}_{-\frac{kp}{N}}(\Sigma ^{(2)}),
\end{equation}
with $\Sigma^{(2)}$ possibly a submanifold of the entire spacetime. We denote by $\alpha_m(L^k) = \frac{kp}{N} \pmod{1}$ the \emph{magnetic coupling} of a line.

\begin{figure}[ht!]
  \centering
  \includegraphics[width=\linewidth]{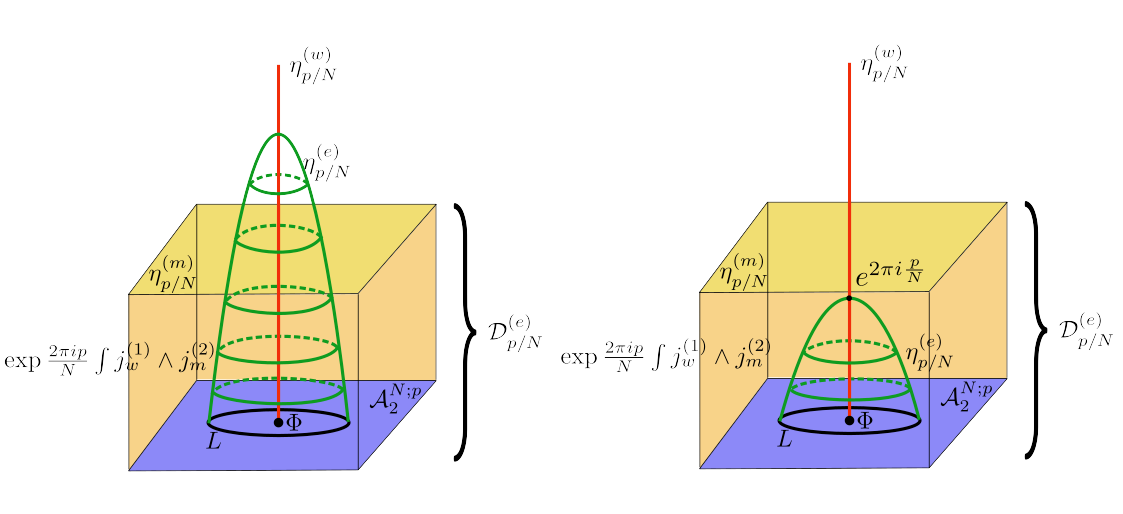}
  \caption{Definition of a noninvertible electric defect as a nontrivial composition of a noninvertible interface and a topological one. The shift defects enjoy an analogous construction.}
  \label{fig:Noninvertible slab}
\end{figure}

However, the slab action weights each intersection of magnetic defects $\eta^{(m)}_{\frac{j}{N}}$ and $\eta^{(m)}_{\frac{j'}{N}}$ with a phase:
\begin{equation}
    \exp(\frac{2\pi i jj'p^{-1}_N}{N}),
\end{equation}
and thus, by topological invariance, the endlines need to braid with the opposite sign (see \cref{fig:Noninvertible slab}).
In theory, there is still a freedom, in choosing a generator of the $\Z_N$ group of lines, which is reflected in the dualities \cref{MinimalDualities} of the minimal theories.

This paragraph can be summarized by saying that:
\begin{proposition}
    The line $L \in \operatorname{End}(\mathds{1}_{\mathcal{D}^(s)_{\frac{p}{N}}})$ lives in the magnetic twisted sector $p\in \Z_N^{(m)}$ (i.e.~is not a genuine defect, but is only defined as the boundary of a magnetic symmetry with open support), but has also magnetic charge $[1]\in \widehat{\Z}_N\sim \Z_N$ (i.e.~is acted upon by other magnetic defect, due to the braiding with the other $L^i$ lines) where the identification:
\begin{equation}
    \Z_N\ni [j]\mapsto\qty([k]\mapsto \exp(\frac{2\pi i jk}{N}))\in \widehat{\Z}_N.
\end{equation}
\end{proposition}

\paragraph{Electric symmetry}
A completely analogous argument applies to the noninvertible electric symmetry. Its definition in terms of the minimal $\mathcal{A}_2^{N,p}\left[j_m^{(2)}, j_w^{(1)}\right]$ \ref{eq: electric-definition} implies that magnetic and winding defects with rational angle $\eta^{(m)}_{\frac{q}{M}},\eta^{(w)}_{\frac{q}{M}}$ can end on electric defects $\mathcal{D}^{(s)}_{\frac{p}{N}}$, for $M|N$: magnetic surfaces of the type $\eta^{(m)}_{\frac{q}{N}}$ end on lines $L^{p^{-1}q}$ on the electric defect $\mathcal{D}^{(e)}_{\frac{p}{N}}$ and winding lines $\eta^{(w)}_{\frac{q}{N}}$ end on points $\Phi^{p^{-1}q}$, where $L$ and $\Phi$ are expressed in the Lagrangian description \eqref{eq:A2-lagrangian} as
\begin{equation}
  L = \exp(ipx^{-1}\int c) \qq{and} \Phi = \exp(ipy^{-1} \phi).
\end{equation}
These defects satisfy the fusion rules of two $\Z_N$ groups and link as\footnote{One may choose different generating defects $L' = L^s$ and $\Phi' = \Phi^t$ for any $s,t$ coprime to $N$. We have chosen $L$ and $\Phi$ to be the topological endpoints of $\eta^{(m)}_{\frac{p}{N}}$ and $\eta^{(w)}_{\frac{p}{N}}$ respectively.}:
\begin{equation}
    \ev{L^s(\gamma)\Phi^t (x)} = \exp(\frac{2\pi i pst}{N}).
\end{equation}
Formally, we write:
\begin{equation}
  \begin{aligned}
    \operatorname{Hom}_{\mathcal{D}^{(e)}_{p/N}}(\eta^{(m)}_{\frac{kp}{N}};\mathds{1}_{\mathcal{D}^{(e)}_{p/N}})=L^k_{\mathcal{D}^{(e)}_{p/N}},\\
    \operatorname{Hom}_{\mathcal{D}^{(e)}_{p/N}}(\eta^{(w)}_{\frac{kp}{N}};\mathds{1}_{\mathcal{D}^{(e)}_{p/N}})=\Phi^k_{\mathcal{D}^{(e)}_{p/N}}.
    \end{aligned}
\end{equation}
Both the existence of the junctions and the braiding of the lines are a consequence of the inflow action \eqref{eq:A2-inflow} of the $A_2^{N,p}$ theory. One can, in fact, see the defect $\mathcal{D}^{(e)}_{\frac{p}{N}}$ as the result of shrinking a thin topological slab with $\mathcal{D}^{(e)}_{\frac{p}{N}}$ on the left, the classical theory $\exp(-\frac{2\pi i p}{N}\int j_w^{(1)} \wedge j_m^{(2)})$ in the middle and $\mathcal{A}_2^{N,p}\qty[j_w^{(1)}, j_m^{(2)}]$ on the right. Now, by definition, the topological theory on the right enjoys a $\Z_N^{(0)} \times \Z_N^{(1)}$ and the coupling with the bulk currents implies that the lines $L^k$ implementing the $0$-form symmetry of the minimal $\mathcal{A}_2^{N,p}\qty[j_w^{(1)}, j_m^{(2)}]$ theories are not topological on their own, but only the following combination is:
\begin{equation}
    L^k(\partial\Sigma^{(2)}) \eta^{(m)}_{\frac{kp}{N}}(\Sigma ^{(2)}),
\end{equation}
with $\Sigma^{(2)}$ possibly a submanifold of the entire spacetime; and that the points $\Phi^i$ implementing the $1$-form symmetry are not topological on their own, but only the following combination is:
\begin{equation}
    \Phi^k(\partial\Sigma ^{(1)}) \eta^{(2)}_{\frac{kp}{N}}(\Sigma ^{(1)}),
\end{equation}
with $\Sigma ^{(1)}$ possibly a submanifold of the entire spacetime. We denote by $\alpha_m(L^k) = \frac{kp}{N} \pmod{1}$ and $\alpha_w(\Phi^k) = \frac{kp}{N} \pmod{1}$ the \emph{magnetic coupling} of a line and the \emph{winding coupling} of a point, respectively.

Again, by definition, the partition function of the slab-boundary system is invariant under topological moves of the aforementioned defects. However, the slab action weights each intersection of magnetic and winding defects $\eta^{(m)}_{\frac{k}{N}}$ and $\eta^{(w)}_{\frac{k'}{N}}$ with a phase
\begin{equation}
  \exp(-\frac{2\pi i p^{-1} kk'}{N}),
\end{equation}
and thus, by topological invariance, the endlines needs to braid with the opposite sign.

\subsection{(Noninvertible) higher group}
\label{sec:non-inv-higher-group}
This theory exhibits a (noninvertible generalization of a) higher group structure, which is characterized by a nontrivial interplay between the various layers of the symmetry category.

For simplicity, let's first discuss the invertible version of this structure, working at a coupling level $|K| > 1$ following \cite[appendix~A]{Choi2023noninvertible}. We remind the reader that a network of defects for a continuous (higher form) symmetry is equivalent to (the Pontryagin dual of the holonomies of) a flat background. If we impose the flatness conditions $G^{(4)} = 0$ and $H^{(3)} = 0$, \cref{eq:3group background} takes the form:
\begin{equation}
    \dd{B_w^{(3)}} = -\frac{K}{2} B_{e}^{(2)} \wedge B_{e}^{(2)}, \quad
    \dd{B_m^{(2)}} = -K B_s^{(1)} \wedge B_{e}^{(2)},
\end{equation}
which implies the following constraints on the allowed defect configurations:
 \begin{itemize}
 \item The one-dimensional intersection of an electric defect $\tilde{\eta}^{(e)}_j$ and a shift defect $\tilde{\eta}^{(s)}_k$ sources a magnetic defect.
   This emission is what makes the junction topological. Physically, a deformation $\Sigma^{(3)} \to (\Sigma^{(3)})'= \Sigma^{(3)} + \partial V^{(4)}$ of the support of the shift defect $\tilde{\eta}^{(s)}_k(\Sigma^{(3)})$ implements a transformation
      \begin{equation}
          \theta \to \theta + \frac{2\pi k}{K}
      \end{equation}
      in the region $V^{(4)}$. From \eqref{eq: tilde-invertible-definition} one sees that this modifies $\tilde{\eta}^{(e)}_j(\Sigma^{(2)})$ by a factor
      \begin{equation}
        \exp\qty(-\frac{2\pi i kj}{K} \int_{\Sigma^{(2)} \cap V^{(4)}} F)
        = \eta^{(m)}_{-jk/K}(\Sigma^{(2)} \cap V^{(4)}).
      \end{equation}
      Since $\partial(\Sigma^{(2)} \cap V^{(4)}) = \Sigma^{(2)} \cap (\Sigma^{(3)})' - \Sigma^{(2)} \cap \Sigma^{(3)}$, this agrees with the shift of a magnetic defect $\eta^{(m)}_{-jk/K}(M^{(2)})$ where $M^{(2)}$ is such that $\partial M^{(2)} = \Sigma^{(2)} \cap \Sigma^{(3)}$. Hence the configuration
      \begin{equation}
        \tilde{\eta}^{(s)}_k(\Sigma^{(3)})
        \tilde{\eta}^{(e)}_j(\Sigma^{(2)})
        \eta^{(m)}_{-jk/K}(M^{(2)})
      \end{equation}
      is topological.
    \item By the same argument, the zero-dimensional intersection of two electric defects $\tilde{\eta}^{(e)}_j$ and $\tilde{\eta}^{(e)}_k$ sources a winding defect $\tilde{\eta}^{(w)}_{-jk/K}$. This emission is what makes the junction topological.
 \end{itemize}

On the other hand, at $K = 1$ there is an analogous relationship between the non-invertible counterparts \cite{yokokura2022noninvertible,Choi2023noninvertible}:
\begin{itemize}
    \item The one dimensional intersection of an electric defect $\mathcal{D}^{(e)}_{p/N}$ and a shift defect $\mathcal{D}^{(s)}_{p'/N'}$ sources a magnetic defect $\eta^{(m)}_{-(pp')/(NN')}$. This emission is what makes the junction topological.
    \item The zero dimensional intersection of two electric defects $\mathcal{D}^{(e)}_{p/N}$ and $\mathcal{D}^{(e)}_{p'/N'}$ sources a winding defect $\eta^{(w)}_{-(pp')/(NN')}$. This emission is what makes the junction topological.
\end{itemize}

\section{Non-topological operators and their junctions}\label{sec: non-top-junction}
The definition of 't Hooft lines and axion strings suffers from some subtleties in the presence of a non-trivial coupling $K \neq 0$. As a result, the spectrum of the theory is enriched by both topological and non-topological junctions. A full account is given in  \cite[section~5]{Choi2023noninvertible}; what follows is a quick summary.

\begin{figure}[t!]
  \centering
  \subfloat[Subfigure 1 list of figures text][]{
    \includegraphics[width=0.5\textwidth]{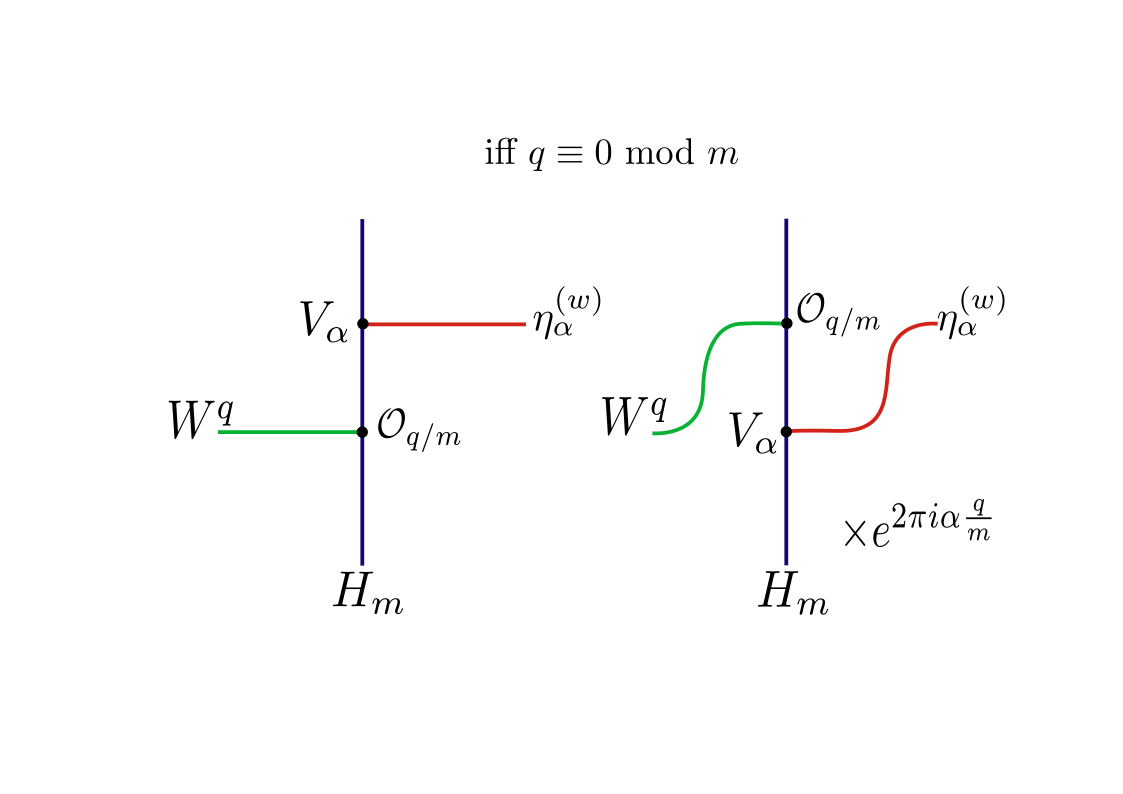}\label{fig: Hooft junctions}
  }%
  ~
  \subfloat[Subfigure 1 list of figures text][]{
    \includegraphics[width=0.5\textwidth]{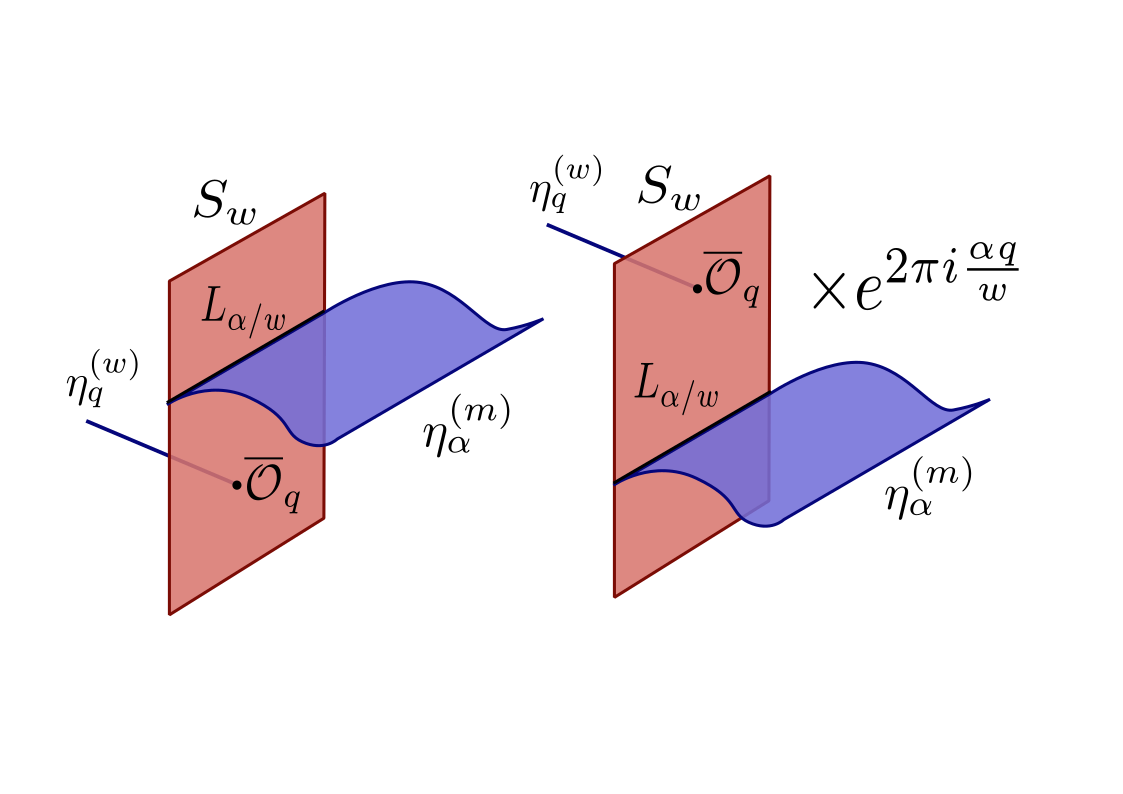}\label{fig: String junctions}
  }
  \caption{Junctions involving non-topological operators: 't Hooft lines host endpoints of Wilson lines (for compatible charges) and winding defects (\ref{fig: Hooft junctions}) and they link nontrivially; axion strings host endpoints of Wilson lines (for compatible charges) and magnetic defects (\ref{fig: String junctions}) and they link nontrivially.}
  \label{fig: GlobalNonTopJunction}
\end{figure}

\paragraph{'t Hooft lines}
The naive definition of an 't Hooft line $H_m(\gamma)$, imposing a boundary condition $d\left(dA\right)=2\pi m\delta(\gamma)$, is not gauge invariant in the presence of a nontrivial coupling $K$. Under a gauge transformation of the form $\theta \to \theta + 2\pi$ the naive 't Hooft line gets dressed by a Wilson line $W_{Km}$ via the Witten effect. Formally, the wordline theory suffers from a gauge anomaly,
and we can obtain a well-defined 't Hooft line by adding local degrees of freedom that cancel it. Whichever choice of theory is made, it must have $U(1)^{(-1)}$
and $U(1)^{(0)}$ symmetries coupled to the bulk winding and magnetic symmetries such that the bulk defects end topologically on the worldline defects.
As we will show later, these topological junctions are needed in order to coherently implement the noninvertible electric and shift actions.

Moreover, the quantum mechanical theory on $H_m$ hosts non-topological points $\mathcal{O}_j$ with charge $j$ under the $U(1)^{(0)}$ symmetry. Due to the coupling of the $U(1)^{(0)}$ symmetry to the bulk winding symmetry, and the noninvertible action of the electric defects, one can show that these points acquire a bulk electric charge $jm$ (see the discussion on the non-invertible symmetry action later on, as well as \cref{fig:Consistency tHooft}) and one can thus coherently end $W_{jm}$ on the $\mathcal{O}_j$ defects of $H_m$.

\paragraph{Axion strings}
Analogously, string worldsheets $S_w$ enjoy a nontrivial worldsheet action which allows both magnetic defects $\eta^{(m)}_\alpha$ (for all $\alpha$) and Wilson lines $W_q$ (if and only if $q\equiv0$ mod $w$) to end on it. The endpoints of the magnetic defect act as symmetry defects for a $U(1)^{(0)}$ symmetry on the string worldsheet, whose charged objects are exactly the endpoints of the Wilson lines.
Moreover, the theory on $S_w$ again hosts non-topological points $\overline{\mathcal{O}}_j$ that acquire a bulk electric charge $jw$ (see \cref{fig:Consistency Axion}); one can thus coherently end $W_{jw}$ on them.


\section{Noninvertible actions}
\label{sec: Noninvertible-actions}
Let us now discuss the non-invertible part of the action of the electric and shift symmetries. In order to derive these actions we will extensively use results from the previous chapters.

\paragraph{Shift symmetry}
First of all, we use \cref{Minimal-3d-half-gauging} to derive a construction of the shift defect as a result of 
half space gauging \cite{Choi2022noninvertible}:
\begin{equation*}
\begin{aligned}
    \mathcal{D}^{(s)}_{\frac{p}{N}}[\partial X^{(4)}]=\int [\mathcal{D}\Phi][\mathcal{D}b][\mathcal{D}c]&\exp\left(S_{\text{axion}}+\frac{2\pi ip}{N}\int_{\partial X^{(4)}} j^{(3)}_s\right)\\
     &\times\exp\left(\int_{X^{(4)}}\frac{iN}{2\pi}b\wedge dc+\frac{iN(p)^{-1}_N}{4\pi}b\wedge b+\frac{i}{2\pi}b\wedge F\right),
\end{aligned}
\end{equation*}
where we observe that Axion Maxwell is self-dual under the higher gauging of $\Z_N^{(m)}$ with  discrete torsion $p$ in the same way massless QED is (once again, the differences between the two theories do not involve the magnetic symmetry).
Now, if move this construction across a 't Hooft line $H_m$, its magnetic charge makes ill-defined in the region where we perform the magnetic gauging. It will thus live in the twisted magnetic sector of the corresponding charge, i.e.~the shift defect will act as:
\begin{equation}
    \mathcal{D}_{p/N}^{(s)}:\quad H(\gamma)\mapsto H(\gamma)\exp\left(\frac{2\pi ip}{N}\int_{\Sigma^{(2)}}\frac{F}{2\pi}\right)=H(\gamma)\mapsto H(\gamma)\eta^{(m)}_{\frac{p}{N}}(\Sigma^{(2)}),
\end{equation}
where $\gamma$ is one component of $\partial\Sigma^{(2)}$.
\begin{figure}[ht!]
  \centering
  \includegraphics[width=0.8\linewidth]{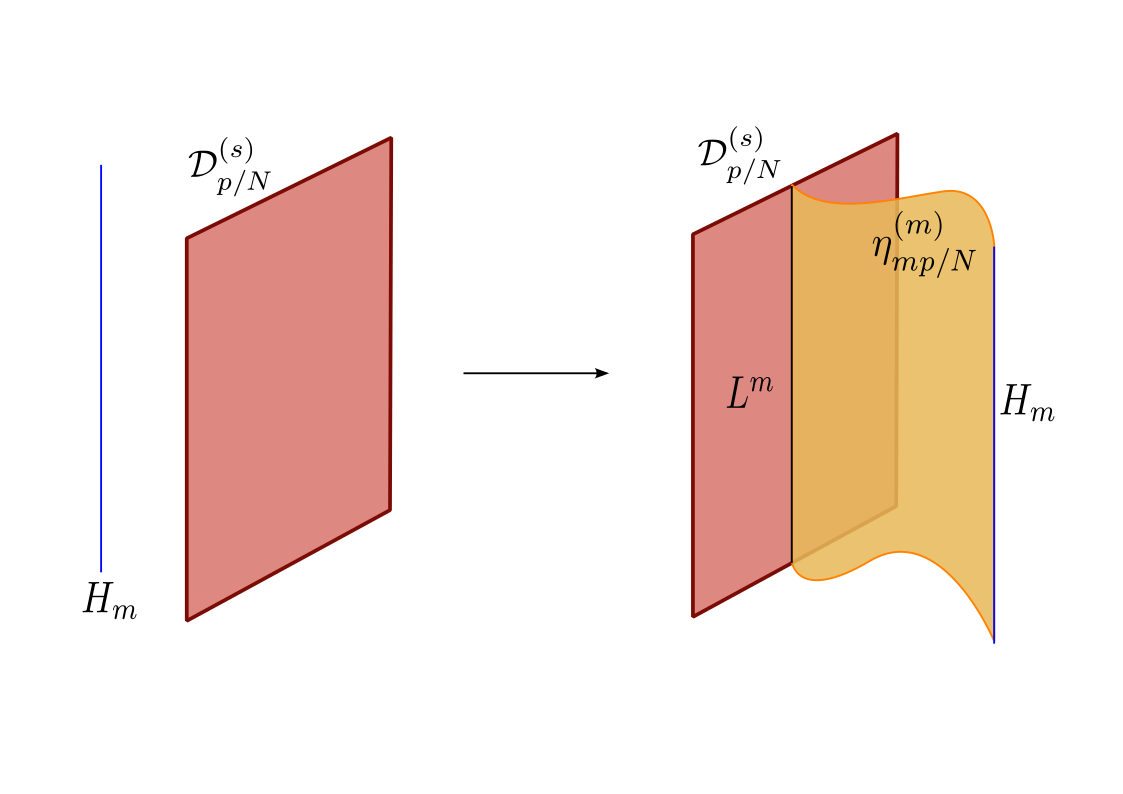}
  \caption{The shift defect $\mathcal{D}^{(s)}_{p/n}$ acts non-inversely on the 't Hooft lines $H_m$ of the theory and attaches them to magnetic defects $\eta^{(m)}_{pm/N}$.}
  \label{fig: ShiftHooftAction}
\end{figure}
One could have arrived at the same result, by noting that the shift defect $\mathcal{D}_{p/N}^{(s)}$ implements the transformation $\theta\to\theta + \frac{2\pi i p}{N}$, which provide electric charge to magnetically charged lines via the Witten effect.
Alternatively, using the half-gauge construction, we can argue that the 't Hooft line picks up a phase $\exp\qty(\frac{2\pi i k}{N})$ when it crosses a component $\eta^{(m)}_{\frac{k}{N}}$ of the network of defects implementing the magnetic gauge, while the $\eta^{(m)}_{\frac{p}{N}}$ factor picks up the opposite contribution due to the discrete torsion, making the combination well-defined even in the presence of the network. Transforming genuine operator into non-genuine ones is a hallmark of non-invertibility.

\begin{figure}[ht!]
    \centering
\includegraphics[width=0.8\linewidth]{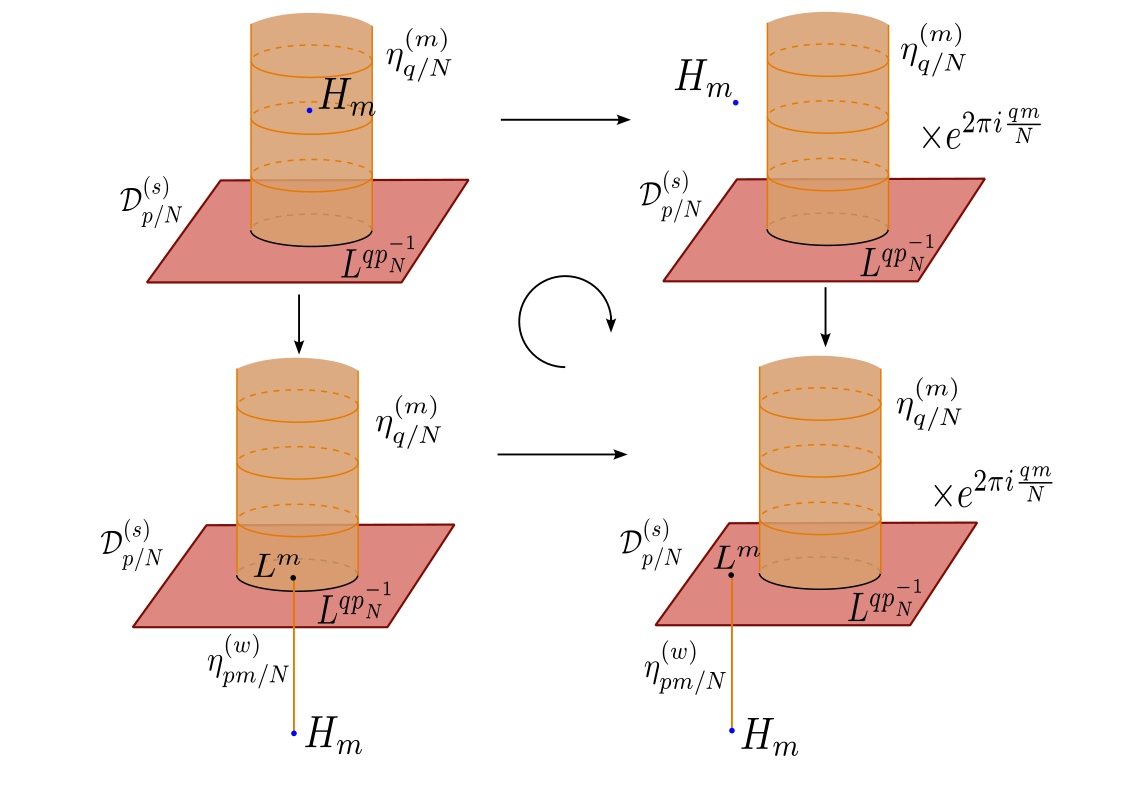}
    \caption{Consistency relation between the braiding of defects $L_i$ living on the shift symmetry operator $\mathcal{D}^{(s)}_{p/N}$ and the magnetic action on 't Hooft lines $H_m$.}
    \label{fig:Shift Symmetry}
\end{figure}

As mentioned before, this action needs topological junctions between $\eta^{(m)}_{\frac{1}{N}}$ and  $\mathcal{D}^{(s)}_{\frac{p}{N}}$ and between $\eta^{(m)}_{\frac{1}{N}}$ and $H_m$ to be well-defined: the other component of $\partial\Sigma^{(2)}$ is a submanifold of the shift defect and hosts $L$. Moreover, assuming these junctions exist, the braiding of the endlines $L^i$ is automatically implied by the consistency under the topological moves depicted in \cref{fig:Shift Symmetry}.

\paragraph{Electric symmetry}
\Textcite{Choi2022noninvertible} showed that the electric defect enjoys a construction by half higher gauging:
\begin{equation*}
  \begin{aligned}
    &\mathcal{D}^{(e)}_{\frac{p}{N}}[\partial X^{(3)}] =
    \int [\mathcal{D}\Phi][\mathcal{D}u][\mathcal{D}v]
      \exp\left[\frac{2\pi ip}{N}\int_{\partial X^{(3)}} j^{(2)}_e \right. \\
    &\quad {} + \left.\frac{iN}{2\pi} \int_{X^{(3)}} \qty(u^{(1)} \wedge \dd v^{(1)} + u^{(2)} \wedge \dd v^{(0)} + u^{(1)} \wedge u^{(2)} + u^{(1)} \wedge F - p\, u^{(2)} \wedge \dd\theta)\right],
  \end{aligned}
\end{equation*}
where we recall from \eqref{eq:self-duality-mixed} that the axion--Maxwell is self-dual under the $1$-condensation of $\Z_N^{(m)} \times \Z_N^{(w)}$ with discrete torsion $p\in \Z_N=\operatorname{Tor}_1^{\Z}\left(H^2(B\Z_N; \Z); H^3(B^2\Z_N; \Z)\right)\subset H^3(B\Z_N\times B^2\Z_N; U(1))$, i.e.~the following condensation defect is trivial:
\begin{equation}
    \mathcal{C}^{(0,m,w)}_{N,N,0,p}\sim \mathds{1}.
\end{equation}

Now, if move this construction across a 't Hooft line $H_m$, its magnetic charge makes ill-defined in the region where we perform the magnetic gauging. It will thus live in the twisted winding sector of the corresponding charge, i.e.~the electric defect will act as:
\begin{equation}
    \mathcal{D}_{p/N}^{(e)}:\quad H(\gamma)\mapsto H(\gamma)\exp\left(\frac{2\pi ip}{N}\int_{\Sigma^{(1)}}\frac{\dd \theta }{2\pi}\right)=H(\gamma)\eta^{(w)}_{\frac{p}{N}}(\Sigma^{(1)}), 
\end{equation}
where one component of $\partial\Sigma^{(1)}$ is a point of the 't Hooft line and supports the operator $\mathcal{O}_p$. As mentioned before, this action needs topological junctions $\Phi^i$ between $\eta^{(w)}_{\frac{1}{N}}$ and $\mathcal{D}^{(e)}_{\frac{p}{N}}$: the other component of $\partial\Sigma^{(1)}$ is a point of the electric defect and hosts $\Phi$.

In the same way, if we move this construction across an axion string $S_w$, its winding charge makes it ill-defined in the region where we perform the winding gauging. It will thus live in the twisted magnetic sector of the corresponding charge, i.e.~the electric defect will act as:
\begin{equation}
    \mathcal{D}_{p/N}^{(e)}:\quad S(\tilde{\Sigma})\mapsto S(\tilde{\Sigma})\exp\left(\frac{2\pi ip}{N}\int_{\Sigma^{(2)}}\frac{F}{2\pi}\right)=S(\tilde{\Sigma})\eta^{(m)}_{\frac{p}{N}}(\Sigma^{(2)}) .
\end{equation}
where one component of $\partial\Sigma^{(2)}$ is a line of the axion string and supports the operator $\mathcal{L}_p$. As mentioned before, this action needs topological junctions $L^i$ between $\eta^{(w)}_{\frac{1}{N}}$ and $\mathcal{D}^{(e)}_{\frac{p}{N}}$: the other component of $\partial\Sigma^{(2)}$ is a submanifold of the electric defect and hosts $L$.

\begin{figure}[ht!]
    \centering
    \subfloat[Subfigure 1list of figures text][]{
    \includegraphics[width=0.5\linewidth]{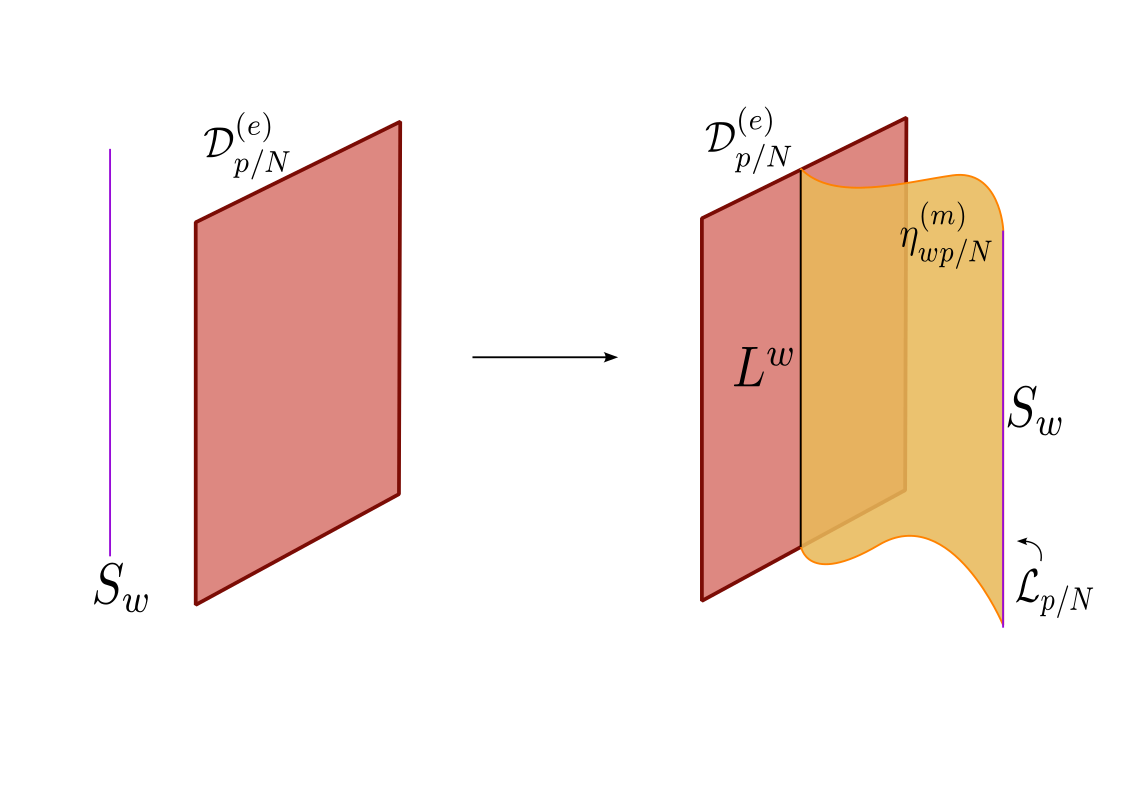}\label{fig: ElectricStringAction}}
    \subfloat[Subfigure 1list of figures text][]{
    \includegraphics[width=0.5\linewidth]{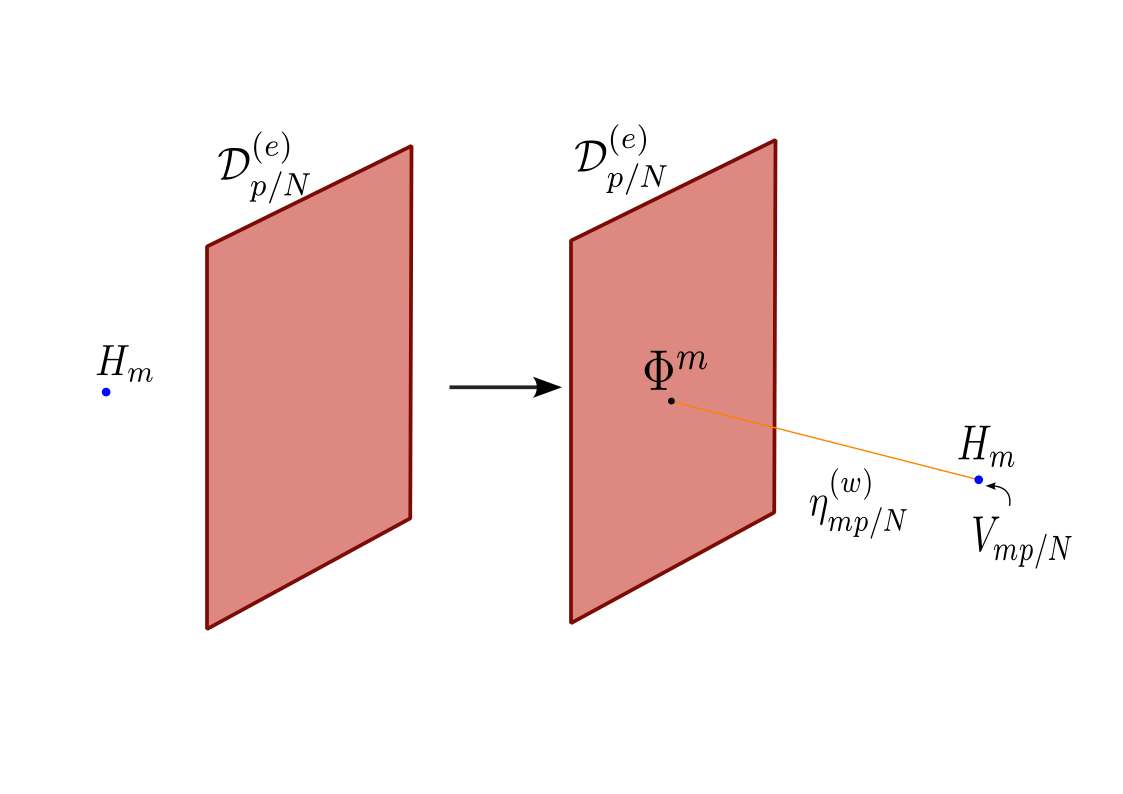}\label{fig: ElectricHooftAction}}
    \caption{The electric defect $\mathcal{D}^{(e)}_{p/n}$ acts non-inversely on the axion strings $S_w$ of the theory attaching them to magnetic defects $\eta^{(m)}_{pw/N}$ (\ref{fig: ElectricStringAction}) and on the 't Hooft lines $H_m$ of the theory attaching them to winding defects $\eta^{(w)}_{pm/N}$ (\ref{fig: ElectricHooftAction}).}\label{fig: Global-ElectricAction}    
\end{figure}

\begin{figure}[ht!]
    \centering
     \includegraphics[width=0.8\linewidth]{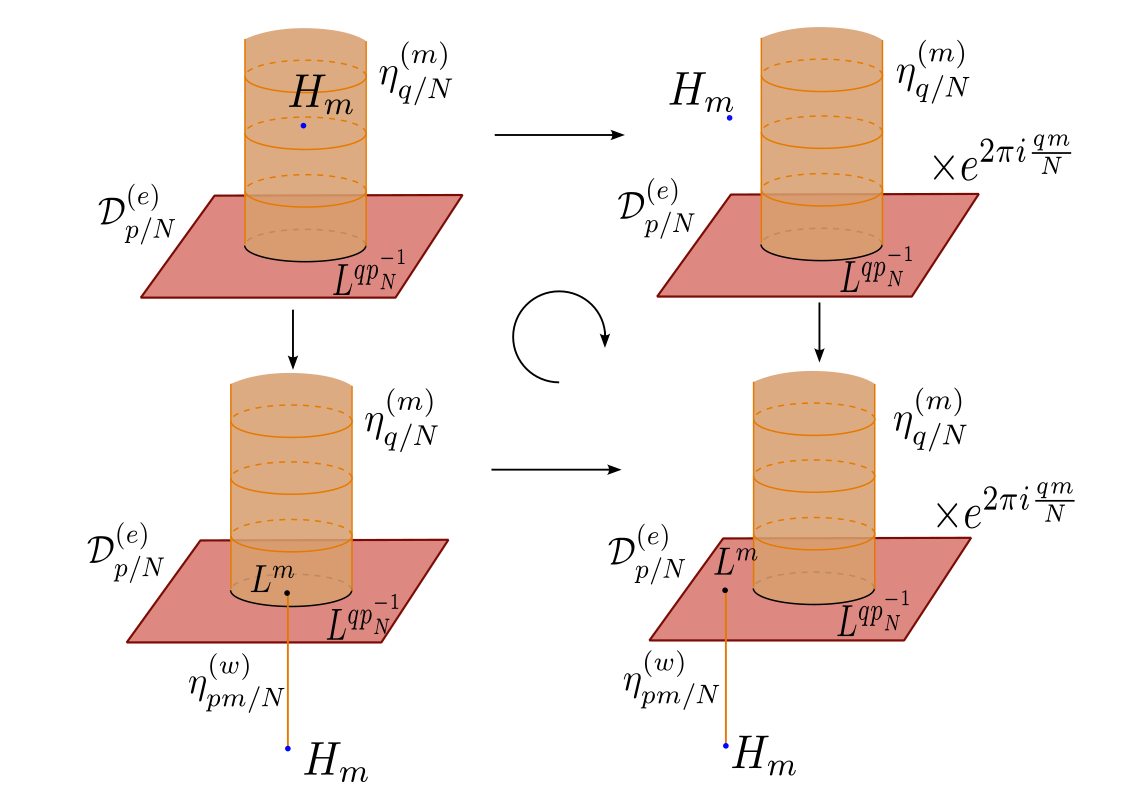}
    \caption{Consistency relation between the linking of defect $L_i, \ \Phi_j$ living on the electric symmetry operator $\mathcal{D}^{(e)}_{p/N}$ and the magnetic action on 't Hooft lines $H_m$.}
    \label{fig:Electric Symmetry1}
\end{figure}

\begin{figure}[ht!]
    \centering
    \includegraphics[width=0.8\linewidth]{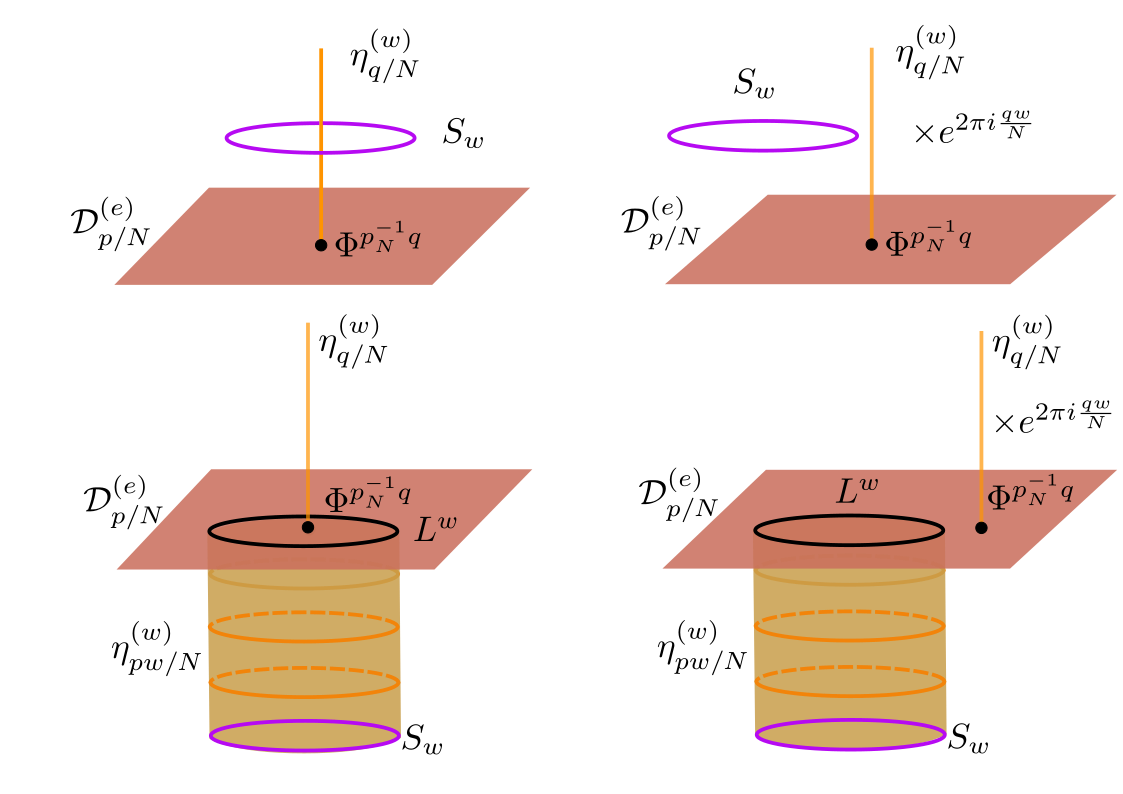}
    \caption{Consistency relation between the linking of defect $L_i, \ \Phi_j$ living on the electric symmetry operator $\mathcal{D}^{(e)}_{p/N}$ and the winding action on axion strings $S_w$.}
    \label{fig:Electric Symmetry2}
\end{figure}

\begin{figure}[ht!]
    \centering
    \includegraphics[width=0.8\linewidth]{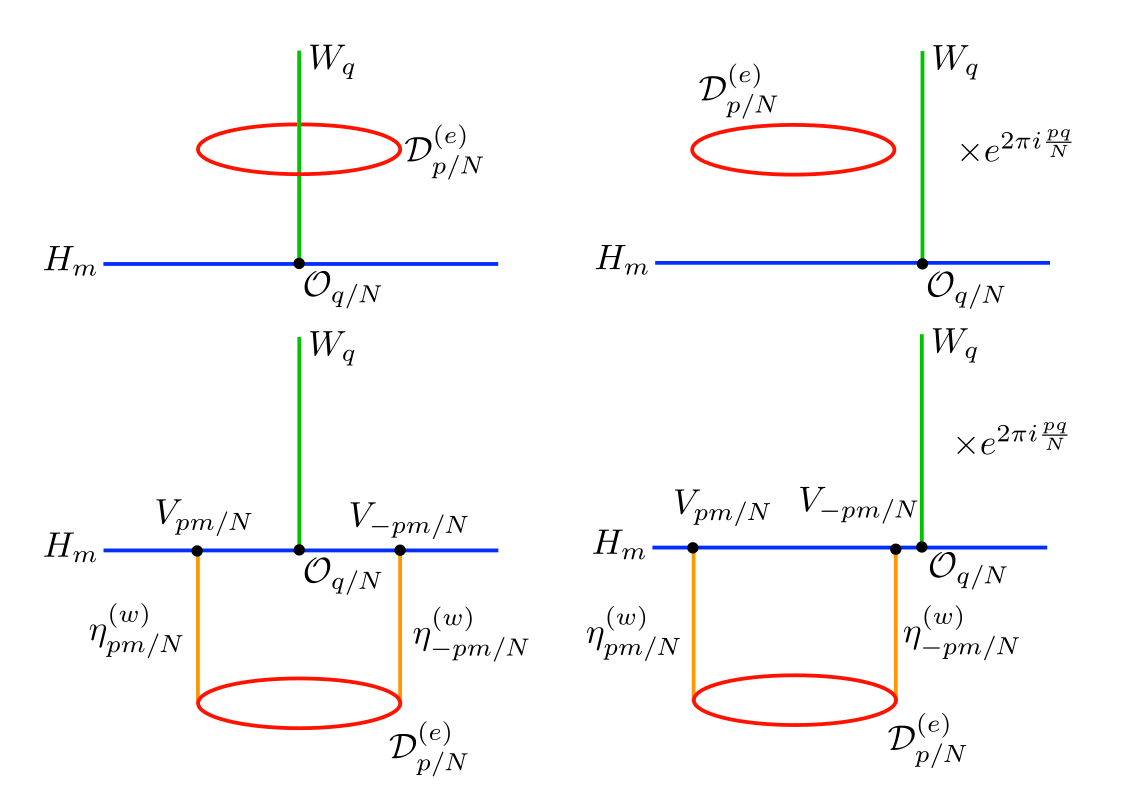}
    \caption{Consistency relation between the braiding of defects $\mathcal{O}_i, \ V_\alpha$ living on the 't Hooft lines $H_m$ and the electric action on Wilson lines $W_q$.}
    \label{fig:Consistency tHooft}
\end{figure}

\begin{figure}[ht!]
    \centering
    \includegraphics[width=0.8\linewidth]{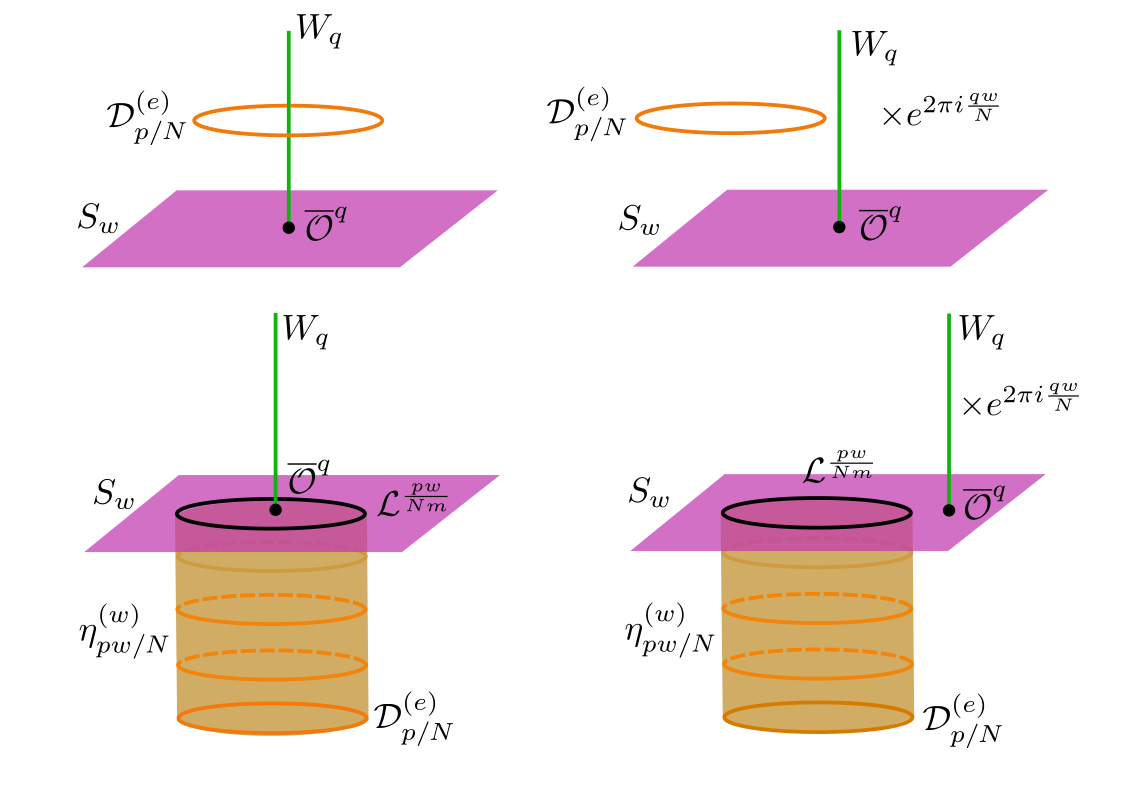}
    \caption{Consistency relation between the linking of defects $\overline{\mathcal{O}}_i, \ \mathcal{L}_\alpha$ living on the axion string $S_w$ and the electric action on Wilson lines $W_q$.}
    \label{fig:Consistency Axion}
\end{figure}

Moreover, assuming these junctions exist, the linking pairing between the lines $L^i$ and the points $\Phi^j$ is automatically implied by the consistency under the topological moves depicted in \cref{fig:Electric Symmetry1,fig:Electric Symmetry2}.
As we mentioned before, this action is also consistent with the existence of junctions involving non-topological operators and the braiding and linking statistics of the observables of the theories living on these non-topological defects (see \cref{fig:Consistency tHooft,fig:Consistency Axion}).

\section{Fusion interfaces}\label{sec: fusion}
In this section, we review the fusion interfaces derived in \cite{Copetti:2023mcq} between noninvertible chiral defects of massless QED adapting the argument to our setup. We then derive analogous interfaces for the fusion of noninvertible electric defects of the axion Maxwell theory.

We will abundantly use the following notation (borrowed by \cite{Copetti:2023mcq}): Given two defects labelled by $\frac{p_{1,2}}{N_{1,2}}$ we define
\begin{equation}
\begin{aligned}
    &M= \gcd(N_1, N_2), \quad  &&L= \lcm(N_1, N_2), \quad  &&K_i= \frac{N_i}{M}\\
   &M'=\gcd(p_1K_2+p_2K_1, L), \quad &&p_3=\frac{p_1K_2+p_2K_1}{M'}, \quad &&N_3=\frac{L}{M'}.
   \end{aligned}
\end{equation}
We will denote with the symbol $(\cdot)^{-1}_N$ the multiplicative inverse of an element of $\Z_N^\times$. Moreover, we will use the modern terminology of gaugeable data as algebra objects (see \cref{Appendix: generalized gauging}).
\subsection{Noninvertible $0$-form symmetry} Once again, due to their analogous definition, the fusion of the shift symmetry is  analogous to that of chiral defects in massless QED computed in \cite{Copetti:2023mcq}:
\begin{proposition}
  There exists a topological interface $m_{M'}$ between $\mathcal{D}^{(s)}_{\frac{p_1}{N_1}}\otimes\mathcal{D}^{(s)}_{\frac{p_2}{N_2}}$ and $\mathcal{D}^{(s)}_{\frac{p_1}{N_1}+\frac{p_2}{N_2}}$ given by the half-defect gauging of the decoupled, anomaly free $\Z_{M'}^{(1)}\subset \Z_{N_1}^{(1)}\times\Z_{N_2}^{(1)}$ subgroup generated by the lines\footnote{While it is not evident form the definition, one can easily prove that $M'|M$, and \textit{a posteriori} define $M'=\gcd(p_1K_2+p_2K_1, M)$ } $(L_1^{(p_1)_{N_1}^{-1}K_1} L_2^{-(p_2)_{N_2}^{-1}K_2})^{\frac{M}{M'}}$:
    \begin{equation}\label{eq: shift fusion}
      \mathcal{D}^{(s)}_{\frac{p_1}{N_1}} \otimes \mathcal{D}^{(s)}_{\frac{p_2}{N_2}} \overset{m_{M^{\prime}}}{\relbar\joinrel\longrightarrow}
      \mathcal{A}^{M/M^{\prime},(p_1^{-1}K_1 + p_2^{-1}K_{2})/M^{\prime}} \mathcal{D}^{(s)}_{\frac{(p_1 K_2 + p_2 K_1)/M^{\prime}}{L/M^{\prime}}}.
    \end{equation}
    Moreover, the line of the resulting defect can be written in terms of the lines of the two fused ones as:
    \begin{equation}
      L_3 = [L_1 L_2] = L_1 L_2 \frac{1}{\sqrt{M'}}
      \sum_{j=0}^{M'-1}(L_1^{(p_1)_{N_1}^{-1}K_1} L_2^{-(p_2)_{N_2}^{-1}K_2})^{j\frac{M}{M'}},
    \end{equation}
    and the line of the TFT coefficient as:
    \begin{equation}
        \tilde{L} = [L_1^{(p_1)_{N_1}^{-1}K_1} L_2^{-(p_2)_{N_2}^{-1}K_2}]
        = \frac{1}{\sqrt{M'}} \sum_{j=0}^{M'-1}
        (L_1^{(p_1)_{N_1}^{-1}K_1}L_2^{-(p_2)_{N_2}^{-1}K_2})^{1+j\frac{M}{M'}}.
    \end{equation}
\end{proposition}
The argument can be summarized as follows:\begin{itemize}
\item The invertible part $\eta^{(s)}$ of the definition of the defects fuse according to the $\mathbb{Q}/\Z$ group law. Thus, all the channels in the fusion $\mathcal{D}^{(s)}_{\frac{p_1}{N_1}}\otimes\mathcal{D}^{(s)}_{\frac{p_2}{N_2}}$ are of the form:
    \begin{equation}
        \mathcal{D}^{(s)}_{\frac{p_1}{N_1}}\otimes\mathcal{D}^{(s)}_{\frac{p_2}{N_2}}\sim \eta^{(s)}_{\frac{p_3}{N_3}}\otimes ?,
    \end{equation}
    times some TFT coefficient. We look for channels in which the $\eta^{(s)}_{p_3/N_3}$ combines with other defects to form $\mathcal{D}^{(s)}_{p_3/N_3}$.
    \item Assuming such a channel exists, the action of the interface on the lines of the original $\mathcal{D}^{(s)}_{\frac{p_1}{N_1}}\otimes\mathcal{D}^{(s)}_{\frac{p_2}{N_2}}$ defect is highly constrained by the coupling with the bulk magnetic symmetry.
    This can be best illustrated by an example. Consider the fusion of the type:
    \begin{equation}
        \mathcal{D}^{(s)}_{\frac{1}{4}}\otimes\mathcal{D}^{(s)}_{\frac{1}{4}} \to \mathcal{C}^{\frac{1}{2}}_{\frac{1}{4};\frac{1}{4}}\mathcal{D}^{(s)}_{\frac{1}{2}},
    \end{equation}
    where we denote with $\mathcal{C}^{\frac{1}{2}}_{\frac{1}{4};\frac{1}{4}}$ the generic TFT coefficient.
    Before the interface, a bulk magnetic defect $\eta^{(m)}_{1/4}$ can end on both of the $\mathcal{D}^{(s)}_{1/4}$. However, there is no such junction between $\eta^{(m)}_{1/4}$ and $\mathcal{D}^{(s)}_{1/2}$.
    Thus, the endlines $L_{1,2}$ of  $\eta^{(m)}_{1/4}$ cannot move across the interface: we will call these lines fractionally/\textit{incoherently} coupled to the bulk.

In the general case, $\mathcal{D}^{(s)}_{\frac{p_1}{N_1}} \otimes \mathcal{D}^{(s)}_{\frac{p_2}{N_2}}$ can host the endlines of a $\Z_L^{(m)}\subset U(1)^{(m)}$ subgroup of the magnetic symmetries, while $\mathcal{D}^{(s)}_{p_3/N_3}$ can only host the endlines of a $\Z_{N_3}^{(m)} = \Z_{\frac{L}{M'}}^{(m)}\subset \Z_L^{(m)}$ subgroup.
The (\textit{improperly coupled}) lines not in the kernel of the map:
\begin{equation}
  \begin{aligned}
    \Z_{N_1}\times\Z_{N_2}&\rightarrow\Z_{M'}\\
    (s,t) &\mapsto (p_1K_1s+p_2K_2t)\ \text{mod}\ M'
  \end{aligned}
\end{equation}
are stuck at the interface.
Even more generally, by the same reasoning, the bulk coupling must match across the interface:
\begin{equation}
  \frac{s_1 p_1}{N_1} + \frac{s_2 p_2}{N_2} = \frac{s_3 p_3}{N_3} \pmod{1},
  \label{eq:3d-charge-match}
\end{equation}
and since there is only one line with a given coupling in
$\mathcal{A}^{N_3,p_3}$, the action of the interfaces is fixed by $\alpha_m$ up to eventual freedom in the action of the interfaces is given by lines in the eventual decoupled coefficient.
\end{itemize}
We know that gauging lines in $3$d theories not only identifies lines differing by an element of the algebra but also selects only the lines that trivially braid with the algebra\footnote{And takes multiple copies of fixed points. More formal details on generalized gauging procedures in $2$ and $3$ dimensions can be found in \ref{Appendix: generalized gauging}.}. The idea is now to look for gaugeable (meaning decoupled from the bulk and spinless) subgroups in the $\Z_{N_1}^{(1)}\times\Z_{N_2}^{(1)}$ spectrum of lines, whose gauging selects exactly the lines which coherently couple to the bulk.

Luckily, there is exactly one gaugeable subgroup and it selects exactly the desired lines.

Moreover, while in the gauged theory the charge of a line doesn't completely characterize the line, one can identify a decomposition of the spectrum/theory in which the redundancy corresponds exactly to the presence of a decoupled TFT coefficient $\mathcal{A}^{N_3,p_3}$.
\begin{remark}
In \cite{Choi2022noninvertible} the authors computed the specific case of $\frac{p_1}{N_1}=-\frac{p_2}{N_2}=\frac{1}{N}$ obtaining:
\begin{equation}
    \mathcal{D}^{(s)}_\frac{1}{N}\otimes\mathcal{D}^{(s)}_{-\frac{1}{N}}
    =\mathcal{C}^{(0,m)}_{N,0}.
\end{equation}
There is a morphism between the condensate and the trivial operator which can either be seen as the Dirichlet boundary condition or the gauging of the quantum symmetry living on the condensation itself, which thus matches our result:
\begin{equation}
    \mathcal{D}^{(s)}_\frac{1}{N}\otimes\mathcal{D}^{(s)}_{-\frac{1}{N}}
    =\mathcal{C}^{(0,m)}_{N,,0} \xrightarrow[\text{Dirichlet}]{\text{m}_{N}}\mathds{1},
\end{equation}
\end{remark}
\begin{example}
    Let's explicitly compute the example:
    \begin{equation}
        \mathcal{D}^{(s)}_{3/8}\otimes \mathcal{D}^{(s)}_{1/8}\stackrel{m_4}{\relbar\joinrel\longrightarrow}\mathcal{A}^{2,1}\mathcal{D}^{(s)}_{1/2}.
    \end{equation}
   We can define the lines $L_\alpha$ and $L_\beta$ as follows:
\begin{equation}
    \begin{aligned}
        &L_\alpha=L_1L_2^5, &&L_\alpha^8=1, &&h[L_\alpha]=\frac{3}{4}, &&\alpha_m(L_\alpha)=0,\\
        &L_\beta=L_1L_2, &&L_\beta^8=1, &&h[L_\beta]=\frac{1}{4}, &&\alpha_m(L_\alpha)=\frac{1}{2},\\
    \end{aligned}
\end{equation}
and it is easy to check that they trivially braid with each other. We would like to say that they generate a subtheory of the form $\mathcal{A}^{8,12}\otimes \mathcal{A}^{8,4}[4 j_m^{(2)}]$, but this statement has two problems. On the one hand, both of the theories are ill-defined since $\gcd(8,4)=4=\gcd(8,12)\neq1$, i.e.~the contains nontrivial lines $L_\alpha^{2j}$ and $L_\beta^{2j}$ undetectable by elements of the subtheories. The second problem is that they have a non-empty intersection given exactly by the invisible lines $L_\alpha^2=L_\beta^2$.

Both problems can be solved by gauging the $\Z_4^{(1)}$ anomaly-free (spinless) decoupled subgroup generated by $L_1^2L_2^2=L_\alpha^2=L_\beta^2$.

Let's look at the effect of the gauging on the lines discussed above:
\begin{itemize}
\item It identifies lines which differ by $L_\alpha^2$. In particular, we have:
  \begin{equation}
    L_\alpha \leadsto [L_\alpha],\quad[ L_\alpha]^2=[\mathds{1}], \quad\quad L_\beta \leadsto [L_\beta],\quad [L_\beta]^2=[\mathds{1}],
  \end{equation}
  where the quantum number of the orbits is well-defined and coincides with the ones of each or their representants (since the gauged algebra is spinless and decoupled).
\item It selects only lines which trivially braid with the generator $\Z_4^{(1)}$ generator $L_1^2L_2^2$. However, we found $L_1^2L_2^2$ exactly by requiring this condition:
  \begin{equation}
    \langle\qty(L_1^2L_2^2)(\gamma_1)L_\alpha(\gamma_2)\rangle=0=\langle\qty(L_1^2L_2^2)(\gamma_1)L_\beta(\gamma_2)\rangle,
  \end{equation}
  and thus, this step has no effect on the lines in question.
\end{itemize}
It follows that now the lines actually generate a $\mathcal{A}^{2,1}\otimes \mathcal{A}^{2,1}[j_m^{(2)}]$ subtheory. As a consequence of the minimality properties (\cref{th: minimality}) of the $\mathcal{A}^{N,p}$ theories we have:
\begin{equation}
  \frac{\mathcal{D}^{(s)}_{1/8}\otimes\mathcal{D}^{(s)}_{1/8}}{\Z_4^{(1)}}=\mathcal{A}^{2,1}\otimes \mathcal{A}^{2,1}[j_m^{(2)}]\otimes\mathcal{T}',
\end{equation}
for some unknown theory $\mathcal{T}'$.
Finally, we can conclude that $\mathcal{T}'=\mathds{1}$, by analysing the effect of gauging on the remaining lines of the original theory. Indeed, we can write a generic line $L_1^sL_2^t$ as:
\begin{equation}
  L_1^sL_2^t= L_\alpha^xL_\beta^y L_1^{s'}L_2^{t'}, \quad \alpha_m(L_1^{s'}L_2^{t'})=\alpha_m(L_1^{s}L_2^{t})\ \operatorname{mod} \frac{1}{2},
\end{equation}
and, since the braiding with $L_1^2L_2^2$ measures exactly the coupling modulo $\frac{1}{2}$:
\begin{equation*}
  \langle\qty(L_1^2L_2^2)(\gamma_1)\qty( L_1^sL_2^t)(\gamma_2)\rangle=\exp\qty(\frac{2\pi \ii(2s+6t)}{8}l(\gamma_1,\gamma_2))=
  \exp\qty(\frac{2\pi \ii\alpha_m(L_1^sL_2^t)}{4}l(\gamma_1,\gamma_2)),
\end{equation*}
lines not in $\mathcal{A}^{2,1}\otimes \mathcal{A}^{2,1}[j_m^{(2)}]$ (i.e.~those for which $(s',t')\neq(0,0)$) aren't gauge invariant and thus do not survive the (second step of the) gauging process.

\end{example}

\subsection{Noninvertible 1-form symmetry}
Let us now discuss the fusion rules of the noninvertible electric symmetry.
The physical picture is similar to the one we discussed in the previous subsection.

First of all, eventual fusion interfaces will be graded by the invertible part:
\begin{equation}
    \mathcal{D}^{(e)}_{\frac{p_1}{N_1}}\otimes\mathcal{D}^{(e)}_{\frac{p_2}{N_2}}\sim \eta^{(e)}_{\frac{p_3}{N_3}}\otimes ?.
\end{equation}
Moreover, assuming the existence of an interface $m$ with a target of the form $\mathcal{D}^{(e)}_{p_3/N_3}$, its action on the line and point defects $L_i, \Phi_j$ localized on $\mathcal{D}^{(e)}_{\frac{p}{N}}$ is highly constrained by the requirement of a coherent coupling to the winding and magnetic bulk symmetries:
\begin{equation}\label{eq:2d-charge-match}
    \alpha_{m}(L_1^sL_2^t)=\alpha_{m}(m\cdot (L_1^sL_2^t)), \quad\quad
    \alpha_{w}(\Phi_1^s\Phi_2^t)=\alpha_{w}(m\cdot (\Phi_1^s\Phi_2^t)).
\end{equation}
However, the coupling of a generic line (point) live in a bigger $\Z_L$ group then that of the lines (point) living on $\mathcal{D}^{(e)}_{\frac{p_3}{N_3}}$.

More explicitly, the condition \eqref{eq:2d-charge-match} becomes
\begin{equation}
  s_1 p_1 K_2 + s_2 p_2 K_1 = s_3 (p_1 K_2 + p_2 K_1) \pmod{L}.
  \label{eq:2d-charge-match-2}
\end{equation}

By Bézout's identity, the right-hand side of \eqref{eq:2d-charge-match-2} spans
$M'\Z$ with $M' = \gcd(p_1 K_2 + p_2 K_1, L)$, while the left-hand side spans $\gcd(p_1 K_2, p_2 K_1, L)\Z = \Z$.\footnote{The gcd is always $1$: Since $L = K_1 N_2$, $p_1 K_2 \Z + p_2 K_1 \Z + L\Z = p_1 K_2 \Z + K_1 (p_2 \Z + N_2 \Z) = p_1 K_2 \Z + K_1 \Z = \Z$ because $K_1$ is coprime to both $p_1$ and $K_2$.} So if $M' \neq 1$, there are lines $L_1^{s_1} \otimes L_2^{s_2}$ whose charge cannot be matched by any line $L_3^{s_3}$.

\begin{figure}[ht!]
  \centering
  \includegraphics[width=.8\textwidth]{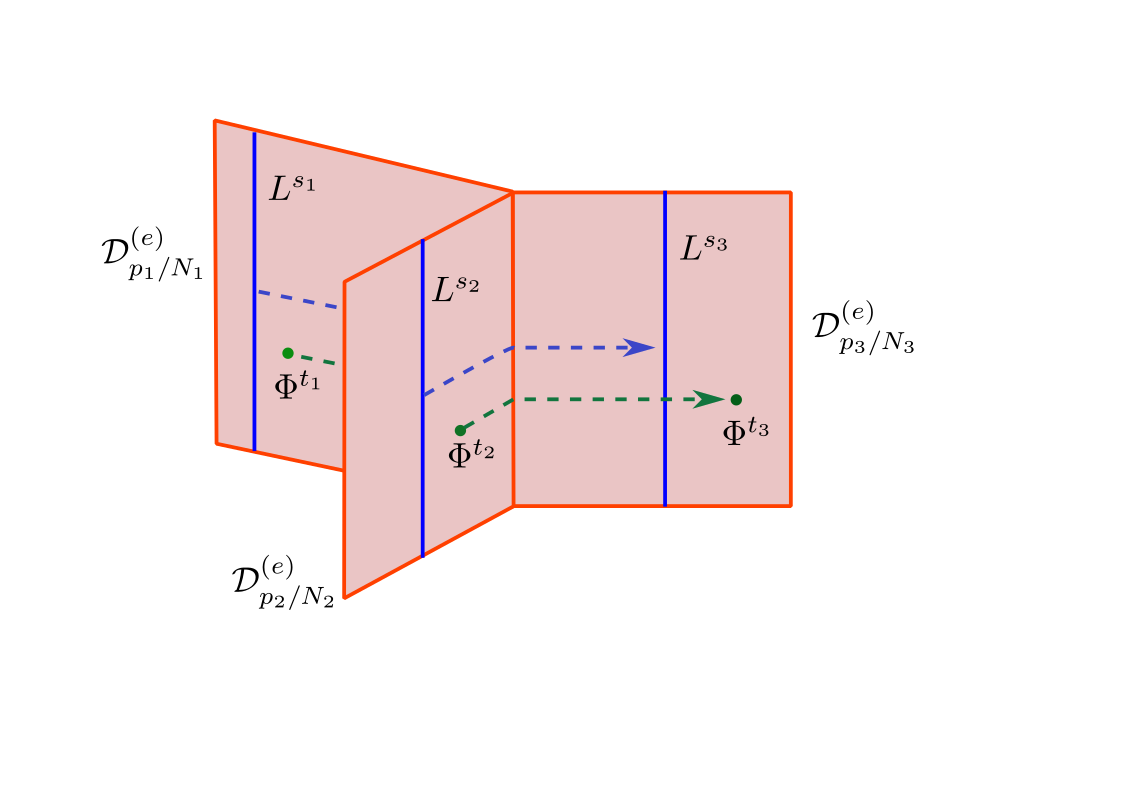}
  \caption{The coupling of the local operators on the electric defects need to be preserved across the fusion interface.}
  \label{fig:bulk-charge-matching-2d}
\end{figure}
\begin{remark}
    Since the behaviour of line and point defects is completely symmetrical under the exchange:
\begin{equation}
    L\longleftrightarrow\Phi, \quad \quad (m)\longleftrightarrow (w),
\end{equation}
we will often discuss only one of the two defects.
\end{remark}
\begin{theorem}
    There exists a topological interface $m_{M',M}$ between $\mathcal{D}^{(e)}_{\frac{p_1}{N_1}}\otimes\mathcal{D}^{(e)}_{\frac{p_2}{N_2}}$ and $\mathcal{D}^{(e)}_{\frac{p_1}{N_1}+\frac{p_2}{N_2}}$ given by the half-defect gauging of the decoupled, anomaly free $\Z_{M'}^{(0)}\subset \Z_{N_1}^{(0)}\times\Z_{N_2}^{(0)}$ subgroup of lines generated by $(L_1^{(p_1)_{N_1}^{-1}K_1}L_2^{-(p_2)_{N_2}^{-1}K_2})^{\frac{M}{M'}}$ and the $\Z_{M}^{(1)}\subset \Z_{N_1}^{(1)}\times\Z_{N_2}^{(1)}$ subgroup of points generated by $\Phi_1^{(p_1)_{N_1}^{-1}K_1}\Phi_2^{-(p_2)_{N_2}^{-1}K_2}$:
    \begin{equation}\label{eq: electric fusion}
        \mathcal{D}^{(e)}_{\frac{p_1}{N_1}}\otimes\mathcal{D}^{(e)}_{\frac{p_2}{N_2}}\overset{m_{M^{\prime}; M}}{\relbar\joinrel\relbar\joinrel\longrightarrow}\mathcal{D}^{(e)}_{\frac{(pK_{2}+p^{\prime}K_{1})/M^{\prime}}{K/M^{\prime}}}\otimes \operatorname{DW}(\Z_{M'}).
    \end{equation}
    where we denoted with $\operatorname{DW}(\Z_{M'})$ a $\Z_{M'}$ gauge theory decoupled from the bulk. 
    
    Moreover, the lines (and points) of the resulting defect can be written in terms of the lines (and points) of the two fused ones as:
    \begin{equation}
    \begin{aligned}
        L_3=[L_1L_2]=L_1L_2\frac{1}{\sqrt{M'}}\sum_{j=0}^{M'-1}(L_1^{(p_1)_{N_1}^{-1}K_1}L_2^{-(p_2)_{N_2}^{-1}K_2})^{j\frac{M}{M'}},\\
        \Phi_3=[\Phi_1\Phi_2]=\Phi_1\Phi_2\frac{1}{\sqrt{M}}\sum_{j=0}^{M-1}(\Phi_1^{(p_1)_{N_1}^{-1}K_1}\Phi_2^{-(p_2)_{N_2}^{-1}K_2})^j.
        \end{aligned}
    \end{equation}
\end{theorem}

The technical proof of the statement relies on the following to lemmas.
\begin{lemma}
    In the theory $\mathcal{A}_2^{N_1,p_1}\otimes\mathcal{A}_2^{N_2,p_2}$ resulting from the fusion of two minimal ones, lines (respectively points) of bulk charge multiple of $\frac{p_3}{N_3}$ forms a $(\Z_L\times \Z_M)/ \Z_{M'}$ subgroup generated by:
    \begin{equation}
        L_\alpha=L_1L_2; \quad \quad L_\beta=L_1^{(p_1^{-1})_{N_1}K_1}L_2^{-(p_2^{-1})_{N_2}K_2}.
    \end{equation}
    Moreover, points and lines of different types do not interact with each other:
    \begin{equation}
        [L_\alpha(\gamma),\Phi_\beta(x)] = 0 = [L_\beta(\gamma),\Phi_\alpha(x)].
    \end{equation}
\end{lemma}
The commutation relation is easily checked. It is also easy to show that:
\begin{equation}
  \alpha_m(L_\alpha)=\frac{p_3}{N_3}, \quad \operatorname{ord}(L_\alpha)=MK_1K_2=L;\quad\quad
  \alpha_m(L_\beta)=0, \quad  \operatorname{ord}(L_\beta)=M.
\end{equation}

Let's show that there are exactly $M$ decoupled lines (and thus $M$ lines for each fixed coupling). In fact, the magnetic coupling can be seen as a homeomorphism:
\begin{equation}
  \begin{split}
    \Z_{N_1}\times\Z_{N_2}&\xrightarrow{\alpha_m}\Z_{L}\\
    (u,v)&\rightarrow up_1K_2+vp_2K_1 .
  \end{split}
\end{equation}
This morphism is surjective. In fact, we have:
\begin{equation}
  \alpha_m\left(L_1^{x(p_1^{-1})_{N_1}}L_2^{y(p_2^{-1})_{N_2}}\right)=xK_2+yK_1,
\end{equation}
and, by Bézout's theorem, $\operatorname{span}_\Z(K_1,K_2)=\Z$.
Thus, the kernel is a subgroup of $\Z_{N_1}\times\Z_{N_2}$ of order $\frac{L}{N_1N_2}=M$. Since $L_\beta$ is an element of the kernel and $\operatorname{ord}(L_\beta)=M$, it generates the whole kernel.

It is also easy to check that lines of the type $L_\alpha^{i\frac{L}{M'}}$ are decoupled as well. This fact implies that $L_\alpha^{\frac{L}{M'}}=L_\beta^c$ for some $c\in \Z_M$. We can be even mere precise: since $\operatorname{ord}(L_\alpha^{\frac{L}{M'}})=M'$, $c$ has to be of the form $c=c'\frac{M}{M'}$ with $c'\in \Z_{M'}^\times$. In definitive, they organize themselves in a group of the form:
\begin{equation}
  \label{eq:right gauge}
  \frac{\Z_L\times\Z_M}{\langle(\frac{L}{M'},-c')\rangle}
\end{equation}
where $(\frac{L}{M'}, -c')$ is the generator of a $\Z_{M'}$ subgroup. We can thus conclude with a counting argument, as we described a subgroup of $\alpha_m^{-1}(\Z_{\frac{L}{M'}})$ which has exactly the right number $\frac{LM}{M'}$ of elements.

\begin{lemma}
In a generic $2$-dimensional theory with $\Z_n^{(0)}\times\Z_m^{(1)}$ symmetries\footnote{See \ref{Appendix: generalized gauging} for more details.}:
\begin{itemize}
    \item Gauging a $\Z_{n'}^{(0)}\subset\Z_n^{(0)}$ subgroup identifies lines which differ by an element of this subgroup (which, in particular, reduces the symmetry $\Z_n^{(0)}\to \Z_{\frac{n}{n'}}^{(0)}$); and selects the points with charge multiple of $n'$ (the "gauge invariant" ones)\footnote{More precisely, the point operators charged under $\Z_{n'}^{(0)}$ become twisted sector operators of the gauged theory, twisted by the generating line of the dual symmetry defined next.}.

    Moreover there is a new quantum symmetry $\hat{\Z}_{n'}^{(0)}$ whose topological defects are Wilson lines for the gauged $\Z_{n'}^{(0)}$. The operator with charge $k$ under the dual symmetry are exactly those that in the original theory are in the twisted sector of the topological operator $\eta_k$ implementing the original  $\Z_{n'}^{(0)}$ symmetry.
    \item Gauging a $\Z_{m'}^{(1)}\subset\Z_m^{(1)}$ subgroup with discrete torsion $k\in H^2(B^2\Z_{m'};U(1))=\Z_{m'}$ identifies points which differ by an element of this subgroup (which, in particular, reduces the symmetry $\Z_m^{(1)}\to \Z_{\frac{m}{m'}}^{(1)}$); and selects the lines with charge multiple of $m'$\footnote{Again, this means that the lines charged under $\Z_{m'}^{(1)}$ are in ``twisted sectors'' of the gauged theory. Since the dual symmetry is in this case a $(-1)$-form symmetry, it is more natural to say that they are domain walls.}.
\end{itemize}
\end{lemma}

By putting together these lemmas one can easily show that the interface implementing the half-space gauging of the $\Z_{M'}^{(0)}\times\Z_{M'}^{(1)}$ discussed in \cref{eq:right gauge} provides a morphism:
\begin{equation}
    \mathcal{D}^{(e)}_{\frac{p_1}{N_1}}\otimes\mathcal{D}^{(e)}_{\frac{p_2}{N_2}}\overset{\tilde{m}_{M^{\prime}; M'}}{\relbar\joinrel\relbar\joinrel\longrightarrow}\mathcal{A}_2^{\frac{M}{M'}}\mathcal{D}^{(e)}_{\frac{p_3}{N_3}}\otimes \operatorname{DW}(\Z_{M'}).
\end{equation}
where the line defects of the gauged theory are exactly those implementing the dual symmetry. Indeed, since in the original theory there are no twisted sector operators (as an immediate consequence of the fact that all the defects are topological and nontrivially acting), in the gauged one there are no charged operators under the dual symmetry\footnote{This can also be seen from the presence of twisted sector operators in the gauged theory; these are topological endpoints of the dual symmetry operator.} and the resulting theory factorizes. Moreover, the dual symmetry lines are naturally decoupled from the bulk, meaning they are genuine lines of the defect theory, not operators defined only at the intersection of the electric defect with some bulk surface defect. Their endpoints, however, are coupled to the bulk; the twisted sector operators are the charged operators of the original theory, coupled to the bulk winding lines $\eta^{(\mathrm{w})}$. As invertible topological interfaces (isomorphisms), they essentially show that the dual symmetry line is nothing but the winding line restricted to the gauged defect. From now on, we will consider only configurations without dual symmetry lines.

Moreover, since the theory $\mathcal{A}_2^{\frac{M}{M'}}$ is not simple, one can further project down to one of its components gauging its $\Z_{\frac{M}{M'}}^{(1)}$ symmetry. The desired morphism is given by the composition of the two:
\DisableQuotes\[\begin{tikzcd}
	{\mathcal{D}^{(e)}_{\frac{p_1}{N_1}}\otimes\mathcal{D}^{(e)}_{\frac{p_2}{N_2}}} && {\mathcal{A}_2^{\frac{M}{M'}}\otimes\mathcal{D}^{(e)}_{\frac{p_3}{N_3}}} \\
	\\
	&& {\mathcal{D}^{(e)}_{\frac{p_3}{N_3}}}
	\arrow["{\tilde{m}_{M',M'}}", from=1-1, to=1-3]
	\arrow["{m_{M',M}}"', curve={height=12pt}, from=1-1, to=3-3]
	\arrow["{\pi_i}", from=1-3, to=3-3]
\end{tikzcd}\]\EnableQuotes
which actually corresponds to gauging the full $\Z_M^{(1)}$ from the beginning.\footnote{See at \cref{Appendix: decomposition} for more details on decomposition and on the composition of two gauging procedures with discrete torsion.}

\begin{remark}
    It is important to note that, because of the bulk coupling, a general operator built from lines and points of the minimal theory is technically a junction between the theory itself and the bulk symmetry defects to which the theory is coupled. In the case under investigation, however, we explicitly dealt with decoupled defects, i.e.~proper (higher) endomorphisms.
\end{remark}
\begin{remark}
In \cite{Choi2023noninvertible} the authors computed the specific case of $\frac{p_1}{N_1}=-\frac{p_2}{N_2}$ obtaining
\begin{small}
\begin{equation}
\begin{aligned}
    \mathcal{D}^{(e)}_\frac{p}{N}\otimes\mathcal{D}^{(e)}_{-\frac{p}{N}}
    &=\int[D\phi D\bar{\phi} Dc D\bar{c}]\exp\left[i\oint_{\Sigma^{(2)}}\left(\frac{N}{2\pi}\phi dc-\frac{N}{2\pi}\bar{\phi}d\bar{c}+\frac{p}{2\pi}\theta d(c-\bar{c})+\frac{1}{2\pi}(\phi-\bar{\phi})dA\right)\right]\\
    &=\int[D\phi Dc^{\prime}]_{\Sigma^{(2)}}\exp\left[i\oint_{\Sigma^{(2)}}\left(\frac N{2\pi}\phi dc^{\prime}+\frac p{2\pi}\theta dc^{\prime}\right)\right]\\&\times\int[D\phi^{\prime} D\bar{c}]_{\Sigma^{(2)}}\exp\left[i\oint_{\Sigma^{(2)}}\left(\frac N{2\pi}\phi^{\prime}d\bar{c}+\frac1{2\pi}\phi^{\prime}dA\right)\right]\\&=\mathcal{C}^{(1,m,w)}_{N,N,0,0},
\end{aligned}
\end{equation}
\end{small}%
where we used the explicit description of $\mathcal{A}_2^{N,p}$ discussed in \cref{remark: shao-gauge} and in the second step we defined $c'=c-\overline{c}$ and $\phi'=\phi-\overline{\phi}$.

There is a morphism between the condensate and the trivial operator which can either be seen as the Dirichlet boundary condition or the gauging of the quantum symmetry living on the condensation itself, which thus matches our result:
\begin{equation}
    \mathcal{D}^{(e)}_\frac{p}{N}\otimes\mathcal{D}^{(e)}_{-\frac{p}{N}}
    =\mathcal{C}^{(1,m,w)}_{N,N,0,0} \xrightarrow[\text{m}_{N.N}]{\text{Dirichlet}}\mathds{1}.
\end{equation}
\end{remark}

\begin{example}
    Let's explicitly compute the example (neglecting the decoupled sector):
    \begin{equation}\label{ex: electric-fusion}
        \mathcal{D}^{(e)}_{3/8}\otimes \mathcal{D}^{(e)}_{1/8}\xrightarrow{m_{4,8}}\mathcal{D}^{(e)}_{1/2},
    \end{equation}
    obtained both in one step or as a composition of the form:
    \begin{equation}
        \mathcal{D}^{(e)}_{3/8}\otimes \mathcal{D}^{(e)}_{1/8}\xrightarrow{m_{4,4}}\mathcal{A}_2^2\mathcal{D}^{(e)}_{1/2}\xrightarrow{\pi_2}\mathcal{D}^{(e)}_{1/2}.
    \end{equation}
   We can define the lines $L_\alpha$ and $L_\beta$ and points $\Phi_\alpha$ and $\Phi_\beta$ as follows:
\begin{equation}
    \begin{aligned}
        &L_\alpha=L_1L_2^5, &&L_\alpha^8=1, &&\alpha_m(L_\alpha)=0,\\
        &L_\beta=L_1L_2, &&L_\beta^8=1, &&\alpha_m(L_\alpha)=\frac{1}{2},\\
         &\Phi_\alpha=\Phi_1\Phi_2^5 ,&&\Phi_\alpha^8=1,&&\alpha_w(\Phi_\alpha)=0,\\
        &\Phi_\beta=\Phi_1\Phi_2, &&\Phi_\beta^8=1  ,&&\alpha_w(\Phi_\beta)=\frac{1}{2},
    \end{aligned}
\end{equation}
   and it is easy to check that $\alpha$- and $\beta$-defects trivially link with each other. We would like to say that they generate a subtheory of the form $\mathcal{A}_2^{8,12}\times \mathcal{A}_2^{8,4}[j_w^{(1)},j_m^{(2)}]$, but this statement has two problems. On the one hand, both of the theories are ill-defined since $\gcd(8,4)=4=\gcd(8,12)\neq1$, i.e.~the contains nontrivial lines $L_\alpha^{2j}$ and $L_\beta^{2j}$ and points $\Phi_\alpha^{2j}$ and $\Phi_\beta^{2j}$ undetectable by elements of the subtheories. The second problem is that they have a non-empty intersection given exactly by the invisible lines $L_\alpha^2=L_\beta^2$ and points $\Phi_\alpha^2=\Phi_\beta^2$.

   Both problems can be solved by gauging the $\Z_4^{(0)}\times \Z_4^{(1)}$ anomaly-free decoupled subgroup generated by $L_1^2L_2^2=L_\alpha^2=L_\beta^2$ and $\Phi_1^2\Phi_2^2=\Phi_\alpha^2=\Phi_\beta^2$  .

   Let's look at the effect of the gauging on the defects discussed above:
   \begin{itemize}
       \item It identifies lines which differ by $L_\alpha^2$. In particular, we have:
       \begin{equation}
           L_\alpha \leadsto [L_\alpha],\quad [L_\alpha]^2=[\mathds{1}], \quad\quad L_\beta \leadsto [L_\beta],\quad [L_\beta]^2=[\mathds{1}],
       \end{equation}
       where the quantum number of the orbits is well-defined and coincides with the ones of each or their representatives (since the gauged algebra is decoupled).
        \item It identifies points which differ by $\Phi_\alpha^2$. In particular, we have:
       \begin{equation}
           \Phi_\alpha\leadsto [\Phi_\alpha],\quad [\Phi_\alpha]^2=[\mathds{1}], \quad \quad\Phi_\beta \leadsto [\Phi_\beta],\quad [\Phi_\beta]^2=[\mathds{1}],
       \end{equation}
       where the quantum number of the orbits is well-defined and coincides with the ones of each or their representatives (since the gauged algebra is decoupled).
       \item It selects only lines which trivially braid with the $\Z_4^{(1)}$ generator $\Phi_1^2\Phi_2^2$. However, as we said, $\Phi_1^2\Phi_2^2$ braids trivially with $\alpha$ lines since $\Phi_1^2\Phi_2^2=\Phi_\alpha^2$ and with $\beta$ line since $\Phi_1^2\Phi_2^2=\Phi_\alpha^2$. This step has no effect on the lines in question.
       \item It selects only points which trivially braid with the $\Z_4^{(0)}$ generator $L_1^2L_2^2$. However, as we said, $L_1^2L_2^2$ braids trivially with $\alpha$ points since $L_1^2L_2^2=L_\alpha^2$ and with $\beta$ points since $L_1^2L_2^2=L_\alpha^2$. This step has no effect on the points in questions.
       \end{itemize}
    It follows that now the $\alpha,\beta$ defects actually generate a $\mathcal{A}_2^{2}\times \mathcal{A}_2^{2,1}[j_w^{(1)},j_m^{(2)}]$ subtheory.

    Finally, we can conclude that this exhausts all the defects of the gauged theory $\frac{\mathcal{D}^{(e)}_{1/8}\otimes \mathcal{D}^{(e)}_{1/8}}{\Z_4^{(0)}\times \Z_4^{(1)}}$, by analysing the effect of gauging on the remaining lines and points of the original theory. Indeed, we can write a generic defect as:
    \begin{equation}
    \begin{aligned}
    L_1^sL_2^t= L_\alpha^xL_\beta^y L_1^{s'}L_2^{t'}, \quad &\quad\alpha_m(L_1^{s'}L_2^{t'})=\alpha_m(L_1^{s'}L_2^{t'})\ \operatorname{mod} \frac{1}{2}\\
        \Phi_1^s\Phi_2^t=\Phi_\alpha^x\Phi_\beta^y \Phi_1^{s'}\Phi_2^{t'},\quad & \quad \alpha_w(\Phi_1^{s'}\Phi_2^{t'})=\alpha_w(\Phi_1^{s'}\Phi L_2^{t'})\ \operatorname{mod} \frac{1}{2},
    \end{aligned}
    \end{equation}
    and, since the linking with $L_1^2L_2^2$ or $\Phi_1^2\Phi_2^2$ measures exactly the coupling modulo $\frac{1}{2}$:
    \begin{equation*}
    \begin{aligned}
    \langle\qty(L_1^2L_2^2)(\gamma)\qty( \Phi_1^s\Phi_2^t)(x)\rangle=\exp\qty(\frac{2\pi \ii(2s+6t)}{8}l(\gamma, x))=
    \exp\qty(\frac{2\pi \ii\alpha_w(\Phi_1^s\Phi_2^t)l(\gamma, x)}{4})\\
     \langle\qty(\Phi_1^2\Phi_2^2)(x)\qty( L_1^sL_2^t)(\gamma)\rangle=\exp\qty(\frac{2\pi \ii(2s+6t)}{8}l(\gamma, x))=
     \exp\qty(\frac{2\pi \ii\alpha_m(L_1^sL_2^t)}{4}l(\gamma, x)),\
     \end{aligned}
    \end{equation*}
    defects not in $\mathcal{A}_2^{8}\times \mathcal{A}_2^{8,4}[j_w^{(1)},j_m^{(2)}]$ of the theory (i.e.~those for which $(s',t')\neq(0,0)$) aren't gauge invariant and thus do not survive the (second step of the) gauging process.

    Finally, there is an interface from the $\mathcal{A}_2^{2}$ factor to the trivial theory by gauging the $\Z_2^{(1)}$ symmetry generated by $[\Phi_\beta]$:
    \begin{itemize}
        \item $[\Phi_\beta]$ is the only point operator, and the gauging process makes it trivial.
        \item The only line operator $[L_\beta]$  is not gauge invariant:
        \begin{equation}
            \langle[L_\beta](\gamma)[\Phi_\beta](x)\rangle=(-1)^{\operatorname{link}(\gamma, x)},
        \end{equation}
        and doesn't survive the gauging process.
    \end{itemize}
     One could have achieved the same results by noting that, since $\Phi_\beta$ is decoupled to begin with, we are allowed to gauge the full $\Z_8^{(1)}$ in one step; and since it doesn't braid with $L_\alpha^2=L_2^2L_2^2$ we could have simultaneously gauged $\Z_4^{(0)}\times\Z_8^{(1)}$ to begin with. This differs from before in two ways:
    \begin{itemize}
        \item Points are now more finely identified, and the only nontrivial one is given by $[\Phi_\alpha]$ or order $2$.
        \item Lines $L_\beta$ are no longer gauge invariant.
    \end{itemize}
    Moreover, the defects not labelled by either $\alpha$ or $\beta$ are not gauge invariant, for the same reason as before.

    We therefore just constructed an interface of the desired form \ref{ex: electric-fusion}.
\end{example}

\section{Associators}\label{sec: associator}
In order to fully specify the symmetry category $\mathcal{C}$ one has to provide the data of the associativity of the fusion operation $\otimes$. 

\subsection{Noninvertible 0-form symmetry}
Let's study the $F$-symbol of the shift symmetry, which amounts to a review of \cite[section~3.4]{Copetti:2023mcq} by analogy.
\bea\label{eq: Disegno-F-bubble-shift}
\begin{tz}[scale=0.5]
\draw[slice] (0,-0.5) to [out=up, in=\dl] (1,1);
\draw[slice] (2,-0.5) to [out=up, in=\dr] (1,1);
\draw[slice] (1,1) to [in=\dl, out=up] (2,2.5);
\draw[slice] (4,-0.5) to [out=up, in=\dr] (2,2.5) to (2,3.5);
\draw[slice] (0,-0.5) to [in=\ul, out=down] (2,-3.5) to (2,-4.5);
\draw[slice] (2,-0.5) to [out=down, in = \ul] (3,-2);
\draw[slice] (4,-0.5) to [out=down, in = \ur] (3,-2);
\draw[slice] (3,-2) to [out=down, in = \ur] (2,-3.5);
\node[dot, label=left:$\mathbb{A}_{12}$] at (1,1) {}; \node[dot, label=right:$\mathbb{A}_{(12)3}$] at (2,2.5) {}; \node[dot, label=right:$\mathbb{A}_{23}$] at (3,-2) {}; \node[dot, label=left:$\mathbb{A}_{1(23)}$] at (2,-3.5) {};
\node[left] at (3,-0.5) { $\mathcal{D}^{(s)}_{\frac{p_2}{N_2}}$};
\node[left] at (0,-0.5) { $\mathcal{D}^{(s)}_{\frac{p_1}{N_1}}$};
\node[right] at (4,-0.5) { $\mathcal{D}^{(s)}_{\frac{p_3}{N_3}}$};
\node[left] at (1.5,2.25) { $\mathcal{D}^{(s)}_{\frac{p_{12}}{N_{12}}}$};
\node[right] at (2.5,-3.25) { $\mathcal{D}^{(s)}_{\frac{p_{23}}{N_{23}}}$};
\node[above] at (2,3.5) { $\mathcal{D}^{(s)}_{\frac{p}{N}}$};
\node[below] at (2,-4.5) { $\mathcal{D}^{(s)}_{\frac{p}{N}}$};
\node at (7,-0.5) {$=$};
\draw[slice] (8,-4.5) to (8,3.5);
\node[above] at (8,3.5) { $\mathcal{D}^{(s)}_{\frac{p}{N}}$};
\node[below] at (8,-4.5) { $\mathcal{D}^{(s)}_{\frac{p}{N}}$};
\draw[fill=blue] (8,-0.5) circle (0.1); \node[right] at (8.5,-0.5) {$\boldsymbol{F}^{(s)}\begin{bmatrix}\mathbb{A}_{12}&\mathbb{A}_{(12)3}\\\mathbb{A}_{23}&\mathbb{A}_{1(23)}\end{bmatrix}_{p_1/N_1,p_2/N_2,p_3/N_3}$};
\end{tz}
\eea
In order to understand the theory of the $F$-symbol, we need to analyse the interfaces it is made of. Being gauging interfaces of lines in $3$-dimensional defects, they serve multiple roles (see \cref{Appendix: generalized gauging}):
\begin{itemize}
\item they identify lines which differ by elements of the gauged algebra;
\item they constrain to the interface the lines which braid nontrivially with the algebra object;
\item they allow the gauged lines to end topologically on the interface.
\end{itemize}
The $F$-symbol bubble has gauging interfaces on both sides: on the left-hand side the successive gauge of $\mathbb{A}_{23}$ and $\mathbb{A}_{1(23)}$ leads to gauge of $\mathbb{A}^-=\mathbb{A}_{23}\rhd\mathbb{A}_{1(23)}$; on the right-hand side the successive gauge of $\mathbb{A}_{12)}$ and $\mathbb{A}_{(12)3}$ leads to gauge of $\mathbb{A}^+=\mathbb{A}_{12}\rhd\mathbb{A}_{(12)3}$. Denoting with $\mathcal{C}=\mathcal{C}_1 \boxtimes\mathcal{C}_2\boxtimes \mathcal{C}_3$ the total symmetry category of $\mathcal{D}^{(s)}_{\frac{p_1}{N_1}}\otimes \mathcal{D}^{(s)}_{\frac{p_2}{N_2}}\otimes \mathcal{D}^{(s)}_{\frac{p_3}{N_3}}$ and with $\mathbb{A}^{+-}=\mathbb{A}^{+}\cap \mathbb{A}^{-}$ the subgroup gauged on both sides of the bubble, once the $F$-bubble it is shrunk we will be left with (see \cref{fig: Fsandwitch}, and \cite{Copetti:2023mcq} for details):
\begin{figure}
    \centering
    \includegraphics[width=0.8\linewidth]{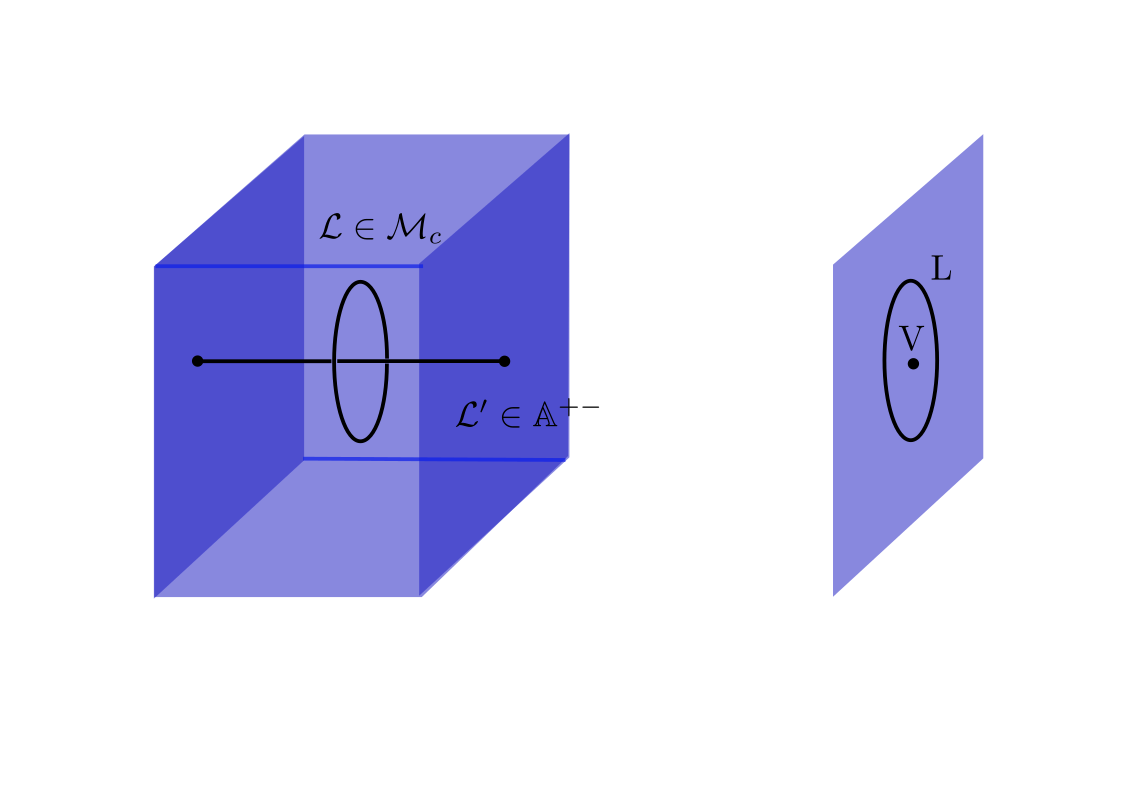}
    \caption{The $F$-symbol bubble can be compressed along the transversal direction, leading to a setup similar to that of the SymTFT. The actual $F$-symbol is obtained by shrinking along the slab direction: topological lines enable on both boundaries give birth to topological point operators, and their braiding with the other lines of the theory descends to a linking. In our case, the blue region hosts the theory $\mathcal{C}_1 \boxtimes\mathcal{C}_2\boxtimes \mathcal{C}_3$; the interfaces implement the half-gauging of $\mathbb{A}^+$ and $\mathbb{A}^-$, and endable lines are in bijection with simple elements of $\mathbb{A}^{+-}$.}
    \label{fig: Fsandwitch}
\end{figure}
\begin{itemize}
\item a set $\mathfrak{M}$ of $\frac{|\mathcal{C}|}{|\mathbb{A}^{+}||\mathbb{A}^{-}|}$ deconfined (orbits of) lines, allowed to escape the $F$-symbol;
\item A set $\mathfrak{M}_c$ of $|\mathbb{A}^{+-}|$ confined (orbits of) lines, constrained to live on the $F$-symbol. Since these lines descend from those which get stuck at the fusion interfaces, they will be coupled to a bigger bulk symmetry subgroup $\Z_{N_{123}}$ (where $N_{123} = \lcm(N_1,N_2,N_3)$) than the $\Z_N$ one which couples to the lines of the defect $\mathcal{D}_{\frac{p}{N}}$: the lines of $\mathfrak{M}_c$ couple to the quotient $\Z_{\frac{N_{123}}{N}}:=\Z_J$;
\item a set $\mathfrak{V}$ of vertex operators labelled by the simple components of $\mathbb{A}^{+-}$.
\end{itemize}
Moreover, the original braiding $\textbf{B}_{i,j}$ between the lines of $\mathcal{C}$ descends to a trivial pairing $\textbf{B}'_{i,j}=0$ between $\mathfrak{M}$ and $\mathfrak{V}$; and a non-degenerate $\widetilde{\textbf{B}}_{i,j}$ pairing between $\mathfrak{M}_c$ and $\mathfrak{V}$. Due to the presence of topological point operators, this theory thus decomposes:
\begin{equation}\label{eq: shift-F}
\boldsymbol{F}^{(s)}[\mathbb{A}^+\mathbb{A}^-]_{\mathcal{D}^{(s)}_{p_1/N_1},\mathcal{D}^{(s)}_{p_2/N_2},\mathcal{D}^{(s)}_{p_3/N_3}} = \bigoplus_{q=0}^{N-1} \left(\bigoplus_{i:\alpha_m(\pi_i)=q} \boldsymbol{F}^{(s)}_{\pi_i}\right) \otimes \eta_{q/N}^{(m)}=d_0\mathcal{E}\bigoplus_{q=0}^{J-1} \eta_{\frac{q}{N_{123}}}^{(m)},
\end{equation}
where all the lines in $\mathfrak{M}_c$ serve the role of domain walls, inducing a magnetic label to the different universes, which are now just given by Euler counterterms $\mathcal{E}$. The last equality follows from the fact that twisted sectors of different labels are equally populated by $d_q=d_0$ defects.

\subsection{Noninvertible $1$-form symmetry}
Let's now study the $F$-symbol of the electric symmetry. There is a strong analogy to that of the shift symmetry or the axial symmetry of \cite{Copetti:2023mcq}, but there are important differences. The $F$-symbol is now a one-dimensional topological quantum mechanics rather than a 2d TQFT, and the bulk contains both line and point operators that descend to the interface.
\bea\label{eq: Disegno-F-bubble-Electric}
\begin{tz}[scale=0.5]
\draw[slice] (0,-0.5) to [out=up, in=\dl] (1,1);
\draw[slice] (2,-0.5) to [out=up, in=\dr] (1,1);
\draw[slice] (1,1) to [in=\dl, out=up] (2,2.5);
\draw[slice] (4,-0.5) to [out=up, in=\dr] (2,2.5) to (2,3.5);
\draw[slice] (0,-0.5) to [in=\ul, out=down] (2,-3.5) to (2,-4.5);
\draw[slice] (2,-0.5) to [out=down, in = \ul] (3,-2);
\draw[slice] (4,-0.5) to [out=down, in = \ur] (3,-2);
\draw[slice] (3,-2) to [out=down, in = \ur] (2,-3.5);
\draw[fill=blue] (1,1) circle (0.1); \node[left] at (1,1) {$(\tilde{\mathbb{A}}_{12},\pi_{12})$};\draw[fill=blue] (2,2.5) circle (0.1); \node[right] at (2,2.5) {$(\tilde{\mathbb{A}}_{(12)3},\pi_{(12)3})$};\draw[fill=blue] (3,-2) circle (0.1); \node[right] at (3,-2) {$(\tilde{\mathbb{A}}_{23}, \pi_{23})$};\draw[fill=blue] (2,-3.5) circle (0.1); \node[left] at (2,-3.5) {$(\tilde{\mathbb{A}}_{1(23)},\pi_{1(23)})$};
\node[left] at (3,-0.5) { $\mathcal{D}^{(e)}_{\frac{p_2}{N_2}}$};
\node[left] at (0,-0.5) { $\mathcal{D}^{(e)}_{\frac{p_1}{N_1}}$};
\node[right] at (4,-0.5) { $\mathcal{D}^{(e)}_{\frac{p_3}{N_3}}$};
\node[left] at (1.5,2.25) { $\mathcal{D}^{(e)}_{\frac{p_{12}}{N_{12}}}$};
\node[right] at (2.5,-3.25) { $\mathcal{D}^{(e)}_{\frac{p_{23}}{N_{23}}}$};
\node[above] at (2,3.5) { $\mathcal{D}^{(e)}_{\frac{p}{N}}$};
\node[below] at (2,-4.5) { $\mathcal{D}^{(e)}_{\frac{p}{N}}$};
\node at (7,-0.5) {$=$};
\draw[slice] (8,-4.5) to (8,3.5);
\node[above] at (8,3.5) { $\mathcal{D}^{(e)}_{\frac{p}{N}}$};
\node[below] at (8,-4.5) { $\mathcal{D}^{(e)}_{\frac{p}{N}}$};
\draw[fill=blue] (8,-0.5) circle (0.1); \node[right] at (8.5,-0.5) {$\boldsymbol{F}^{(e)}\begin{bmatrix}(\tilde{\mathbb{A}}_{12},\pi_{12})&(\tilde{\mathbb{A}}_{(12)3},\pi_{(12)3})\\(\tilde{\mathbb{A}}_{23}, \pi_{23})&(\tilde{\mathbb{A}}_{1(23)},\pi_{1(23)})\end{bmatrix}_{p_1/N_1,p_2/N_2,p_3/N_3}$};
\end{tz}
\eea

In this case, the $F$-symbol is composed of gauging interfaces of both lines and points, which implement the following operations:
\begin{itemize}
\item they identify lines and points which differ by elements of the gauged algebras;
\item they constrain to the interface the lines which link nontrivially with the point algebra object, and vice versa the point nontrivially charged under the algebra of lines;
\item they allow the gauged lines to end topologically on the interface.
\end{itemize}
The $F$-symbol bubble has gauging interfaces on both sides: on the left-hand side the successive gauging of $\tilde{\mathbb{A}}_{23}$ and $\tilde{\mathbb{A}}_{1(23)}$ algebra of lines and $\pi_{23}$ and $\pi_{1(23)}$ algebra of points leads to gauging of $\tilde{\mathbb{A}}^-=\tilde{\mathbb{A}}_{23}\rhd\tilde{\mathbb{A}}_{1(23)}$ and $\pi^-=\pi_{23}\rhd\pi_{1(23)}$; on the right-hand side the successive gauging of $\tilde{\mathbb{A}}_{12)}$ and $\tilde{\mathbb{A}}_{(12)3}$ algebra of lines and $\pi_{12}$ and $\pi_{(12)3}$ algebra of points leads to gauging of $\tilde{\mathbb{A}}^+=\tilde{\mathbb{A}}_{12}\rhd\tilde{\mathbb{A}}_{(12)3}$ and $\pi^+=\pi_{12}\rhd\pi_{(12)3}$. Denoting with $\mathcal{C}=\mathcal{C}_1 \boxtimes\mathcal{C}_2\boxtimes \mathcal{C}_3$ the total symmetry category of the defect $\mathcal{D}^{(e)}_{\frac{p_1}{N_1}}\otimes\mathcal{D}^{(e)}_{\frac{p_2}{N_3}}\otimes\mathcal{D}^{(e)}_{\frac{p_2}{N_3}}$ and with $\tilde{\mathbb{A}}^{+-}=\tilde{\mathbb{A}}^{+}\cap \tilde{\mathbb{A}}^{-}$ and $\pi^{+-}=\pi^+\cap\pi^-$ the algebras gauged on both sides, once the $F$-bubble it is shrunk we will be left with:
\begin{itemize}
\item a set $\tilde{\mathfrak{M}}$ of $\frac{|\mathcal{C}||\tilde{\mathbb{A}}^{+-}|}{|\tilde{\mathbb{A}}^{+}||\tilde{\mathbb{A}}^{-}||\pi^{+-}|}$ deconfined (orbits of) lines and $\frac{|\Omega\mathcal{C}||\pi^{+-}|}{|\pi^{+}||\pi^{-}||\tilde{\mathbb{A}}^{+-}|}$ deconfined (orbits of) points, allowed to escape the $F$-symbol. The cardinality is obtained by first looking at the neutral lines $\frac{|\mathcal{C}|}{|\pi^{+-}|} = \left.\frac{|\mathcal{C}|}{|\pi^+|} \frac{|\mathcal{C}|}{|\pi^-|}\middle/\frac{|\mathcal{C}|}{|\pi^-||\pi^-|/|\pi^{+-}|}\right.$ (those neutral w.r.t.~$\pi^{+-}$, or alternatively those neutral w.r.t.~either $\pi^+$ or $\pi^-$, in the same way as in \cite{Copetti:2023mcq}) and then identifying those who differ by algebra objects, i.e.~dividing by $\frac{|\tilde{\mathbb{A}}^{+}||\tilde{\mathbb{A}}^{-}|}{|\tilde{\mathbb{A}}^{+-}|}$, and vice versa for points;
\item a set $\tilde{\mathfrak{M}}_c$ of $|\pi^{+-}|$ confined lines, constrained to live on the $F$-symbol,
  where the cardinality is obtained by dividing the total number of line orbits $\frac{|\mathcal{C}||\tilde{\mathbb{A}}^{+-}|}{|\tilde{\mathbb{A}}^{+}||\tilde{\mathbb{A}}^{-}|}$ by the number of deconfined ones $\frac{|\mathcal{C}||\tilde{\mathbb{A}}^{+-}|}{|\tilde{\mathbb{A}}^{+}||\tilde{\mathbb{A}}^{-}||\pi^{+-}|}$.
Since these lines descend from those which get stuck at the fusion interfaces, they will be coupled to a bigger bulk symmetry subgroup $\Z_{N_{123}}$ than the $\Z_N$ one which couples to the lines of the defect $\mathcal{D}_{\frac{p}{N}}$: the lines of $\tilde{\mathfrak{M}}_c$ couples to the quotient $\Z_{\frac{N_{123}}{N}}=\Z_J$;
\item a set $\mathfrak{P}_c$ of $|\tilde{\mathbb{A}}^{+-}|$ confined points, constrained to live on the $F$-symbol.
Since these points descend from those which get stuck at the fusion interfaces, they will be coupled to a bigger bulk symmetry subgroup $\Z_{N_{123}}$ than the $\Z_N$ one which couples to the points of the defect $\mathcal{D}_{\frac{p}{N}}$: the points of $\tilde{\mathfrak{M}}_c$ couples to the quotient $\Z_{\frac{N_{123}}{N}}=\Z_J$;
\item a set $\mathfrak{V}$ of vertex operators labelled by the simple components of $\tilde{\mathbb{A}}^{+-}$.
\end{itemize}
Moreover the original linking $\textbf{B}$ between the lines and points of $\mathcal{C}$ descends to a trivial pairing between deconfined points and $\mathfrak{P}_c\times\mathfrak{V}$, and a non-degenerate pairing $\widetilde{\textbf{B}}$ between $\mathfrak{P}_c$ and $\mathfrak{V}$, or equivalently a torus algebra:
\begin{equation}
    V\Phi=\widetilde{\textbf{B}}(\Phi, V) \Phi V.
\end{equation} This topological quantum mechanics thus decomposes as:
\begin{equation}\label{eq: electri-F}
\boldsymbol{F}^{(e)}[\tilde{\mathbb{A}}^+\pi^+;\tilde{\mathbb{A}}^-\pi^-]_{\mathcal{D}^{(e)}_{p_1/N_1},\mathcal{D}^{(e)}_{p_2/N_2},\mathcal{D}^{(e)}_{p_3/N_3}}=\bigoplus_{q=0}^{N-1}\left(\bigoplus_{i:\alpha_w(\pi_i)=q}\boldsymbol{F}^{(e)}_{\pi_i}\right)\otimes\eta_{q/N}^{(w)}=d_0\bigoplus_{q=0}^{J-1} \eta_{\frac{q}{N_{123}}}^{(w)} ,
\end{equation}
where the coupling of the confined points induces a winding label on the different universes, and the last equality follows from the fact that twisted sectors of different labels are equally populated by $d_q=d_0$ defects.

The last equality of \eqref{eq: electri-F} holds in the absence of operator insertions on the $F$-symbol. Inserting confined operators gives rise to a family of twisted interfaces attached to bulk symmetry defects. In particular, there is a family of topological interfaces
\begin{equation}
    \boldsymbol{F}^{(e)}[\tilde{\mathbb{A}}^+\pi^+;\tilde{\mathbb{A}}^-\pi^-]_{\mathcal{D}^{(e)}_{p_1/N_1},\mathcal{D}^{(e)}_{p_2/N_2},\mathcal{D}^{(e)}_{p_3/N_3}}
    \otimes L \otimes \eta^{(m)}_{\alpha_m(L)}
    = d_0\bigoplus_{q=0}^{J-1} \eta_{\frac{q}{N_{123}}}^{(w)}
    \otimes L \otimes \eta^{(m)}_{\alpha_m(L)}
\end{equation}
where $L$ is a confined line attached to a magnetic surface $\eta^{(m)}_{\alpha_m(L)}$, and similarly for confined points.

\subsection{Detecting the $F$-symbol}\label{par: detectingF}
\begin{figure}[ht!]
    \centering
    \includegraphics[width=0.8\linewidth]{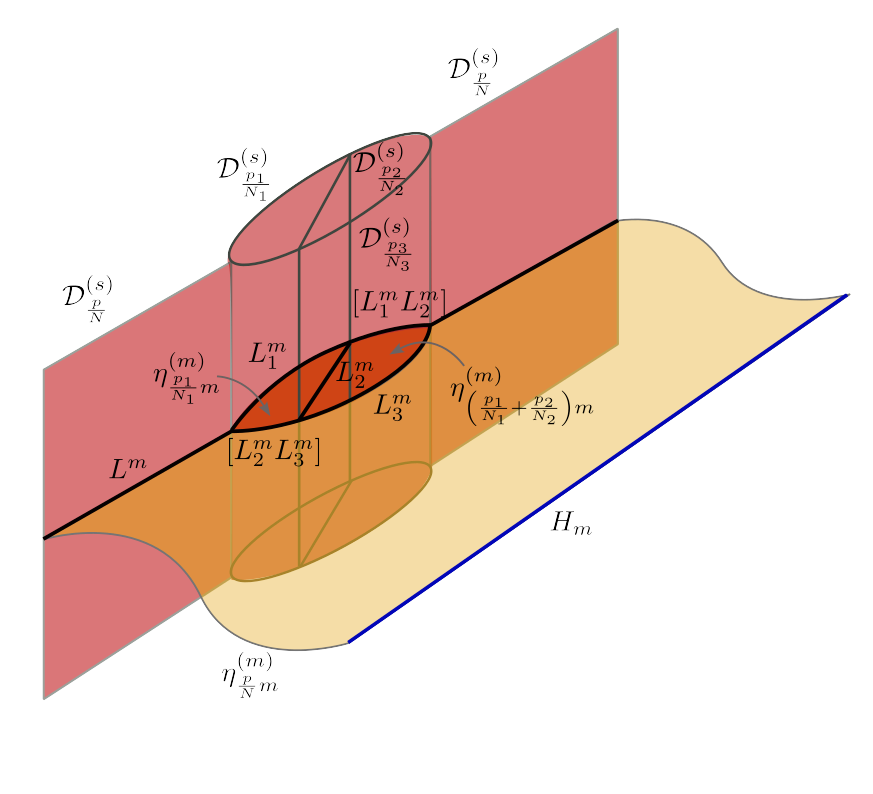}
    \caption{Detecting the shift $F$-symbol: acting transversally on a 't Hooft line $H_m$ leads to a coherent insertion of lines on the $F$-symbol bubble of $\mathcal{D}^{(s)}_{\frac{p}{N}}$. When $N|m$ there is no magnetic defect extending between the 't Hooft line and the bubble, leaving a genuine defect on it.}
    \label{fig:Hooft Shift Detection}
\end{figure}

\begin{figure}[ht!]
    \centering
    \subfloat[Subfigure 1 list of figures text][]{
        \includegraphics[width=0.5\linewidth]{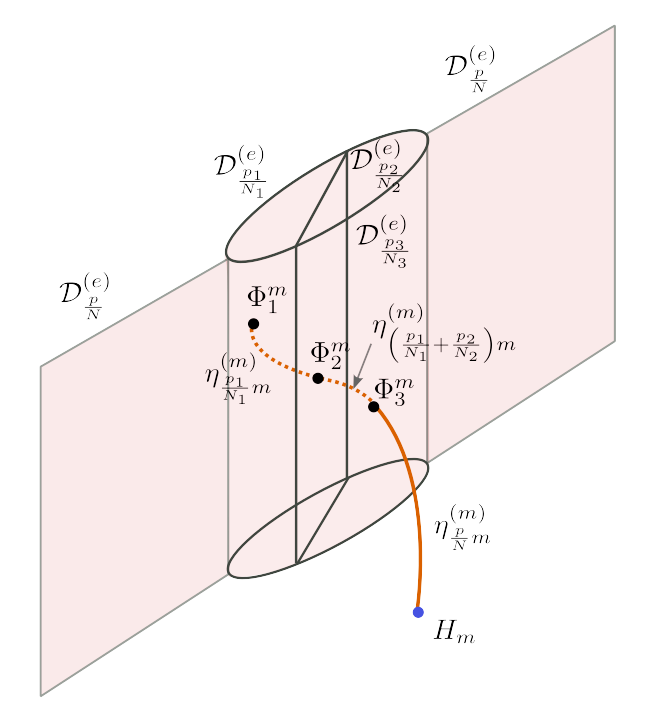}\label{fig: Hooft Electric detection 1}
    }%
    ~ 
    \subfloat[Subfigure 2 list of figures text][]{
        \includegraphics[width=0.5\linewidth]{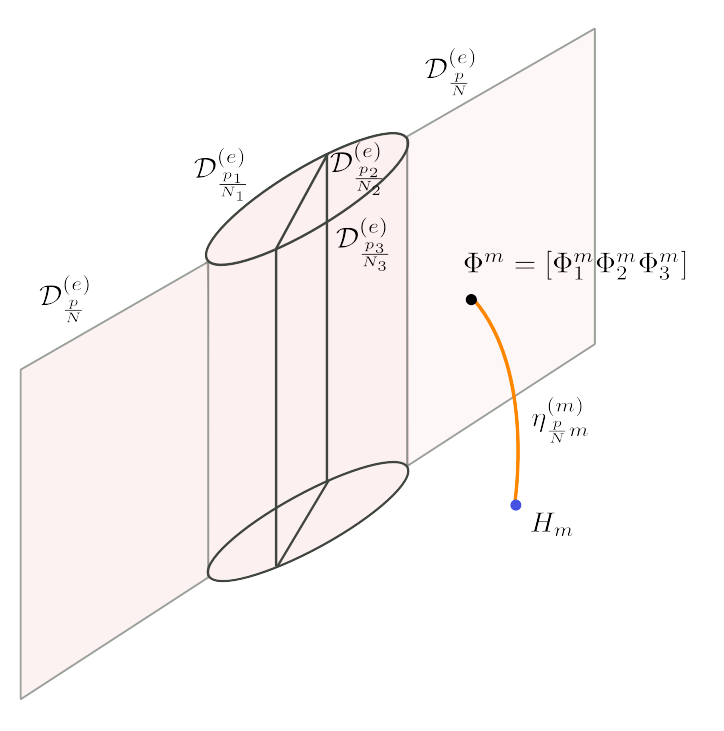}\label{fig: Hooft Electric detection 2}
    }
    \caption{Detecting the electric $F$-symbol: acting on a 't Hooft line $H_m$ leads to an insertion of 
 a pointlike topological operator on the $F$-symbol bubble of $\mathcal{D}^{(e)}_{\frac{p}{N}}$. When $N|m$ there is no winding defect extending between the 't Hooft line and the bubble, leaving a genuine defect on it.}\label{fig: Global Hooft Electric detection}
\end{figure}

We would now like to use the previously defined defects to produce new predictions. As a first step, we study the effect of inserting a configuration of defects forming the $F$-symbol and letting it act on the various operators of the theory:
\begin{itemize}
    \item \textbf{'t Hooft lines transverse to the shift $F$-symbol:} Sweeping a shift $F$-symbol bubble through an 't Hooft line $H_m$ attaches it to a magnetic defect $\eta^{(m)}_{\frac{mp}{N}}$, which itself attaches to lines on the bubble. The coherence at each junction of the labelling of the lines left behind on the bubble,
    \begin{equation}
        (L_i)_{\mathcal{D}^{(s)}_{p_i/N_i}}\otimes (L_j)_{\mathcal{D}^{(s)}_{p_j/N_j}}\leadsto m_{M'_{ij}} \cdot (L_i\otimes L_j)_{\mathcal{D}^{(s)}_{p_i/N_i+p_j/N_j}},
    \end{equation}
    and the coherence of the magnetic label at each region,
    \begin{equation}
        \eta^{(m)}_\alpha\xrightarrow{L^s_{\mathcal{D}^{(s)}_{\frac{p}{N}}}}\eta^{(m)}_{\alpha+s\frac{p}{N}}
    \end{equation}
    is exactly due to the fact that the fusion interfaces were built using the principle of consistent coupling (see \cref{fig:Hooft Shift Detection}).

    Choosing $m=N$, the 't Hooft line doesn't get attached to any magnetic defect, and the nontrivial configuration is localized only on the $F$-symbol. Using the result \eqref{eq: shift-F} we obtain:
    \begin{equation}
      \boldsymbol{F}^{(s)}[\mathbb{A}^{+} \mathbb{A}^{-}]_{p_{1}/N_{1},p_{2}/N_{2},p_{3}/N_{3}}\otimes H_N=H_N\otimes d_{0}\bigoplus_{q=0}^{J-1}e^{\frac{2\pi iq}{J}} \eta^{(m)}_{\frac{q}{N_{123}}}.
    \end{equation}
    Upon shrinking the support of the $F$-symbol, what is left is the factor
    \begin{equation}
      d_{0}\sum_{q=0}^{J-1}e^{\frac{2\pi iq}{J}} =
      \begin{cases} d_0 & J = 1 \\ 0 & J > 1, \end{cases}
    \end{equation}
    so that the 't Hooft line distinguishes between a trivial ($J = 1$) and nontrivial ($J > 1$) $F$-symbol.

  \item \textbf{'t Hooft lines parallel to the shift $F$-symbol:} Now shrinking the bubble will lead to deconfined lines of the theory of the $F$-symbol.

\item \textbf{Axion strings transverse to the electric $F$-symbol:} This case is precisely analogous to that of 't Hooft lines and the shift $F$-symbol. Sweeping an electric bubble through an axion string $S_w$ again attaches it to a magnetic defect $\eta^{(m)}_{\frac{mp}{N} }$ which itself attaches to lines on the bubble.
  Choosing $w=N$ the string doesn't get attached to any magnetic defect and the nontrivial configuration is localized only on the $F$-symbol. Using the result \cref{eq: electri-F} we obtain:
    \begin{equation}
      \boldsymbol{F}^{(e)}[\tilde{\mathbb{A}}^+\pi^+;\tilde{\mathbb{A}}^-\pi^-]_{\mathcal{D}^{(e)}_{p_1/N_1},\mathcal{D}^{(e)}_{p_2/N_2},\mathcal{D}^{(e)}_{p_3/N_3}}\otimes S_N=S_N\otimes d_{0}\bigoplus_{q=0}^{J-1}e^{\frac{2\pi iq}{J}} \eta^{(w)}_{\frac{q}{N_{123}}},
    \end{equation}
    and shrinking the (now one-dimensional) support of the $F$-symbol again leaves behind the factor
    \begin{equation}
      d_{0}\sum_{q=0}^{J-1}e^{\frac{2\pi iq}{J}} =
      \begin{cases} d_0 & J = 1 \\ 0 & J > 1, \end{cases}
    \end{equation}
    so that an axion string distinguishes between a trivial ($J = 1$) and nontrivial ($J > 1$) $F$-symbol.

    \item \textbf{Axion strings parallel to the electric $F$-symbol:} now shrinking the bubble will lead to deconfined lines of the theory of the $F$-symbol (whose topological nature is guaranteed by the same coherence across junctions as before, see \cref{fig: Hooft Electric detection 1,fig: Hooft Electric detection 2}).

    \item \textbf{'t Hooft lines and the electric $F$-symbol:} now shrinking the bubble will lead to deconfined points of the theory of the $F$-symbol.

\end{itemize}

\subsection{Constraints and dualities}
\begin{figure}[ht!]
  \centering
  \subfloat[Subfigure 1 list of figures text][]{
          \includegraphics[width=0.6\linewidth]{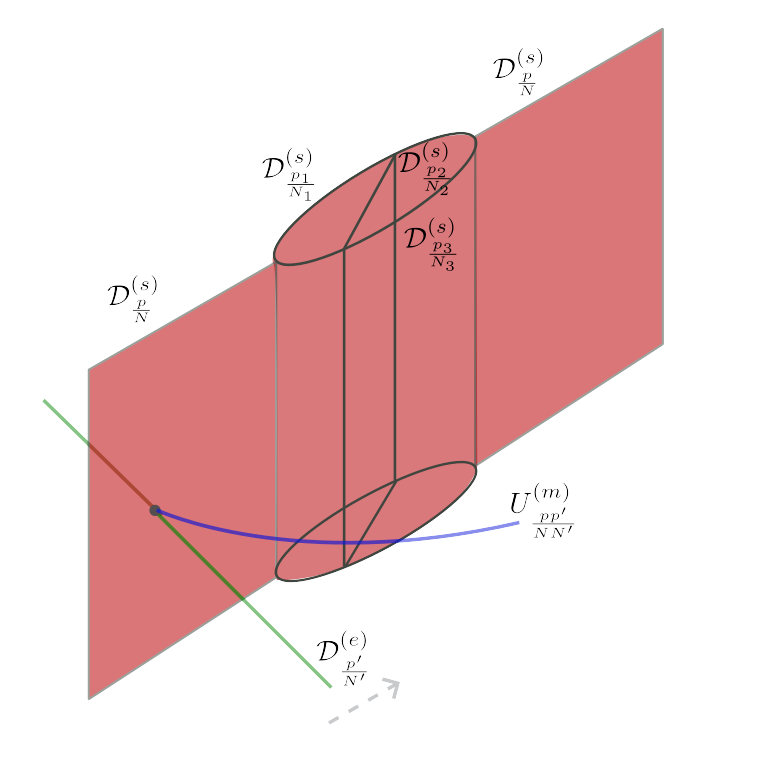}\label{fig: Higher group consistency 1}
  }%
  \\
  \subfloat[Subfigure 2 list of figures text][]{
      \includegraphics[width=0.7\linewidth]{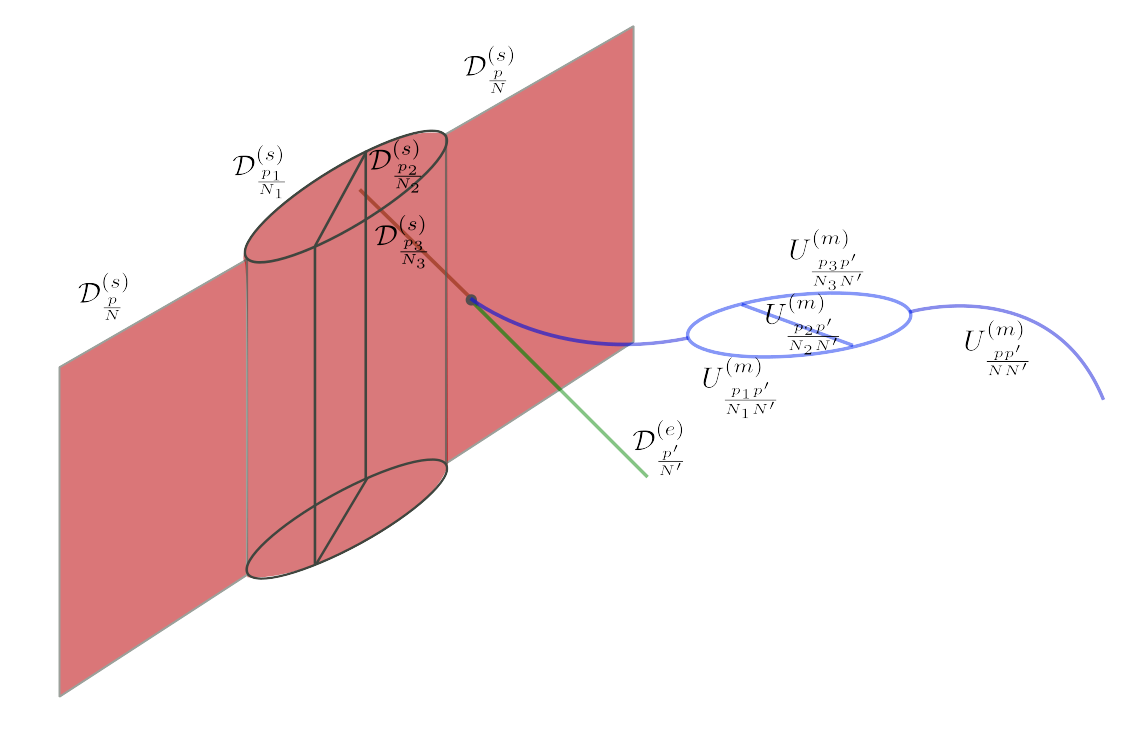}\label{fig: Higher group consistency 2}
  }
  \caption{The presence of a (non-invertible) higher group structure implies that moving an electric symmetry defect through an $F$-symbol of the shift symmetry, produces a $F$-symbol of the magnetic symmetry (which is trivial in our case).}\label{fig: Higher group consistency}
\end{figure}

\paragraph{Interplay with the (non-invertible) higher group structure}\label{par: consistency 3group}
In this section, we are going to explore how the nontrivial higher group structure affects associativity.
\begin{itemize}
\item We remember that the one-dimensional intersection of an electric defect $\mathcal{D}^{(e)}_{p/N}$ and a shift defect $\mathcal{D}^{(s)}_{p'/N'}$ sources a magnetic defect $\eta^{(m)}_{(pp')/(NN')}$.
    If we move this configuration across an $F$-symbol for the shift defect, the sourced magnetic defect will split and recompose according to the shift $F$-symbol bubble (see \cref{fig: Higher group consistency}). After shrinking, we will end up with a magnetic $F$-symbol.
    However, since the $F$-symbol of the magnetic symmetry is trivial, this part of the higher group doesn't lead to new constraints.

    Similarly, moving the configuration across the electric $F$-symbol, will have the same effect.
  \item Similarly, recall that the zero-dimensional intersection of two electric defects $\mathcal{D}^{(e)}_{p/N}$ and $\mathcal{D}^{(e)}_{p'/N'}$ sources a winding defect $\eta^{(w)}_{(pp')/(NN')}$. Moving this configuration across an electric $F$-symbol gives rise to a winding $F$-symbol bubble (see \cref{fig: Higher group consistency}), which again is trivial and does not lead to new constraints.
\end{itemize}

\paragraph{Rotational constraints}\label{par: framing}
In this paragraph, we are going to derive some dualities between different $F$-symbol for the electric symmetry (and more generally higher-form symmetries) coming from topological manipulations.

\begin{figure}[ht!]
    \centering
    \includegraphics[width=0.8\linewidth]{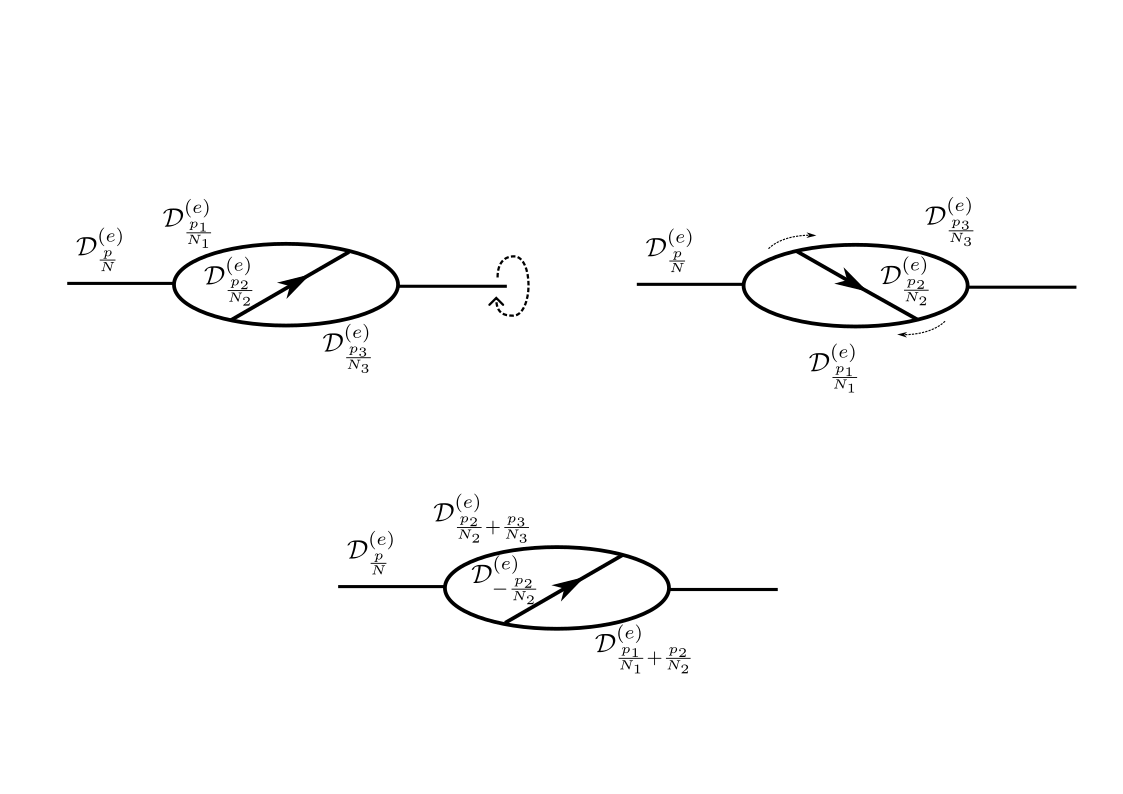}
    \caption{Topological manipulations leading to dualities between electric $F$-symbols.}
    \label{fig:Fduality}
\end{figure}

Let us consider a $F$-bubble and perform a $180^\circ$ rotation on a plane orthogonal to $\mathcal{D}^{(e)}_{p/N}$. After swapping the relative position of the fusions $\mathcal{D}^{(e)}_{\frac{p_1}{N_1}}\otimes\mathcal{D}^{(e)}_{\frac{p_2}{N_2}}$ and $\mathcal{D}^{(e)}_{\frac{p_2}{N_2}}\otimes \mathcal{D}^{(e)}_{\frac{p_3}{N_3}}$, the rotated configuration will differ from the original one in the following way (see \cref{fig:Fduality}):
\begin{equation}
    \mathcal{D}^{(e)}_{\frac{p_1}{N_1}}\leftrightarrow\mathcal{D}^{(e)}_{\frac{p_2}{N_2}+\frac{p_3}{N_3}}, \quad \mathcal{D}^{(e)}_{\frac{p_3}{N_3}}\leftrightarrow\mathcal{D}^{(e)}_{\frac{p_1}{N_1}+\frac{p_2}{N_2}}, \quad \mathcal{D}^{(e)}_{\frac{p_2}{N_2}}\rightarrow\overline{\mathcal{D}^{(e)}_{\frac{p_2}{N_2}}}=\mathcal{D}^{(e)}_{-\frac{p_2}{N_2}}
\end{equation}
Finally, after shrinking the bubble we will be left with the following defect:
\begin{equation}
    \boldsymbol{F}^{(e)}_{(\frac{p_2}{N_2}+\frac{p_3}{N_3}),-p_{2}/N_{2},(\frac{p_1}{N_1}+\frac{p_2}{N_2})},
\end{equation}
and thus with the identity\footnote{The identity could actually be spoiled by phase factors which detects the framing dependence of the defects. While we are not able to formally prove it, we believe both the electric and the shift defects to have trivial frame anomaly.}:
\begin{equation}\label{eq: Fduality}
    \boldsymbol{F}^{(e)}_{p_{1}/N_{1},p_{2}/N_{2},p_{3}/N_{3}}=\boldsymbol{F}^{(e)}_{(\frac{p_2}{N_2}+\frac{p_3}{N_3}),-p_{2}/N_{2},(\frac{p_1}{N_1}+\frac{p_2}{N_2})}.
\end{equation}

\begin{figure}[ht!]
    \centering
    \includegraphics[width=0.8\linewidth]{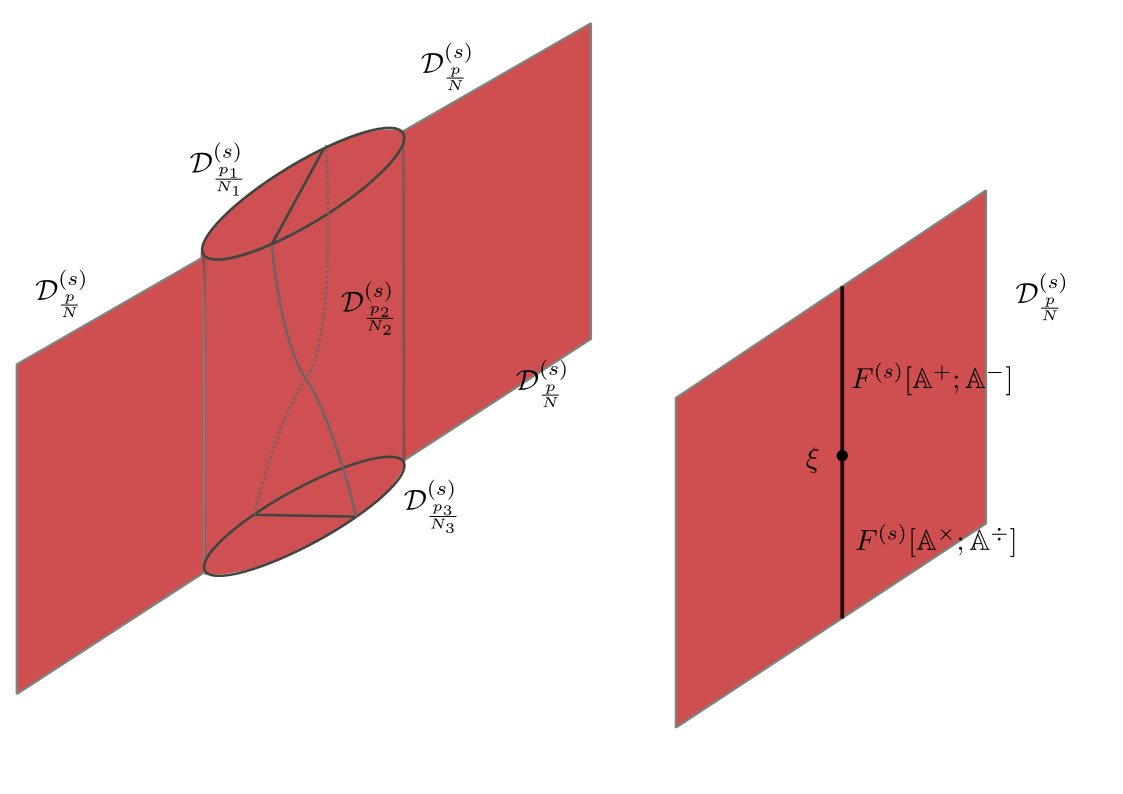}
    \caption{Topological manipulation leading to the existence of morphisms between $F$-symbols for the shift symmetry.}
    \label{fig:Higher Fduality}
\end{figure}

A similar argument also applies to the shift symmetry. While, we cannot rotate across any axes, as this would make the defect sweep the whole spacetime, we can still modify the bubble by swapping the relative position of the fusions $\mathcal{D}^{(s)}_{\frac{p_1}{N_1}}\otimes\mathcal{D}^{(s)}_{\frac{p_2}{N_2}}$ and $\mathcal{D}^{(s)}_{\frac{p_2}{N_2}}\otimes \mathcal{D}^{(s)}_{\frac{p_3}{N_3}}$ (see \cref{fig:Higher Fduality}). This new bubble can be read as:
\begin{equation}
     \boldsymbol{F}^{(s)}[\mathbb{A}^{\times}\mathbb{A}^{\divisionsymbol}]_{\mathcal{D}^{(s)}_{p_1/N_1},\overline{\mathcal{D}^{(s)}_{p_2/N_2}},\mathcal{D}^{(s)}_{p_3/N_3}},
\end{equation}
where $\mathbb{A}^{\times}=\overline{\mathbb{A}_{23}}\rhd\mathbb{A}_{(12)3}$ and $\mathbb{A}^{\divisionsymbol}=\overline{\mathbb{A}_{12}}\rhd\mathbb{A}_{1(23)}$.
Even more precisely, one could develop the swap across the "time" axis. Collapsing this configuration will lead to a pointlike interface of the form (see \cref{fig:Higher Fduality}):
\begin{equation}
     \boldsymbol{F}^{(s)}[\mathbb{A}^+\mathbb{A}^-]_{\mathcal{D}^{(s)}_{p_1/N_1},\mathcal{D}^{(s)}_{p_2/N_2},\mathcal{D}^{(s)}_{p_3/N_3}}\xrightarrow{\xi}\boldsymbol{F}^{(s)}[\mathbb{A}^{\times}\mathbb{A}^{\divisionsymbol}]_{\mathcal{D}^{(s)}_{p_1/N_1},\overline{\mathcal{D}^{(s)}_{p_2/N_2}},\mathcal{D}^{(s)}_{p_3/N_3}}.
\end{equation}

\subsection{Example}
We conclude the section with an explicit computation of an electric $F$-symbol, and the check of the corresponding duality.
We are going to explicitly compute the $F$-symbol for the following configuration:
\begin{equation}\mathds{1}\xrightarrow{\quad \overline{m}_{4,4 } \quad} \mathcal{D}^{(e)}_{1/4} \otimes \mathcal{D}^{(e)}_{1/4} \otimes \mathcal{D}^{(e)}_{1/2}  \xrightarrow{\quad m_{2,2}\circ m_{2,4} \quad} \;\mathds{1}.\end{equation}
as well as verify the proposed duality \ref{eq: Fduality}.

One can compute the fusion interfaces, which are going to define the $F$-symbol:
\begin{equation}
\boldsymbol{F} ^{(e)}\begin{bmatrix}
\{(L_1 L_2)^2, (\Phi_1 \Phi_2)^2\}& \{(L_1 L_2L_3), (\Phi_1 \Phi_2\Phi_3)\} \\
\{\mathds{1}, \mathds{1}\} & \{(L_1 L_2L_3), (\Phi_1 \Phi_2\Phi_3)\}
\end{bmatrix}_{1/4, \, 1/4, \, 1/2},
\end{equation}
computing the difference between a gauging of a $\Z_4^{(0)}\times\Z_4^{(1)}$ subgroup and the composition of two $\Z_2^{(0)}\times\Z_2^{(1)}$ gaugings. We have $\tilde{\mathbb{A}}^+=\tilde{\mathbb{A}}^-$ and $\pi^+=\pi^+$. 
The quantum mechanics of the $F$-symbol hosts the following pointlike operators:
\begin{equation}
\begin{aligned}
      V=(L_1 L_2L_3), \quad V^4=\mathds{1},\\
    \Phi=\prescript{}{\pi^-}{\Phi_1\otimes \mathds{1}\otimes\mathds{1}}_{\pi^+}\,\quad \Phi^4=\mathds{1}.
    \end{aligned}  
\end{equation}

We conclude that the $F$-symbol is a winding condensate:
\begin{equation}
    \boldsymbol{F} ^{(e)}\begin{bmatrix}
\{(L_1 L_2)^2, (\Phi_1 \Phi_2)^2\}& \{(L_1 L_2L_3), (\Phi_1 \Phi_2\Phi_3)\} \\
\{\mathds{1}, \mathds{1}\} & \{(L_1 L_2L_3), (\Phi_1 \Phi_2\Phi_3)\}
\end{bmatrix}_{1/4, \, 1/4, \, 1/2}=\mathcal{C}^{(2,w)}_4.
\end{equation}

Let us now perform the topological manipulation described in \cref{par: framing}, and compute the $F$-symbol corresponding to:
\begin{equation}
    \mathds{1}\xrightarrow{\quad \overline{m}_{4,4 } \quad} \mathcal{D}^{(e)}_{1/4} \otimes \mathcal{D}^{(e)}_{1/4} \otimes \mathcal{D}^{(e)}_{1/2}  \xrightarrow{\quad m_{2,2}\circ m_{2,4} \quad} \;\mathds{1}.
\end{equation}
The $F$-symbol in this case is computed as:
\begin{equation}
    \boldsymbol{F} ^{(e)}\begin{bmatrix}
\{(L_\beta L_{\overline{2}})^2, (\Phi_\beta \Phi_{\overline{2}})^2\}& \{(L_\beta L_{\overline{2}}L_\alpha), (\Phi_\beta \Phi_{\overline{2}}\Phi_\alpha)\} \\
\{\mathds{1}, \mathds{1}\} & \{(L_\beta L_{\overline{2}}L_\alpha), (\Phi_\beta \Phi_{\overline{2}}\Phi_\alpha)\}\end{bmatrix},
\end{equation}
where we denoted the defect of the theory living on $\mathcal{D}^{(s)}_{12}=\mathcal{D}^{(s)}_{12}$ with the subscript $\alpha$, the defect of the theory living on $\mathcal{D}^{(s)}_{23}=\mathcal{D}^{(s)}_{12}$ with the subscript $\beta$ and the defect of the theory living on $\overline{\mathcal{D}}^{(s)}_{2}=\mathcal{D}^{(s)}_{-\frac{1}{4}}$ with the subscript $\overline{2}$.
Once again we have $\tilde{\mathbb{A}}^+=\tilde{\mathbb{A}}^-$ and $\pi^+=\pi^+$ and the spectrum is given by:
\begin{equation}
    \begin{aligned}
      V=(L_\beta L_{\overline{2}}L_\alpha), \quad V^4=\mathds{1},\\
    \Phi=\prescript{}{\pi^-}{\Phi_\beta\otimes \mathds{1}\otimes\mathds{1}}_{\pi^+},\quad \Phi^4=\mathds{1},
    \end{aligned}
\end{equation}
which can be obtained by the previous one by the relabelling:
\begin{equation}
    1\leftrightarrow\beta, \quad2\leftrightarrow\overline{2}, \quad 3\leftrightarrow\alpha,
\end{equation}
which immediately verifies the claimed identity between the two.

\section*{Acknowledgments}
We thank Shani Meynet for discussions and collaboration during the initial stages of this project. MDZ thanks Christian Copetti, Kantaro Ohmori and Yifan Wang for useful discussions.
MDA thanks Kenichi Konishi, Mihail Mintchev, Alessandro Vichi for their insightful remarks.

\section*{Declarations}

\paragraph{Funding.} The work of MDZ and ERG has received funding from the European Research Council (ERC) under the European Union’s Horizon 2020 research and innovation program (grant agreement No. 851931) and from the Simons Foundation (grant \#888984, Simons Collaboration on Global Categorical Symmetries). MDZ also acknowledges support from the VR project grant No. 2023-05590 and the VR Centre for Geometry and Physics (VR grant No. 2022-06593). 
The work of MDA has received funding from the MacCracken Fellowship and by the Erasmus+ traineeships program. Most of the work was done while MDA was affiliated with the Scuola Normale Superiore and the University of Pisa.

\paragraph{Conflict of interest.} The authors have no competing interests to declare that are relevant to the content of this
article. The authors have no financial or proprietary interests in any material discussed in this article.

\paragraph{Data availability statement.} This manuscript has no associated data.

\numberwithin{theorem}{section}
\begin{appendices}
\section{Cohomology of classifying spaces}\label{Appendix: cohomology}
In this appendix, we report some results regarding the singular cohomology of some classifying spaces. Physically, they compute (some) bosonic SPTs, and thus classify anomalies and discrete torsions for fixed symmetry types.

Since in the text we often deal with low dimensional theories with a $\Z_n^{(0)}\times\Z_n^{(1)}$ symmetries, we will focus on the cohomology of $B\Z_n\times B^2\Z_n$.

Using our knowledge of the cohomology groups of $B\Z_n$ and $B^2\Z_n$ \cite{Benini_2019, MacLane},
\begin{equation}
\begin{split}
    H^\bullet(B\Z_n; \Z)&=\Z [c_1]/(nc_1)\\
    H^i(B^2\Z_n;\Z)&=\{\Z; 0;0; \Z_n; 0; \Z_{(2)n}; 
    ... \}
\end{split}
\end{equation}
(where the notation $(2)n$ means $2n$ if $n$ even and $n$ if $n$ is odd),
and the Künneth formula
\begin{equation}
    H^i(X\times Y; \Z)= \bigoplus_{p+q=i}H^p(X; \Z)\otimes H^q(Y; \Z) \bigoplus_{p+q=i+1}\operatorname{Tor}_1^{\Z}\left(H^p(X; \Z), H^q(Y; \Z)\right),
\end{equation} we can conclude:
\begin{equation}
    H^i(B\Z_n\times B^2\Z_n; \Z)=\{\Z; 0; \Z_n; \Z_n; \Z_n^2; \Z_{(2)n}\times \Z_n; \Z_n^2\times \Z_{(2)n}\times H^6(B^2\Z_n; \Z); ...\}.
\end{equation}
We can actually be more precise, and keep  track where the various contributions come from. For example:
        \begin{equation}
        \begin{split}
             H^3(B\Z_n\times B^2\Z_n; \Z)&=  H^3(B^2\Z_n; \Z)\\
              H^4(B\Z_n\times B^2\Z_n; \Z)&= H^4(B\Z_n; \Z)\times\operatorname{Tor}_1^{\Z}\left(H^2(B\Z_n; \Z); H^3(B^2\Z_n; \Z)\right)\\&= \Z_n\times\Z_n.
              \end{split}
        \end{equation}
Or be even more general, computing the cohomologies of $B\Z_n\times B^2\Z_m$:
        \begin{equation}
            H^i(B\Z_n\times B^2\Z_m; \Z)=\{\Z; 0; \Z_n; \Z_m; \Z_n\times \Z_{(n,m)}; \Z_{(2)m}\times \Z_{(n,m)};...\}
        \end{equation}
        where we abbreviated $\gcd(\cdot,\cdot)$ with $(\cdot,\cdot)$.

Given a cocycle $\omega$ representing a cohomology class of degree $d+1$ computed above, and a background for the symmetry $\Z_n^{(0)}\times\Z_n^{(1)}$, viewed as a map $f:X\to B\Z_n\times B^2\Z_m$ the corresponding topological theory is given by \cite{dijkgraaf1990topological}:
\begin{equation}
    \int_{X^{(d)}}f^*\omega
\end{equation}
which, for specific examples computed above, takes the form:
\begin{equation}
    \begin{aligned}
        H^2(B\Z_n\times B^2\Z_m)=\Z_n \ni [k] &\leadsto S_k[B^{(1)}]=\frac{2\pi i k}{n}\int_{X^{(1)}}B^{(1)}\\
        H^3(B\Z_n\times B^2\Z_m)=\Z_m \ni [j]&\leadsto S_k[B^{(2)}]=\frac{2\pi i j}{m}\int_{X^{(2)}}B^{(2)}\\
        H^4(B\Z_n\times B^2\Z_m)=\Z_n\times\Z_{(n,m)}\ni[(k,j)]&\leadsto S_k[B^{(1)},B^{(2)}]=\frac{2\pi i k}{n}\int_{X^{(3)}}B^{(1)}\wedge\mathfrak{P}(B^{(1)})\\&\quad\quad+\frac{2\pi i j}{(n,m)}\int_{X^{(3)}}B^{(1)}\wedge B^{(2)},
    \end{aligned}
\end{equation}
where we used the following identifications \cite{Wang_2015}:
\begin{equation}
    \operatorname{Tor}^\Z\leftrightarrow \wedge d, \quad \otimes_\Z\leftrightarrow \wedge
\end{equation}
and we now interpreted the backgrounds as $H^i(X;\Z_n)$ cocycles.

\section{Generalized gauging and categorical review}\label{Appendix: generalized gauging}
In this appendix, we review aspects of generalized gauging in $2$d and $3$d theories; i.e.~in fusion and modular tensor categories\footnote{We acknowledge that our definitions and statements may be sloppy at times as we do not aim to the level of rigor of a math paper.}
as discussed in \cite{Fuchs_2002,Bhardwaj_2018,Barkeshli:2014cna} (see e.g.~\cite{Antinucci:2023ezl} for another review); and of gauging topological point defects (i.e.~decomposing) in generic dimensions, as discussed in \cite{sharpe2022introduction}. In each of these cases, the general physical idea to keep in mind is that gauging should be thought of as the insertion of a mesh of topological defects,
which is itself topological. The question thus reduces to asking:
\begin{itemize}
    \item What data are needed to label the components of the network in such a way that the partition function is well-defined and independent of the the specific triangulation chosen,
    \item What are the allowed objects, well-defined in the presence of such a mesh (and thus invariant under its topological manipulations),
    \item What are the ingredients needed to define an interface between two regions of spacetime, one of which is filled with the mesh.
\end{itemize}
\subsection*{Gauging lines in 2d}
Here we review the gauging of $0$-form symmetries in 2d, thoroughly treated by \textcite{Bhardwaj_2018}.
\begin{definition}[(Multi)fusion category]
  A \textit{multifusion} category $\mathcal{C}$ is:
\begin{itemize}
    \item A monoidal category: it is provided the data of a tensor product $\otimes: \mathcal{C}\times\mathcal{C}$ governing the fusion of topological defects, an identity object $\mathds{1}\in \operatorname{Ob}\mathcal{C}$ which is the trivial defect, and natural isomorphisms called associators $\alpha_{a,b,c}\colon(a\otimes b)\otimes c \to a\otimes (b\otimes c)$ satisfying various conditions insuring well definition of topological manipulations.
    \item Rigid: each object $a$ has right (left) dual $\overline{a}$ given by the orientation reversal, and evaluation and coevaluation morphisms:
    \begin{equation}
        \operatorname{ev}^L_a: a\otimes\overline{a}\to \mathds{1}; \quad \operatorname{coev}^L_a:\mathds{1}\to  \overline{a}\otimes a 
    \end{equation}
    implementing this reversal; satisfying:
\DisableQuotes\[\begin{tikzcd}
	a \\
	& {a\otimes \overline{a}\otimes a} \\
	a
	\arrow["{\mathds{1}_a\otimes \operatorname{coev^L_a}}", from=1-1, to=2-2]
	\arrow["{\mathds{1}_a}"', from=1-1, to=3-1]
	\arrow["{\operatorname{ev^L_a}\otimes\mathds{1}_a}", from=2-2, to=3-1]
\end{tikzcd}\]\EnableQuotes
    and its dual.
    \item Linear: there is a direct sum operation $\oplus: \mathcal{C}\times \mathcal{C}\to \mathcal{C}$ and hom-spaces are vector spaces. These data both implement linear superposition. Physically, the reason why the set of objects doesn't also acquire the structure of a vector space arises from the fact that lines play both the role of operators and of defect: while it makes sense multiplying an operator by a phase, there is no clear interpretation of the twisted Hilbert space of $cL$ for $c\in \mathbb{C}\setminus \Z$.
    \item Semisimple: there is a set of simple objects $\{x_i\}_{i\in \iI}$ satisfying:
    \begin{equation}
       x_i\otimes x_j=\delta_{i,j}\mathbb{C}; \quad \bigoplus_i\operatorname{Hom}(V;x_i)\otimes\operatorname{Hom}(x_i;W)\simeq \operatorname{Hom}(V;W)
    \end{equation}
    such that each object $V$ (and hom-space) decomposes in terms of simple ones:
    \begin{equation}
         V=\bigoplus_i x_i^{\oplus n_i}
    \end{equation}
\end{itemize}
with finitely many isomorphism classes of simple objects. Typically, one also requires that the endomorphisms of the unit object form just the ground field, i.e.~the absence of genuine topological pointlike operators (see \ref{Appendix: decomposition} for further discussion on this point), leading to the definition of a \textit{fusion} category.

\end{definition}
\paragraph{Gaugeable data}
\begin{definition}[Monoid in a monoidal category] A monoidal object $(\mathbb{A},\mu,\eta) $ in a monoidal category $\mathcal{C}$ is a triple:
\begin{equation}
    \mathbb{A}\in\operatorname{Ob}( \mathcal{C}), \mu\in \operatorname{Hom}_\mathcal{C}(\mathbb{A};\mathbb{A}),\eta\in \operatorname{Hom}_\mathcal{C}(\mathds{1};\mathbb{A}) ,
\end{equation}
where $\mu, \eta$ are respectively called the multiplication and unit morphisms, and satisfying:
\begin{itemize}
    \item The unit law:
\DisableQuotes\[\begin{tikzcd}
	{\mathds{1}\otimes\mathbb{A}} && {\mathbb{A}\otimes\mathbb{A}} && {\mathbb{A}\otimes\mathds{1}} \\
	\\
	&& {\mathbb{A}}
	\arrow["{\mu\otimes\mathds{1}_\mathbb{A}}", from=1-1, to=1-3]
	\arrow["{l_\mathbb{A}}", curve={height=6pt}, from=1-1, to=3-3]
	\arrow["\mu"', from=1-3, to=3-3]
	\arrow["{\mathds{1}_\mathbb{A}\otimes\mu}"', from=1-5, to=1-3]
	\arrow["{r_\mathbb{A}}"', curve={height=-6pt}, from=1-5, to=3-3]
\end{tikzcd}\]\EnableQuotes
    \item The associative law
\DisableQuotes\[\begin{tikzcd}
	& {(\mathbb{A}\otimes\mathbb{A})\otimes\mathbb{A}} && {\mathbb{A}\otimes(\mathbb{A}\otimes\mathbb{A})} \\
	{\mathbb{A}\otimes\mathbb{A}} &&&& {\mathbb{A}\otimes\mathbb{A}} \\
	&& {\mathbb{A}}
	\arrow["{\alpha_{\mathbb{A},\mathbb{A},\mathbb{A}}}", from=1-2, to=1-4]
	\arrow["{\mu\otimes\mathds{1}_\mathbb{A}}"', from=1-2, to=2-1]
	\arrow["{\mathds{1}_\mathbb{A}\otimes\mu}", from=1-4, to=2-5]
	\arrow["\mu"', from=2-1, to=3-3]
	\arrow["\mu", from=2-5, to=3-3]
\end{tikzcd}\]\EnableQuotes
\end{itemize}
    
\end{definition}
By dualizing all the diagrams (i.e. inverting the arrows) one can define a comonoid. Physically, the multiplication implies that we can define a trivalent vertex in which all three of the edges are labelled by the same object; the existence of a unit (and its identity) allows for truncated edges to be shrunk away; and the associativity condition implements the invariance under $F$-moves.
\begin{definition}[Symmetric Frobenius algebra object]\label{AlgObjDef}
    A Frobenius algebra in a monoidal category is a quintuple $(A,\Delta,\epsilon, \mu,\eta)$ such that:
    \begin{itemize}
        \item $(A,\mu,\eta)$ is a monoid with multiplication $\mu:A\otimes A\to A$ and unit $\eta:I\to A$,
        \item $(A,\Delta,\epsilon)$ is a comonoid with comultiplication $\Delta:A\to A\otimes A$ and counit $\epsilon:A\to I$,
        \item and the Frobenius laws hold: 
\DisableQuotes\[\begin{tikzcd}
	& {\mathbb{A}\otimes\mathbb{A}} \\
	{\mathbb{A}\otimes\mathbb{A}\otimes\mathbb{A}} & {\mathbb{A}} & {\mathbb{A}\otimes\mathbb{A}\otimes\mathbb{A}} \\
	& {\mathbb{A}\otimes\mathbb{A}}
	\arrow["{\mathds{1}_\mathbb{A}\otimes \Delta}"', from=1-2, to=2-1]
	\arrow["\mu", from=1-2, to=2-2]
	\arrow["{ \Delta\otimes\mathds{1}_\mathbb{A}}", from=1-2, to=2-3]
	\arrow["{ \mu\otimes\mathds{1}_\mathbb{A}}"', from=2-1, to=3-2]
	\arrow["\Delta", from=2-2, to=3-2]
	\arrow["{\mathds{1}_\mathbb{A}\otimes \mu}", from=2-3, to=3-2]
\end{tikzcd}\]\EnableQuotes
        where we used the associativity conditions to suppress the parenthesis (and compositions with the associator).
    \end{itemize}
    Moreover, it is $\Delta-$separable, or special, if it satisfies:
\DisableQuotes\[\begin{tikzcd}
	\mathbb{A} \\
	& {\mathbb{A}\otimes \mathbb{A}} \\
	\mathbb{A}
	\arrow["\Delta", from=1-1, to=2-2]
	\arrow["{\mathds{1}_\mathbb{A}}"', from=1-1, to=3-1]
	\arrow["\mu", from=2-2, to=3-1]
\end{tikzcd}\]\EnableQuotes
 
    and symmetric if:
   
\DisableQuotes\[\begin{tikzcd}
	& {\mathbb{A}\otimes \mathbb{A} \otimes \mathbb{A}^*} & {\mathbb{A}\otimes \mathbb{A}^*} \\
	\mathbb{A} &&& \mathbb{A} \\
	& {\mathbb{A}^*\otimes \mathbb{A}\otimes \mathbb{A} } & {\mathbb{A}^*\otimes \mathbb{A}}
	\arrow["{\mu \otimes\mathds{1}_{\mathbb{A}^*}}", from=1-2, to=1-3]
	\arrow["{\epsilon\otimes \mathds{1}_{\mathbb{A}^*}}", from=1-3, to=2-4]
	\arrow["{\mathds{1}_\mathbb{A}\otimes\operatorname{coev}^L_\mathbb{A}}", from=2-1, to=1-2]
	\arrow["{\operatorname{coev}^R_\mathbb{A}\otimes\mathds{1}_\mathbb{A}}"', from=2-1, to=3-2]
	\arrow["{\mathds{1}_{\mathbb{A}^*}\otimes\mu }"', from=3-2, to=3-3]
	\arrow["{\mathds{1}_{\mathbb{A}^*}\otimes \epsilon}"', from=3-3, to=2-4]
\end{tikzcd}\]\EnableQuotes
   
\end{definition}
\begin{proposition}
  The gauging data (including discrete torsion) of a $0$-form symmetry of a $2$-dimensional theory $\mathcal{T}$ corresponds to the choice of a symmetric separable special Frobenius algebra object $\AAA$ in the symmetry category $\mathcal{C}$.
\end{proposition}
\paragraph{Half-gauging interface}
Since gauging $\AAA$ in a subregion of spacetime amounts to filling the region with a mesh of $\mathbb{A}$, along with a choice of interface between the theories. Any such half-gauging interface needs to be well-defined in the presence of the $\AAA$-mesh on one side of it. This structure is captures by the following definition:
\begin{definition}[Left modules]
   Given a monoidal category $\mathcal{C}$ and a symmetric Frobenus algebra object $(A,\Delta,\epsilon, \mu,\eta)$ in it; a left $\mathbb{A}$-module is the data $(M; *)$ :
   \begin{equation}
       M\in \operatorname{Ob}(\mathcal{C}); \quad *\in \operatorname{Hom}_\mathcal{C}(\mathbb{A}\otimes M; M)
   \end{equation}
   compatible with the algebra structure of $\mathbb{A}$. They form the $\cC$-module category of $\AAA$-modules $_\AAA\cC$.
\end{definition}
\begin{proposition}
    The topological interfaces between a $2$-dimensional theory $\mathcal{T}$ with symmetry category $\mathcal{C}$ and the gauged theory $\mathcal{T}/\mathbb{A}$ are given by the objects o $_\AAA\cC$.
\end{proposition}
It is important to note that this category is not monoidal: it doesn't make sense to use two interfaces between $\mathcal{T}_1$ and $\mathcal{T}_2$ to build another interface of the same type.
On the other hand, it has a natural $\cC$-module structure by the tensor product in $\cC$, representing the action of the lines of the theory $\tT$ on the interface by fusion on the left. As we will see later, it has also a right module structure over the symmetry category of $\tT/\AAA$.
\paragraph{Resulting spectrum}
As for the interfaces, lines in the gauged theory should be invariant under fusion with a $\mathbb{A}$ line, now both from the left and the right:
\begin{proposition}
    The topological lines of the gauged theory $\mathcal{T}/\mathbb{A}$ are given by the category of $\AAA$-$\AAA$-bimodules $_\AAA\cC_\AAA$.
\end{proposition}
\begin{remark}
    As previously mentioned, the category $_\AAA\cC$ is also a right $_\AAA\cC_\AAA$-module.
\end{remark}

\paragraph{Composition}
In general a two-step consecutive gauging by $\AAA_1\in \cC$ and $\AAA_2\in \prescript{}{\AAA_1}{\cC} _{\AAA_1}$ can always be performed by a simple algebra object in the original category, which we denote with $\AAA_1\rhd\AAA_2$.

This should be thought of as corresponding to the physical intuition that each of the lines forming the algebra $\mathbb{A}_2$ are orbits of lines of the origin theory under the action of $\mathbb{A}_1$: the operation $\rhd$ just represents the passage from equivalence classes (orbits) to the full set of representatives.

\paragraph{Invertible symmetries}In the case of invertible symmetries, we recover familiar results:
    \begin{itemize}
        \item Gaugeable data in $\operatorname{Vec}_G^\omega$ are in one to one correspondence with pairs $(H, \epsilon)$ of a subgroup $H<G$ and a trivialization of the cocycle $i^*_H\omega=d\epsilon \in H^3(H;U(1))$
        \item Denoting with $\mathbb{A}=\bigoplus_{g\in G}g$ we have:
        \begin{equation}
            \prescript{}{\mathbb{A}}{\qty(\operatorname{Vec}_G)}_\mathbb{A}\simeq\operatorname{Rep}(G)
        \end{equation}
    \end{itemize}

\subsection*{Gauging lines in 3d}
If we want to describe topological lines in 3d we need to provide more data to
our symmetry category allowing for the greater freedom of 
topological moves. See \cite{Barkeshli:2014cna,Antinucci:2023ezl} for
physics reviews and \cite{Carqueville:2018sld,Mulevicius:2020bat} for a rigorous mathematical construction.
\begin{definition}[Unitary Modular Tensor Category (UMTC)]
A UMTC is a fusion category which is also:
\begin{itemize}
    \item Braided: there exists a braiding isomorphism
    \begin{equation}
        b_{X,Y}: X\otimes Y\rightarrow Y\otimes X
    \end{equation}
    compatible with the associator and implementing the effect of exchanging the position of lines in space
    \item Ribbon:
    there exists a twist isomorphism
    \begin{equation}
        b_{X,Y}: X\otimes Y\rightarrow Y\otimes X
    \end{equation}
    compatible with the braiding and the rigid structure and quantifying the spin of the lines.
    \item Modular: invertibility of the $S$-matrix
    \begin{equation}
        S_{i,j}:=tr(c_{X_{j},X_{i}}\circ c_{X_{i},X_{j}})
    \end{equation}
    computing the correlation function of Hopf links of simple object; enforcing the faithfulness of the action of the $0$-form symmetries

\end{itemize}
    
\end{definition}
\paragraph{Gaugeable data}
On top of the data of a symmetric Frobenius algebra; now we need the lines of the network to be invariant under braiding and twisting:
\begin{definition}[Commutative Frobenius algebra]
    A Frobenius algebra object $\mathbb{A}$ is said to be commutative if it is invariant under braiding and twisting:
\DisableQuotes\[\begin{tikzcd}
	{\mathbb{A}\otimes\mathbb{A}} && {\mathbb{A}\otimes\mathbb{A}} \\
	\\
	&& {\mathbb{A}}
	\arrow["{b_{\mathbb{A}\otimes\mathbb{A}}}", from=1-1, to=1-3]
	\arrow["\mu"', from=1-1, to=3-3]
	\arrow["\mu", from=1-3, to=3-3]
\end{tikzcd}\]\EnableQuotes
\end{definition}
\paragraph{Resulting spectrum}
Naively, one would define line of the gauged theory as those bimodules (orbits) which also trivially links with the gauged algebra (mutually local). Mathematically this is encapsulated in a single definition:
\begin{definition}[Local module]
A local module in a UMTC $\mathcal{C}$ over a commutative Frobenius algebra $\mathbb{A}$ is a left $\mathbb{A}$ module compatible with braiding:
\DisableQuotes\[\begin{tikzcd}
	{\mathbb{A}\otimes M} && { M\otimes\mathbb{A}} && {\mathbb{A}\otimes M} \\
	\\
	&& M
	\arrow["{b_{\mathbb{A}, M}}", from=1-1, to=1-3]
	\arrow["{*_L}"', from=1-1, to=3-3]
	\arrow["{b_{M,\mathbb{A}}}", from=1-3, to=1-5]
	\arrow["{*_R}", dashed, from=1-3, to=3-3]
	\arrow["{*_L}", from=1-5, to=3-3]
\end{tikzcd}\]\EnableQuotes
where we can provide a structure of bimodule to $M$ by defining the right multiplication $*_R=*_L\circ b_{M, \mathbb{A}}$ as in the diagram.
\end{definition}
\begin{proposition}
        The topological lines of the  gauged theory $\mathcal{T}/\mathbb{A}$ form the category of local modules:
    \begin{equation}
        \mathcal{C}^{\text{loc}}_\mathbb{A}
    \end{equation}
\end{proposition}
\begin{remark}
    Gauging lines in any dimension produces a quantum $0$-form symmetry. In $\dd=2$ this symmetry is implemented by topological lines again and the gauging process leads to the same number of lines in the gauged theory:
\begin{equation}
    \operatorname{dim}(\mathcal{C})=\operatorname{dim}(\prescript{}{\mathbb{A}}{\mathcal{C}}_{\mathbb{A}})
\end{equation}
    On the other hand the number of lines is not invariant under gauging in $\dd=3$ as the dual symmetry is here realized by topological surface operators \cite{Mulevicius:2020bat}:
    \begin{equation}
        \operatorname{dim}\qty(\cC^{\text{loc}}_{\mathbb{A}})=\frac{\operatorname{dim}(\mathcal{C})}{\operatorname{dim}(\mathbb{A})^2},\quad\operatorname{dim}(\mathbb{A})=\sum_{i\in\mathrm{~simple}}N_i\operatorname{dim}(x_{i}).
    \end{equation}
\end{remark}
However, the same process is still invertible \cite{Mulevicius:2022gce}:
\begin{theorem}
   Given an algebra object $\BBB\in \cC$, there exist an algebra object $I_\AAA(\BBB)\in\cC^{\text{loc}}_\AAA$ such that:
\begin{equation}
    \cC^{\text{loc}}_{\BBB}\simeq\qty(\cC^{\text{loc}}_\AAA)_{I_\AAA(\BBB)}
\end{equation}
The invertibility of the gauging process is obtained specializing the previous statement to the case $\BBB=\mathds{1}$
\end{theorem}
\paragraph{Half-gauging interfaces}
Each of the interfaces is a two-dimensional theory whose lines should allow for endpoints of lines in $\mathcal{C}$ from the left and of $\operatorname{Mod}(\mathcal{C})^{\text{loc}}_\mathbb{A}$ from the right:
\begin{proposition}
    The symmetry category of an interface between a $3$-dimensional theory $\mathcal{T}$ with symmetry category $\mathcal{C}$ and the gauged theory $\mathcal{T}/\mathbb{A}$ is given by the category $_\AAA\cC$ itself via the Witt trivialization:
    \begin{equation}
        \cC^{\text{op}}\boxtimes\cC^{loc}_\AAA\simeq Z(_\AAA\cC)
    \end{equation}
\end{proposition}
\paragraph{Compositions} 
In general a two-step consecutive gauging by $\AAA_1\in \cC$ and $\AAA_2\in \prescript{}{\AAA_1}{\cC} _{\AAA_1}$ can always be performed by a simple algebra object in the original category, which we denote with $\AAA_1\rhd\AAA_2$.

 A constructive definition for the operation $\rhd$ is given by the following theorem \cite{Mulevicius:2022gce}:
 \begin{theorem}
     Given an algebra object $\BBB\in \cC^{\text{loc}}_\AAA$, the underline object $U_\AAA(\BBB)$ inherits a canonical algebra structure and:
\begin{equation}
    \cC^{\text{loc}}_{U_\AAA(\BBB)}\simeq\qty(\cC^{\text{loc}}_\AAA)_\BBB
\end{equation}
 \end{theorem}
Using the notation adopted in the main text, this can be rewritten as:
\begin{equation}
    \AAA_1\rhd\AAA_2=U_{\AAA_1}(\AAA_2).
\end{equation}

This corresponds to the physical intuition that each of the lines forming the algebra $\mathbb{A}_2$ are orbits of lines of the origin theory under the action of $\mathbb{A}_1$: the operation $\rhd$ just represents the passage from equivalence classes (orbits) to the full set of representatives.
\subsection*{Gauging points: decomposition}\label{Appendix: decomposition}
\paragraph{Gaugable data} The only topological freedom of a $0$-dimensional mesh is given by fusion, thus:
\begin{proposition}
    The data of a gaugeable point is given by a projector $\pi\otimes\pi=\pi$
\end{proposition}
\paragraph{Resulting spectrum}
We only require observables to be invariant under fusion with the projector:
    \begin{itemize}
        \item Point operators in the gauged theory are given by:
        \begin{equation}
            \operatorname{Ker}\left\{(\mathds{1}-\bullet\otimes \pi):\Omega^{d-1}\mathcal{C} \rightarrow\Omega^{d-1}\mathcal{C}\right\}
        \end{equation}
        \item Codimension $1$ operators in the gauged theory are given by chargeless operators
        
    \end{itemize}

\paragraph{Invertible symmetries: decomposition}
In the case of invertible symmetries $G^{(d-1)}$, with $G$ abelian (visually, there is no order in the fusion of pointlike operators) 
We have pointlike operators $\eta_g$ labelled by group elements $g\in G$. We can use the characters of the group to form $|G|$ projectors\footnote{We will often implicitly choose an identification $\hat{G}\simeq G$ and thus (imprecisely) label the projectors by elements of the group $G$}:
\begin{equation}\label{eq: orthogonal relation}
\begin{aligned}
      \pi_{\hat{g}}:=&\bigoplus_{h\in G}\hat{g}(h)\eta^{(d-1)}_h\\
      \pi_{\hat{g}}\pi_{\hat{h}}=\delta_{\hat{g},\hat{h}}\pi_{\hat{g},\hat{h}}, \quad &\quad \mathds{1}=\bigoplus_{\hat{h}\in \hat{G}}\pi_{\hat{h}}
\end{aligned}
\end{equation}
This process can be restated, noting that $H^d(K(G;d);U(1))=\hat{G}$. Formally, we can see the projector $\pi_{\hat{g}}$ in the theory $\mathcal{T}$ as the projector $\pi_{\hat{e}}$ of the theory $\mathcal{T}\times Z_{\hat{g}}/G^{(d-1)}$ where we donated with $Z_{\hat{g}}$ the SPT:
\begin{equation}
    S_{Z_{\hat{g}}}=\int B_{(d)}^*\hat{g}, \quad \quad \langle \Pi_i h_i\cdot [X] \rangle=\Pi_i \hat{g}(h_i)
\end{equation}

Moreover, \cref{eq: orthogonal relation} implies also a decomposition of the theory.
\begin{equation}\label{eq: decomposition}
    \langle \mathds{1}\dots \rangle_\mathcal{T}=\sum_{\hat{h}\in\hat{G}} \langle \pi_{\hat{h}}\dots \rangle_\mathcal{T}:= \sum_{\hat{h}\in\hat{G}} \langle  \dots\rangle_{\mathcal{T}_{\hat{h}}}\leadsto \mathcal{T}=\bigoplus_{\hat{h}\in\hat{G}}\mathcal{T}_{\hat{h}}
\end{equation}

Actually, \cref{eq: decomposition} is true only in the absence of codimension-one operators. The proof relies on the fact that one can always insert an identity operator; write it as the sum of projectors and use the orthogonality of the projectors to identify the various labels:
\begin{equation}
\begin{aligned}
    \langle\mathcal{O}_1\dots \mathcal{O}_n\rangle_\mathcal{T}&=\langle(\mathds{1}\mathcal{O}_1)\dots (\mathds{1}\mathcal{O}_n)\rangle_\mathcal{T}=\sum_{\hat{h_i}\in\hat{G}} \langle (\pi_{\hat{h_1}}\mathcal{O}_1)\dots (\pi_{\hat{h_n}}\mathcal{O}_n)\rangle_\mathcal{T}\\
    &=\sum_{\hat{h_i}\in\hat{G}}\prod_{i=1}^{n-1}\delta_{\hat{h_i},\hat{h_{i+1}}} \langle (\pi_{\hat{h_1}}\mathcal{O}_1)\dots (\pi_{\hat{h_n}}\mathcal{O}_n)\rangle_\mathcal{T}:= \sum_{\hat{h}\in\hat{G}} \langle  \mathcal{O}_1^{\hat{h}}\dots \mathcal{O}_n^{\hat{h}}\rangle_{\mathcal{T}_{\hat{h}}}
    \end{aligned}
    \end{equation}
where in the last equality we used the topological nature of the operators $\pi_{\hat{h_i}}$ to move them close together and the orthogonality in \cref{eq: decomposition}; and we denoted with $\mathcal{O}^{\hat{h}}$ the projection $\pi_{\hat{h}}\mathcal{O}$ of the operator $\mathcal{O}$ in the theory $\mathcal{T}_{\hat{h}}$.

However, this is not true in the presence of codimension-one operators, as they may divide the spacetime into disconnected components.
\begin{proposition}
    A codimension $1$ operator of charge $\hat{g}\in \hat{G}$ is a domain wall between the theory $\mathcal{T}_{\hat{h}}$ and the theory $\mathcal{T}_{\hat{h}+\hat{g}}$.

    In particular, only chargeless operators descend to proper operators of the gauged theory, and for these kinds of operators the decomposition statement still holds: a chargeless operator $\mathcal{O}$ of the theory $\mathcal{T}$ decomposes into the sum of $|G|$ operators $\pi_{\hat{h}}\mathcal{O}$ each living in their own universe $\mathcal{T}_{\hat{h}}$.
\end{proposition}
Let's explicitly prove this statement for a finite cyclic group $\Z_n$ and leave the generalization to the reader. Gauge invariant codimension-one operators in the theory $\mathcal{T}_k$ are those satisfying:
\begin{equation}\label{eq: explicit decomposition condition}
    \langle\qty(\sum_{s=0}^{n-1}e^{\frac{2\pi i sk  }{n}}\eta^{(d-1)}_{s})\mathcal{O}\qty(\sum_{t=0}^{n-1}e^{\frac{2\pi i tk  }{n}}\eta^{(d-1)}_{t})\rangle\stackrel{?}{=}\langle\mathcal{O}\qty(\sum_{u=0}^{n-1}e^{\frac{2\pi i uk  }{n}}\eta^{(d-1)}_{u})\rangle.
\end{equation}
Starting from the left-hand side, reorganizing the summations and ``bubbling'' operators $\mathds{1}=\eta^{(d-1)}_{-s}\eta^{(d-1)}_{s}$ we obtain:
\begin{equation}
    \langle\qty(\sum_{s=0}^{n-1}e^{\frac{2\pi i sk  }{n}}\eta^{(d-1)}_{s}\mathcal{O}\eta^{(d-1)}_{-s}\eta^{(d-1)}_{s})\qty(\sum_{t=0}^{n-1}e^{\frac{2\pi i tk  }{n}}\eta^{(d-1)}_{t})\rangle=\sum_{t=0}^{n-1}\sum_{s=0}^{n-1}\langle e^{\frac{2\pi i [(s+t)k+sq(\mathcal{O}) ] }{n}}\mathcal{O}\eta^{(d-1)}_{t+s}\rangle
\end{equation}
where we acted on the operator $\mathcal{O}$ using the symmetry $\Z_n^{(d-1)}$. We now rewrite the sum in terms of $x=s+t$:
\begin{equation}
    \sum_{t=0}^{n-1}\sum_{s=0}^{n-1}\langle e^{\frac{2\pi i [(s+t)k+sq(\mathcal{O}) ] }{n}}\mathcal{O}\eta^{(d-1)}_{t+s}\rangle=\sum_{x=0}^{2n-2}\sum_{s=\operatorname{max}\{0, x-n+1\}}^{\operatorname{min}\{n-1, x\}}\langle e^{\frac{2\pi i sq(\mathcal{O})  }{n}}\mathcal{O}\qty(e^{\frac{2\pi i x  k}{n}}\eta^{(d-1)}_{x})\rangle.
\end{equation}
Since $\eta^{(d-1)}_{x}\sim \eta^{(d-1)}_{x+n}$ we can group some of the terms:
\begin{equation}
    \sum_{x=0}^{n-1}\langle\left(\sum_{s=\operatorname{max}\{0, x-n+1\}=0}^{\operatorname{min}\{n-1, x\}=x} e^{\frac{2\pi i sq(\mathcal{O})  }{n}}+\sum_{s=\operatorname{max}\{0, x+1\}=x+1}^{\operatorname{min}\{n-1, x+n\}=n-1} e^{\frac{2\pi i sq(\mathcal{O})  }{n}}\right)\mathcal{O}\qty(e^{\frac{2\pi i x  k}{n}}\eta^{(d-1)}_{x})\rangle
\end{equation}
which is equal to the right-hand side of \cref{eq: explicit decomposition condition} if and only if $q(\mathcal{O})=0$ because of the fact that:
\begin{equation}
    \sum_{s=\operatorname{max}\{0, x-n+1\}=0}^{\operatorname{min}\{n-1, x\}=x} e^{\frac{2\pi i sq(\mathcal{O})  }{n}}+\sum_{s=\operatorname{max}\{0, x+1\}=x+1}^{\operatorname{min}\{n-1, x+n\}=n-1} e^{\frac{2\pi i sq(\mathcal{O})  }{n}}=\sum_{s=0}^d  e^{\frac{2\pi i sq(\mathcal{O})}{n}}=\delta_{q(\mathcal{O}),0}.
\end{equation}

\section{Minimal theories}\label{Appendix: minimal theories}
\subsection{Minimal theory $\mathcal{A}_1^{N,p}$ in 3 dimensions}
\label{app:3d}
In this section we present the definition and some basic properties of the minimal $2+1$d TQFTs $\mathcal{A}^{N,p}$ having a $\Z_N^{(1)}$ symmetry. We refer the reader to \cite{Hsin_2019} for more details.

A generic theory with these properties will be characterized by a set of topological lines $\eta^{(1)}_k$ fusing according to the $\Z_N$ group law. More precisely, these operators might suffer from a framing dependence, i.e.~a spin of the form:
\begin{equation}
     h\qty[\eta^{(1)}_k]=\frac{pk^2}{2N} \ \operatorname{mod}\ 1
\end{equation}
On one hand, this quantum number $p$ corresponds to the anomaly of the symmetry $p\in \Z_N$\footnote{The anomalies are classified by $H^4(\Z_N; U(1))=\Z_{(2)N}$, as explained in \cref{Appendix: cohomology}, but we limit ourselves to spin theories, in which case the $p = N/2$ anomaly is trivial.}. The inflow action of such an anomaly is:
\begin{equation}\label{eq: minimal inflow}
    S_{\text{inflow}}[B^{(2)}]=-\frac{2\pi i p}{N}\int_{X^{(4)}}B^{(2)}\wedge B^{(2)}
\end{equation}
On the other hand, the spin makes the line charged under the same $\Z_N^{(1)}$ symmetry they generate, with charge:
\begin{equation}
    q(\eta^{(1)}_k)=N\qty(h\qty[\eta^{(1)}_1]+h\qty[\eta^{(1)}_k]-h\qty[\eta^{(1)}_{k+1}])\bmod N=-pk ,
\end{equation}
We deduce that the lines by themselves make a nontrivial theory. Moreover, for $\gcd(p,N)=1$ the theory is actually well defined and is denoted with $\mathcal{A}^{N,p}$.

We would like to stress that we can change the labelling of the lines without changing the physical informations of the theory. In particular, the group $\Z_N$ can be generated by each of its invertible elements $\Z_N^\times$. It follows that, given an integer $r$ coprime with $N$, the minimal theories satifies the duality:
\begin{equation}\label{MinimalDualities}
    \mathcal{A}^{N,p}\simeq\mathcal{A}^{N,r^2p}
\end{equation}
implemented by the relabelling $\eta^{(s)}_r\to\tilde{\eta}^{(s)}_1 $.

The importance of such theories, and their name, is given by the following properties \cite{Hsin_2019}:
\begin{theorem}[Minimality]\label{th: minimality}
    Given a three dimensional theory $\mathcal{T}$ with a $\Z_N^{(1)}$ symmetry, whose anomaly inflow is of the form \ref{eq: minimal inflow}, then it is of the form:
    \begin{equation}
        \mathcal{T}=\mathcal{A}^{N,p}\otimes\mathcal{T}'
    \end{equation}
    and thus $\mathcal{A}^{N,p}$ is the minimal theory with those characteristics.
\end{theorem}
Moreover, those theories enjoy another properties that can be used to prove the topological nature of the $\mathcal{D}^{(s)}$ defects:
\begin{theorem}[Half-gauge construction]\label{Minimal-3d-half-gauging}
    The bulk-boundary system $(\mathcal{A}^{N,p}\otimes S_{\text{inflow}}[B^{(2)}])[X^{(4)}]$ can be realixed as a dynamical theory:
    \begin{equation}
        \int_{X^{(4)}}\left(\frac{iN}{2\pi}b\wedge dc+\frac{iN(p)^{-1}_N}{4\pi}b\wedge b+\frac{iN}{2\pi}b\wedge B^{(2)}\right)
    \end{equation}
    on a $4$-dimensional manifold $X^{(4)}$ with boundary.
\end{theorem}
In fact, if we stack this theory with the Axion-Maxwell theory on a closed manifold $Y^{(4)}$ which extends $X^{(4)}$, and we impose the background to match the ($\Z_N$ reduction of the) electromagnetic field strength $B^{(2)}=\frac{F}{N}$ we obtain a definition of the shift defect $\mathcal{D}^{(s)}$ in terms of half space gauging \cite{Choi2022noninvertible}:
\begin{equation}
\begin{aligned}
    \mathcal{D}^{(s)}_{\frac{p}{N}}[\partial X^{(4)}]=\int [\mathcal{D}\Phi][\mathcal{D}b][\mathcal{D}c]\exp\Big(&S_{\text{axion}}+\frac{2\pi ip}{N}\int_{\partial X^{(4)}} j^{(3)}_s+ \int_{X^{(4)}}\frac{iN}{2\pi}b\wedge dc \\ 
    & +\frac{iN(p)^{-1}_N}{4\pi}b\wedge b+\frac{i}{2\pi}b\wedge F\Big)
    \end{aligned}
\end{equation}
Finally, we would like to mention that the $\mathcal{A}^{N,p}[B^{(2)}]$ theories enjoy a Lagrangian description an abelian Chern--Simons theory \cite{vanBeest:2023dbu}:
\begin{equation}\label{continued fraction minimal lagrangian}
    a^tKda + v^ta\wedge B^{(2)}
\end{equation}
where $a$ is a $q$-dimensional vector with components given by $U(1)$ gauge fields, $v$ is a $q\times 1 $ vector of integers (called the charge vector) and:
\begin{equation*}K=\left(\begin{array}{cccccc}
+k_1 & 1 & 0 & 0 & 0 & \cdots \\
1 & -k_2 & 1 & 0 & 0 & \cdots \\
0 & 1 & +k_3 & 1 & 0 & \cdots \\
0 & 0 & 1 & -k_4 & 1 & \cdots \\
\vdots & \vdots & \vdots & \vdots & \vdots & \ddots
\end{array}\right),
\end{equation*}
where the $k_i$ are the coefficient of the continued fraction of $p/N$, namely:
\begin{equation*}
\frac{p}{N}=\frac{1}{k_1+\frac{1}{k_2+\frac{1}{k_3+\cdots}}}.
\end{equation*}
The choice of such a peculiar coupling is motivated by the following properties of the matrix $K$ (which can be proved inductively on the dimension of the continuous fraction, and some algebraic manipulation):
\begin{equation*}
(K^{-1})_{11}=\frac{p}{N}, \quad \operatorname{SNF}(K)= \text{diag}(1,\dots, 1, N),
\end{equation*}
where we denoted with $\operatorname{SNF}(K)$ the Smith normal form of the
matrix $K$.

\subsection{Fusion rules}\label{sec: fusion original}
This subsection is dedicated to an alternative derivation of the following result:
\begin{theorem}
There exists a fusion interface of the form:
\begin{equation}\label{eq: fusion-Minimal-3d}
\qty(\mathcal{A}_1^{N_1,p_1}\otimes\mathcal{A}_1^{N_2,p_2})[F]\xrightarrow[]{m_{M'}}\mathcal{A}_1^{M/M^{\prime},((p_1^{-1})_{N_1}K_1+(p_2^{-1})_{N_2}K_2)/M^{\prime}}\otimes \mathcal{A}_1^{(p_1K_2+p_2K_1)/M^{\prime},K/M}[F],
\end{equation}
with $M=\operatorname{gcd}(N_1,N_2)$, $N_{1,2}=K_{1,2}M$, $M'=\operatorname{gcd}(p_1K_2+p_2K_1,M)$, and $K=\operatorname{lcm}(N_1,N_2)$, implemented by half gauging of a $Z_{M'}^{(1)}$ subgroup of lines.

Moreover, while in general the two theories $\mathcal{A}_1^{N_1,p_1}$ and $\mathcal{A}_1^{N_2,p_2}$ should be coupled to backgrounds whose holonomies are valued in two different groups $\Z_{N_i}$, we focus on the case where both backgrounds descend from the same $U(1)$ field strengths:
\begin{equation}
   B^{(2)}_{1,2}=\frac{F}{N_{1,2}}.
\end{equation}
\end{theorem}
While in \cref{Appendix: generalized gauging} we presented a derivation based on an abstract study of the category of lines, the goal of the next pages is to arrive at the same result with a more familiar Lagrangian approach.

\paragraph{Proof} In this first attempt, we make a slight detour in order to avoid the complicated Lagrangian description \ref{continued fraction minimal lagrangian} of a generic $\mathcal{A}_1^{N,p}$. We begin by reminding that the following duality holds:
\begin{equation*}
U(1)_{pN}\iff\mathcal{A}_1^{pN,1}=\mathcal{A}_1^{N,p}\otimes\mathcal{A}_1^{p,N} \ \text{when}\ \gcd(N,p)=1.
\end{equation*}
where $\mathcal{A}_1^{pN,1}$ has a simple description in terms of a $U(1)_{pN}$ Chern--Simons theory $\mathcal{L}_{pN}=\frac{pN}{4\pi}a\dd a$. The factors in the composition are given by the theories respectively generated by the lines:
$$\mathcal{W}_p(\gamma)=\exp(2\pi i \int_\gamma pa) \quad \text{and} \quad \mathcal{W}_N(\gamma)=\exp(2\pi i \int_\gamma Na).$$
Here we are just rewriting the $\Z_{pN}^{(1)}$ symmetry of $U(1)_{pN}$ as $\Z_p^{(1)}\times\Z_N^{(1)}$, using the Chinese remainder theorem.

Moreover, we are able to describe the coupling of just the second factor to a background $\qty[\frac{F}{N}]\in\HH{2}{X}{\Z_N}$ for the $Z_N^{(1)}$ symmetry, expressed as a closed $2$-form with $\Z_N$ periods $F\in \Omega^2_{\Z_N}(X)$, by adding to the Lagrangian a term:
\begin{equation*}
\mathcal{A}_1^{pN,1}\qty[p\qty(\frac{F}{N})]=\mathcal{A}_1^{N,p}\qty[\frac{F}{N}]\otimes\mathcal{A}_1^{p,N}\iff\mathcal{L}\qty[\frac{F}{N}]=\frac{p}{2\pi}\left(\frac{1}{2}Na\dd a+a\wedge F\right).
\end{equation*}
In the following, we will denote this particular choice of background and coupling for the theory $\mathcal{A}_1^{pN,1}$ with $\tilde{\mathcal{A}}^{pN,1}[F]$.

We are now ready to prove:
\begin{theorem}
There exists a fusion interface of the form:
\begin{equation}\label{eq: fusion-Minimal-3d-II}
\begin{aligned}
\frac{\qty(\tilde{\mathcal{A}}^{p_1N_1,1}\otimes \tilde{\mathcal{A}}^{p_2N_2,1})[F]}{\Z_{M'}^{(1)}}
    \xrightarrow[]{m_{M'}}\\
    \frac{\mathcal{A}_1^{\frac{(p_1K_2+p_2K_1)}{M'},\frac{MK_1K_2}{M'}}\otimes \mathcal{A}_1^{p_1p_2\frac{(p_1K_2+p_2K_1)}{M'},\frac{M}{M'}}}{\Z_{\frac{p_1K_2+p_2K_1}{M'}}^{(1)}} \mathcal{A}_1^{\frac{M}{M'},p_1p_2\frac{(p_1K_2+p_2K_1)}{M'}}&\otimes\mathcal{A}_1^{\frac{MK_1K_2}{M'},\frac{(p_1K_2+p_2K_1)}{M'}}[F].
    \end{aligned}
\end{equation}
implemented by half gauging of a $Z_{M'}^{(1)}$ subgroup of lines.
\end{theorem}
This theorem is physically equivalent to \cref{eq: fusion-Minimal-3d}. In fact, \cref{eq: fusion-Minimal-3d} will be used to derive fusion rules for defects of the form:
\begin{equation}
    \mathcal{D}_{\frac{p}{N}}^{(0)}=\eta_{\frac{p}{N}}^{(0)}\otimes \mathcal{A}_1^{N,p}\qty[\frac{F}{N}],
\end{equation}
where the degrees of freedom of the minimal theories $\mathcal{A}_1^{N,p}$ will live only on its support. The only way in which they will interact with the actual theory $\mathcal{T}$, of which $\mathcal{D}^{(0)}_{\frac{p}{N}}$ is a defect, is via the coupling with the dynamical $\mathcal{T}$-field $F$. Given this setup, the result of \cref{eq: fusion-Minimal-3d-II} can be applied to derive similar fusion rules for the defects:
\begin{equation}
     \qty(D_{\frac{p}{N}}^{(0)})'=\eta_{\frac{p}{N}}^{(0)}\otimes \tilde{\mathcal{A}}^{pN,1}\qty[\frac{F}{N}]= D_{\frac{p}{N}}^{(0)}\otimes \mathcal{A}_1^{p,N},
\end{equation}
which differ from the original ones by the stacking of a $\mathcal{A}_1^{p,N}$ theory completely undetectable by the degrees of freedom.

In particular, also the results of the two theorems differ just by decoupled factors. In the following, we are going to declare equivalent two such defects differing by decoupled TQFT coefficients. Denoting with $\llbracket\cdot \rrbracket$ equivalence classes of defects under the aforementioned relation, both theorems lead to the same fusion rules:
\begin{equation}
    \llbracket\mathcal{D}^{(0)}_{\frac{p_1}{N_1}} \rrbracket\otimes \llbracket\mathcal{D}^{(0)}_{\frac{p_2}{N_2}}  \rrbracket\rightarrow\llbracket\mathcal{D}^{(0)}_{\frac{p_1}{N_1}+\frac
    {p_2}{N_2}}  \rrbracket.
\end{equation}

We are going to work through this result step by step, starting with the simplest case possible and moving toward the most general formula. 

Moreover, in each step, we start with an analysis of the spectrum of the theories. The goal is to organize the lines of the minimal theories $\mathcal{A}_1^{N_i,p_i}$ at the left-hand side of the fusion in such a way that makes evident the presence of the factors on the right-hand side. This analysis is then used to motivate the change of variable used in the computation with the explicit Lagrangian description.

\underline{Case 1: $N_1=N_2=N, \ \operatorname{gcd}(p_1+p_2,N)=1$}. We must reconstruct a $\mathcal{A}_1^{N,p_1+p_2}$ using the lines of the original $\mathcal{A}_1^{N,p_1}\otimes\mathcal{A}_1^{N,p_2}$ factors. A good candidate for its generating line is given by $L=L_1\otimes L_2$ as it has spin $h[L^s]=\frac{(p_1+p_2)s^2}{2N}$, background coupling $\alpha[L^s]=\frac{(p_1+p_2)s}{N}$ and satisfies $L^N=1$.

Recalling the factorization theorem \ref{th: minimality}, there must exist lines (generated by $K=L_1^x\otimes L_2^y$) which trivially braids with $L$ reproducing the decomposition:
\begin{equation}
    \mathcal{A}_1^{p_1N,1}\otimes\mathcal{A}_1^{p_2N,1}=\mathcal{A}_1^{N,p_1+p_2}\otimes \mathcal{T}',
\end{equation}
and if these lines form a consistent theory $\mathcal{A}_1^{M,q}$ this would be another factor of $\mathcal{T}'$:
\begin{equation}
    \mathcal{T}'=\mathcal{A}_1^{M,q}\otimes \mathcal{T}''.
\end{equation}

We are thus looking for a solution $(x,y)$ of the constraint:
\begin{equation}
B(L,K)=\frac{p_1x+p_2y}{N}=0 \quad \operatorname{mod} 1.
\end{equation}
The simplest solution is given by $(x,y)=(p_2, -p_1)$. The corresponding line has spin $h[K^s]=\frac{(p_1x^2+p_2y^2)s^2}{2N}=\frac{p_1p_2(p_1+p_2)s^2}{2N}$ and satisfies $K^N=1$ forming $\mathcal{A}_1^{N,p_1p_2(p_1+p_2)}$ where the consistency condition $\operatorname{gcd}(N,p_1p_2(p_1+p_2))=1$ is automatically satisfied as it is equivalent to the hypothesis $\gcd(p_1+p_2,N)=1$.

Moreover, it is important to note that the lines generated by $K$ have a trivial coupling to the background as $\alpha(K)=\frac{p_1x+p_2y}{N}=0$. Thus, not only do they combine into a theory decoupled from $\mathcal{A}_1^{N,p_1+p_2}$ but their lines are actually well defined and topological on their own.

In order to conclude we should now show a decomposition of the final factor $\mathcal{T}''$ of the form:
\begin{equation}
    \mathcal{T}''=\mathcal{A}_1^{p_1+p_2,N}\otimes\mathcal{A}_1^{p_1p_2(p_1+p_2),N}.
\end{equation}

We avoid this check by the factorization directly at the level of the Lagrangian:
\begin{equation}
\mathcal{L}_{p_1N}\qty[a_1,\frac{F}{N}]+\mathcal{L}_{p_2N}\qty[a_2,\frac{F}{N}]=\frac{1}{4\pi}\left(p_1N a_1\dd a_1+p_2N a_2\dd a_2+2(p_1a_1+p_2a_2)F\right).
\end{equation}
Motivated by the previous discussion, we redefine $(p_1+p_2)\alpha=p_1a_1+p_2a_2$ (which expresses $L$ in terms of $L_1\otimes L_2$) and $p_1p_2(p_1+p_2)\beta=p_2(p_1a_1)+(-p_1)(p_2a_2)$ (which expresses $K$ in terms of $L_1^{p_1}\otimes L_2^{-p_1}$). In terms of these new quantities, the Lagrangian becomes:
\begin{equation*}
\frac{N}{4\pi} p_1p_2(p_1+p_2)\beta d \beta + (p_1+p_2)\qty(\frac{N}{4\pi}\alpha d\alpha+\frac{1}{2\pi}\alpha F) =\mathcal{L}_{p_1p_2(p_1+p_2)N}[\beta;0]+\mathcal{L}_{(p_1+p_2)N}\qty[\alpha,\frac{F}{N}],
\end{equation*}
which shows the desired fusion rule. Alternatively, we can state the duality:\begin{small}
\begin{equation*}
\mathcal{L}_{p_1N}\qty[(p_1+p_2)a_1,\frac{F}{N}]+\mathcal{L}_{p_2N}\qty[(p_1+p_2)a_2,\frac{F}{N}]=\mathcal{L}_{p_1p_2(p_1+p_2)N}[a_1-a_2,0]+\mathcal{L}_{(p_1+p_2)N}\qty[p_1a_1+p_2a_2, \frac{F}{N}].
\end{equation*}\end{small}
However, the prescribed reparametrizations are not invertible, hence we cannot conclude:
\begin{equation}
    \qty(\tilde{\mathcal{A}}^{p_1N,1}\otimes \tilde{\mathcal{A}}^{p_2N,1})[F]\stackrel{?}{\simeq}\mathcal{A}_1^{(p_1+p_2)N,1}\otimes \tilde{\mathcal{A}}^{p_1p_2(p_1+p_2)N,1}[F],
\end{equation}
which was the desired identity. Instead, we have that the transformation:
\begin{equation}
    \begin{bmatrix}a_1\\a_2\end{bmatrix}=\begin{bmatrix}
       -1 &-p_2\\
        -1 &p_1
    \end{bmatrix}\begin{bmatrix}\alpha\\ \beta\end{bmatrix}
\end{equation}
has determinant $p_1+p_2$. Another operation which has the naive effect of a reparametrization with Jacobian $p_1+p_2$ is the gauging of a $\Z_{p_1+p_2}^{(1)}$ symmetry \cite{Hsin_2019}. We thus claim the following identity:
\begin{small}
\begin{equation}
\begin{aligned}
    \qty(\tilde{\mathcal{A}}^{p_1N,1}\otimes \tilde{\mathcal{A}}^{p_2N,1})[F]&\simeq\frac{\mathcal{A}_1^{(p_1+p_2)N,1}\otimes \tilde{\mathcal{A}}^{p_1p_2(p_1+p_2)N,1}[F]}{\Z_{p_1+p_2}^{(1)}}\\&\simeq\frac{\mathcal{A}_1^{(p_1+p_2),N}\otimes \mathcal{A}_1^{p_1p_2(p_1+p_2),N}}{\Z_{p_1+p_2}^{(1)}}\otimes \mathcal{A}_1^{N,p_1p_2(p_1+p_2)}\otimes\mathcal{A}_1^{N,p_1+p_2}[F].
    \end{aligned}
\end{equation}\end{small}%
where the second equality follows from the fact that the lines $\eta^{(1)}_i$ forming the $\Z_{p_1+p_2}^{(1)}$ group cannot be written as $\eta^{(1)}_i=(\eta^{(1)}_i)' L$ with $L\in \mathcal{A}_1^{N,p_1p_2(p_1+p_2)}\otimes\mathcal{A}_1^{N,p_1+p_2} $ since a generic line of $\mathcal{A}_1^{N,p_1p_2(p_1+p_2)}\otimes\mathcal{A}_1^{N,p_1+p_2}$ has order $N$ which is coprime with $p_1+p_2$ by hypothesis.

\underline{Case 2: $N_1\neq N_2,\ \operatorname{gcd}(p_1K_2+p_2K_1,M)=1$}. This is a straightforward generalization of the previous procedure. One starts by identifying the lines in $\mathcal{A}_1^{p_1N_1,1}\otimes \mathcal{A}_1^{p_2N_2,1}$ that are going to form a $\mathcal{A}_1^{MK_1K_2,p_1K_2+p_2K_1}$ factor:
\begin{equation}
L=L_1\otimes L_2 \leadsto  \ L^{MK_1K_2}=1 \ \begin{cases}
    h[L^s]=(\frac{p_1}{2N_1}+\frac{p_2}{2N_2})s^2=\frac{(p_1K_2+p_2K_1)s^2}{2K}, \\ \alpha[L^s]=\frac{(p_1K_2+p_2K_1)s}{MK_1K_2}.
\end{cases}
\end{equation}
And then look for solution $K=L_1^x\otimes L_2^y$ to the constraint:
\begin{equation}
B(L,K)=\frac{p_1x}{2N_1}+\frac{p_2y}{2N_2}=\frac{p_1K_2x+p_2K_1y}{2MK_1K_2}=0\quad \operatorname{mod} 1.
\end{equation}
finding $(x,y)=(K_1 p_2,-K_2p_1)$, which amounts to:
\begin{equation*}
    h[K^s]=\frac{p_1p_2(K_1p_2+K_2p_1)}{2M}s^2, \ \alpha(K^s)=\frac{K_1p_-+K_2p_1}{2M}s, \ 
    K^M=1.
\end{equation*}
Those lines organize themselves into a consistent theory $\mathcal{A}_1^{M, K_1(p_1)^{-1}_{N_1}+K_2(p_2)^{-1}_{N_2}}\simeq\mathcal{A}_1^{M, K_1p_2+K_2p_1} $ (where the equivalence is a consequence of \ref{MinimalDualities} and the consistency of the theories follows from the hypothesis $\operatorname{gcd}(p_1K_2+p_2K_1,M)=1$) which is completely decoupled from the background.

Again, the same derivation can be obtained by the Lagrangian description:
\begin{equation*}
    \mathcal{L}_{p_1N_1}\qty[a_1,\frac{F}{N_1}]+\mathcal{L}_{p_2N_2}\qty[a_2,\frac{F}{N_2}]=\frac{1}{4\pi}\left(p_1N_1 a_1\dd a_1+p_2N_2 a_2\dd a_2+2(p_1a_1+p_2a_2)F\right).
\end{equation*}
After an adequate redefinition of the variables $(p_1K_2+p_2K_1)\alpha=p_1a_1+p_2a_2 $ and $p_1p_2(p_1K_2+p_2K_1)\beta=p_2(p_1K_1a_1)-p_1(p_2K_2a_2)$, 
which leads to:
\begin{equation*}
\begin{aligned}\label{Case2}
\frac{1}{4\pi}\left( MK_1K_2(p_1K_2+p_2K_1)\alpha d\alpha + Mp_1p_2(p_1K_2+p_2K_1)\beta d \beta + 2(p_1K_2+p_2K_1)\alpha F\right) =\\=\mathcal{L}_{p_1p_2(p_1K_2+p_2K_1)M}[\beta;0]+\mathcal{L}_{(p_1K_2+p_2K_1)MK_1K_2}\qty[\alpha; \frac{F}{MK_1K_2}].
\end{aligned}
\end{equation*}
Once again, this can be restated as:
\begin{equation*}
\begin{aligned}
\mathcal{L}_{p_1N_1}\qty[(p_1K_2+p_2K_1)a_1;\frac{F}{N_1}]+\mathcal{L}_{p_2N_2}\qty[(p_1K_2+p_2K_1)a_2;\frac{F}{N_2}]=\\
=\mathcal{L}_{p_1p_2(p_1K_2+p_2K_1)M}\qty[K_1a_1-K_2a_2;0]+\mathcal{L}_{(p_1K_2+p_2K_1)MK_1K_2}\qty[p_1a_1+p_2a_2; \frac{F}{MK_1K_2}].
\end{aligned}
\end{equation*}
However, the prescribed reparametrizations are not invertible, hence we cannot conclude:
\begin{equation}
   \qty(\tilde{\mathcal{A}}^{p_1N_1,1}\otimes \tilde{\mathcal{A}}^{p_2N_2,1})[F]\stackrel{?}{\simeq}\mathcal{A}_1^{(p_1K_2+p_2K_1)K_1K_2M,1}\otimes \tilde{\mathcal{A}}^{p_1p_2(p_1K_2+p_2K_1)M,1}[F].
\end{equation}
which was the desired identity. Instead, we have that the transformation:
\begin{equation}
    \begin{bmatrix}a_1\\a_2\end{bmatrix}=\begin{bmatrix}
        -K_2 & -p_2\\
         -K_1 & p_1
    \end{bmatrix}\begin{bmatrix}\alpha\\ \beta\end{bmatrix}
\end{equation}
has determinant $p_1K_2+p_2K_1$. Another operation which has the naive effect of a reparametrization with Jacobian $p_1K_2+p_2K_1$ is the gauging of a $\Z_{p_1K_2+p_2K_1}^{(1)}$ symmetry \cite{Hsin_2019}. We thus claim the following identity:
\begin{small}
\begin{equation*}
\begin{aligned}
    \qty(\tilde{\mathcal{A}}^{p_1N_1,1}\otimes \tilde{\mathcal{A}}^{p_2N_2,1})[F]\simeq\frac{\mathcal{A}_1^{(p_1K_2+p_2K_1)MK_1K_2,1}\otimes \tilde{\mathcal{A}}^{p_1p_2(p_1K_2+p_2K_1)M,1}[F]}{\Z_{p_1K_2+p_2K_1}^{(1)}}\\
    \simeq\frac{\mathcal{A}_1^{(p_1K_2+p_2K_1),MK_1K_2}\otimes \mathcal{A}_1^{p_1p_2(p_1K_2+p_2K_1),M}}{\Z_{p_1K_2+p_2K_1}^{(1)}} \mathcal{A}_1^{M,p_1p_2(p_1K_2+p_2K_1)}\otimes\mathcal{A}_1^{MK_1K_2,(p_1K_2+p_2K_1)}[F].
    \end{aligned}
\end{equation*}
\end{small}
where the second equality follows from the fact that the lines $\eta^{(1)}_i$ forming the $\Z_{p_1K_2+p_2K_1}^{(1)}$ group cannot be written as $\eta^{(1)}_i=(\eta^{(1)}_i)' L$ with $L\in \mathcal{A}_1^{M,p_1p_2(p_1K_2+p_2K_1)}\otimes\mathcal{A}_1^{MK_1K_2,(p_1K_2+p_2K_1)} $ since a generic line of $\mathcal{A}_1^{M,p_1p_2(p_1K_2+p_2K_1)}\otimes\mathcal{A}_1^{MK_1K_2,(p_1K_2+p_2K_1)}$ has order dividing $MK_1K_2$ which is coprime with $(p_1K_2+p_2K_1)$ by hypothesis.

\underline{Case 3: $N_1\neq N_2, \  \operatorname{gcd}(p_1K_2+p_2K_1,M)=M'\neq1$} we finally consider the general case.  The difference with the previous cases is given by the fact that:
\begin{equation}
\begin{aligned}
\mathcal{A}_1^{(p_1K_2+p_2K_1)M,1}&\neq\mathcal{A}_1^{M,p_1K_2+p_2K_1}\otimes
\mathcal{A}_1^{p_1K_2+p_2K_1,M},\\
\mathcal{A}_1^{(p_1K_2+p_2K_1)MK_1K_2,1}&\neq\mathcal{A}_1^{MK_1K_2,p_1K_2+p_2K_1}\otimes
\mathcal{A}_1^{p_1K_2+p_2K_1,MK_1K_2}.
\end{aligned}
\end{equation}

Physically, what is happening is that the spectrum of the lines contains a $\Z_{M'}$ subgroup of lines which are spinless and thus trivially braids with any line of the spectrum. 

These lines lead to a non-unitary braiding matrix and thus an ill-defined theory. However, we can cure this behaviour by gauging this subgroup.

At the level of the Lagrangian description, however, we can proceed exactly like before, except that after having obtained the expression \ref{Case2} we perform the gauging of the diagonal $\Z_{M'}$ subgroups. This is equivalent to the substitution $\alpha\to \alpha/M', \ \beta\to \beta/M'$:
\begin{equation*}
\begin{aligned}
\frac{1}{4\pi}\left( \frac{MK_1K_2}{M'}\frac{(p_1K_2+p_2K_1)}{M'}\alpha d\alpha + \frac{M}{M'}p_1p_2\frac{(p_1K_2+p_2K_1)}{M'}\beta d \beta + 2\frac{(p_1K_2+p_2K_1)}{M'}\alpha F\right)\\
=\mathcal{L}_{\frac{p_1p_2(p_1K_2+p_2K_1)}{M'}\frac{M}{M'}}[\beta,0]+\mathcal{L}_{\frac{(p_1K_2+p_2K_1)}{M'}\frac{K}{M'}}[\alpha, F].
\end{aligned}
\end{equation*}
Now the reparametrization is given by:
\begin{equation}
    \begin{bmatrix}
        M'&0\\0&M'
    \end{bmatrix}\begin{bmatrix}a_1\\a_2\end{bmatrix}=\begin{bmatrix}
        -K_2 & -p_2\\
         -K_1 & p_1
    \end{bmatrix}\begin{bmatrix}\alpha\\ \beta\end{bmatrix}
\end{equation}
where the gauging of the diagonal $\Z_{M'}^{(1)}$ is made evident. We thus conclude:

\begin{small}
\begin{equation*}
\begin{aligned}
    \frac{\qty(\tilde{\mathcal{A}}^{p_1N_1,1}\otimes \tilde{\mathcal{A}}^{p_2N_2,1})[F]}{\Z_{M'}^{(1)}}
    \simeq\\
    \frac{\mathcal{A}_1^{\frac{(p_1K_2+p_2K_1)}{M'},\frac{MK_1K_2}{M'}}\otimes \mathcal{A}_1^{p_1p_2\frac{(p_1K_2+p_2K_1)}{M'},\frac{M}{M'}}}{\Z_{\frac{p_1K_2+p_2K_1}{M'}}^{(1)}} \mathcal{A}_1^{\frac{M}{M'},p_1p_2\frac{(p_1K_2+p_2K_1)}{M'}}&\otimes\mathcal{A}_1^{\frac{MK_1K_2}{M'},\frac{(p_1K_2+p_2K_1)}{M'}}[F].
    \end{aligned}
\end{equation*}
\end{small}

\subsection{Minimal \texorpdfstring{$\mathcal{A}_2^{N,p}$}{TEXT} theories in 2 dimensions}
In this section, we introduce the definition and some basic properties of the minimal $(1+1)$d TQFTs $\mathcal{A}_2^{N,p}$ having a $\Z_N^{(0)}\times\Z_N^{(1)}$ symmetry, similarly to the well known minimal $(2+1)$-dimensional theories $\mathcal{A}_1^{N,p}$.

A generic theory with this property will be characterized by a set of topological lines $\eta^{(0)}_k$ and a set of topological points $\eta^{(1)}_k$ both fusing according to the $\Z_N$ group law. On top of that, a generic local observable $\mathcal{O}$ of the theory has a well-defined quantum number given by its charge $q_0(\mathcal{O})\in \hat{\Z}_N$ under the $\Z_N^{(0)}$ symmetry, while a generic line operator $\mathcal{W}$ of the theory has a well-defined quantum number given by its charge $q_1(\mathcal{W})\in \hat{\Z}_N$ under the $\Z_N^{(1)}$ symmetry.

We would like to characterize these theories by the linking of those defects: 
\begin{equation}\label{defect_braiding}
    \eta^{(0)}_k(\gamma)\eta^{(1)}_{k'}(x)=\exp(\frac{2\pi i p kk' \langle\gamma, x \rangle}{N}),
\end{equation}
representing an anomaly $p\in \Z_N=\operatorname{Tor}_1^{\Z}\left(H^2(B\Z_N; \Z); H^3(B^2\Z_N; \Z)\right)\subset H^3(B\Z_N\times B^2\Z_N; U(1))$ as described in \cref{Appendix: cohomology}, with inflow:
\begin{equation}
    S_{\text{inflow}}[B^{(1)};B^{(2)}]=-\frac{2\pi i p}{N}\int_{X^{(4)}}B^{(1)}\wedge B^{(2)}.
\end{equation}

As in the $(2+1)$-dimensional case, these defects alone allow for nontrivial correlation functions. However, when $\operatorname{gcd}(N,p)=M\neq 1$, the lines $\eta^{(0)}_{\frac{N}{M}}$ and points $\eta^{(1)}_{\frac{N}{M}}$ generate two $\Z_M$ subgroups undetectable by other topological operators. On the other hand, if $\operatorname{gcd}(N,p)= 1$, the topological defects form a consistent theory that we denote with $\mathcal{B}^{N,p}$.
\begin{example}
    An explicit Lagrangian for the $\mathcal{B}^{N}$ is given by the BF action at a coupling $N$:
    \begin{equation}
        \frac{iN}{2\pi} \phi\wedge dc
    \end{equation}
\end{example}
We would like to stress that we can change the labelling of the lines and points without changing the physical information of the theory. In particular, the group $\Z_N$ can be generated by each of its invertible elements $\Z_N^\times$. It follows that, given two integers $r,s$ coprime with $N$, the minimal theories satisfies the duality:
\begin{equation}
    \mathcal{B}^{N,p}\simeq\mathcal{B}^{N,rsp}
\end{equation}
implemented by the relabelling:
\begin{equation}
    \eta^{(0)}_r\to\tilde{\eta}^{(0)}_1, \quad \eta^{(1)}_s\to\tilde{\eta}^{(1)}_1 .
\end{equation}
In particular, since we assumed $p$ to be itself invertible, we conclude (for example by choosing $(r,s)=(p^{-1}_n,1)$)that the only physical information is given by $N$:
    \begin{equation}
        \mathcal{B}^{N,p}\sim \mathcal{B}^{N,1}.
    \end{equation}
We denote the theory $\mathcal{B}^{N,1}$ with $\mathcal{A}_2^N$.

We would like the $\mathcal{A}_2^N$ theories to satisfy a property analogous to \cref{th: minimality}. While we are not aware of such result, we will still refer to these theories as \textit{minimal} as they are the $1+1$d theories with the smallest dimensional Hilbert space having a $\Z_N^{(0)}\times\Z_N^{(1)}$ symmetry with nontrivial mixed anomaly.
\begin{remark}
    Since the $\mathcal{B}^{N,p}$ theories enjoy a $d-1=1$-form symmetry, they decompose. The $N$ universes $\mathcal{T}_i$ are invertible Euler counterterms \cite{Huang_2021}:
\begin{equation}
    \mathcal{B}^{N,p}=\bigoplus_{k=0}^{N-1}\mathcal{T}_k, \quad \quad \mathcal{T}_k(\Sigma_g)=(\operatorname{dim}(\eta^{(0)}_k))^{2-2g},
\end{equation}
where each of the topological lines $\eta^{(0)}_k$ becomes a domain wall.
\end{remark}
Moreover, those theories enjoy another property that will be used to prove the topological nature of the noninvertible electric defects $\mathcal{D}^{(e)}$ of the axion--Maxwell:
\begin{theorem}\label{Minimal-2d-half-gauging}
    The bulk-boundary system $(\mathcal{A}^{N}\otimes S_{\text{inflow}}[B^{(1)},B^{(2)}])[X^{(3)}]$ can be realised as a \textit{dynamical} theory:
    \begin{equation}
       -\frac{iN}{2\pi}\int_{X^{(3)}}\left(u^{(1)}dv^{(1)}+u^{(2)}dv^{(0)}+u^{(1)}u^{(2)}+u^{(1)}B^{(2)}-u^{(2)}B^{(1)}\right),
    \end{equation}
   with boundary, and $U(1)$ gauge fields $u^{(i)},\ v^{(i)}$ of degree $i$ with Dirichlet boundary conditions for the $v^{(i)}$.
\end{theorem}

Abstractly, we have coupled the $\mathcal{B}^{N,1}$ theory living on $\partial X^{(3)}$ to a global $\Z_N^{(d-3)}\times \Z_N^{(d-2)}$ symmetry of a given higher dimensional theory $\widetilde{\mathcal{T}}^{(d)}$ by imposing $\eta^{(0)}_i$ and $\eta^{(1)}_i$ to be in some twisted sector of the bulk symmetry defects.

More generally, given a pair $(a,b)\in \Z_N^2$ we denote with $\mathcal{B}^{N, p;a,b}$ the coupled $\mathcal{B}^{N, p}$ theory obtained by imposing the topologically invariant defects to be:
\begin{equation}
    \left(\eta_a^{(d-3)}\right)(\Sigma) \eta^{(0)}_1(\partial \Sigma); \quad \quad \left(\eta_b^{(d-2)}\right)(\gamma) \eta^{(1)}_1(\partial \gamma).
\end{equation}
Once again, we can analyse the base independent information and conclude that the coupled theories satisfy the following dualities:
    \begin{equation}
        \mathcal{B}^{N, p; a , b}\equiv\mathcal{B}^{N, pst; as , bt};
    \end{equation}
   for $s,t$ integers coprime with $N$. In particular, by choosing $(s,t)=(p^{-1}_na^{-1}_n,a)$, we conclude:
    \begin{equation}
        \mathcal{B}^{N, p; a , b}\equiv\mathcal{B}^{N, 1; 1 , ba};
    \end{equation}
    or, more symmetrically, by choosing $(s,t)=(a^{-1}_n,p^{-1}_na)$, we conclude:
    \begin{equation}
        \mathcal{B}^{N, p; a , b}\equiv\mathcal{B}^{N, 1; 1 ,p^{-1}_nab}.
    \end{equation}
We denote the theory $\mathcal{B}^{N,p; p,p}$ with $\mathcal{A}_2^{N,p}$, since this is the most symmetric choice.

\begin{remark}\label{remark: shao-gauge}
    In \cite{Choi2023noninvertible} the authors define the noninvertible electric defects using a different presentation for $\mathcal{A}_2^{N,p}=\mathcal{B}^{N,1;a,b}$\footnote{More specifically, they chose $a=1$, $b=p$.} which has the advantage of enjoying a simple Lagrangian presentation:
    \begin{equation}\label{eq: coupled BF lagrangian}
  \mathcal{L}_{\mathcal{B}^{N,1;a,b}}\qty[B^{(1)} ,B^{(2)}] =
  \frac{\ii N}{2\pi} \phi \dd{c} + a\frac{\ii N}{2\pi} c \wedge B^{(1)}
  + b\frac{\ii N}{2\pi}\phi B^{(2)},
\end{equation}
where $\phi$ and $c$ are respectively $0$ and $1$-form gauge fields, while $A$ and $B$ are currents for the $\Z_N^{(D-2)}\times \Z_N^{(D-3)}$ bulk symmetry defects.

The topological defects for this bulk-boundary system are given by:
\begin{equation}
  \eta^{(D-3)}_b(\Sigma) \eta^{(0)}_1(\partial \Sigma)
  = \exp \ii \qty(b \int_\Sigma B+ \int_{\partial \Sigma} c),
\end{equation}
and
\begin{equation}
  \eta^{(D-2)}_a(\gamma) \eta^{(1)}_1(\partial \gamma)
  = \exp \ii \qty(a \int_{\gamma} B^{(1)} + \phi(\partial \gamma)).
\end{equation}
In fact, they are invariant under $\Sigma \mapsto \Sigma + \Gamma +
\partial V^{(3)}$ (as above) and $\gamma \mapsto \gamma' = \gamma + C + \partial
\Sigma^{(2)}$ with $C \subset M^{(2)}$ and $\partial C = \partial \gamma' -
\partial \gamma$, as it is easily checked by imposing the equations of motion:
\begin{equation}
    d\phi+aB^{(1)}=0, \quad dc+bB^{(2)}=0.
\end{equation}

Let us now provide an alternative derivation of the fusion rules of the electric defects \ref{eq: electric fusion}

Technically, the two theories $\mathcal{B}^{N_1,1;1,p_1}$ and $\mathcal{B}^{N_2,1;1,p_2}$ are coupled to background whose holonomy is valued in two different groups $\Z_{N_i}$. We will avoid the technicality of this more general setup by focusing to the case where both backgrounds descend from the same $U(1)$ field strengths:
\begin{equation}
    B^{(1)}_{1,2}=\frac{\dd \theta}{N_{1,2}}, \quad B^{(2)} _{1,2}=\frac{F}{N_{1,2}}.
\end{equation}
The above hypothesis simplify the coupling terms by removing the $N_{1,2}$ dependence:
\begin{equation}
    a\frac{\ii N}{2\pi} c \wedge B^{(1)}
  + b\frac{\ii N}{2\pi}\phi B^{(2)}=a\frac{\ii }{2\pi} c \wedge \dd \theta
  + b\frac{\ii }{2\pi}\phi F.
\end{equation}
We denote the compound system coupled in this particular way with $\qty(\mathcal{B}^{N_1,1;1,p_1}\otimes\mathcal{B}^{N_2,1;1,p_2})[\dd \theta, F]$.

The action of a given a combined system $\qty(\mathcal{B}^{N_1,1;1,p_1}\otimes\mathcal{B}^{N_2,1;1,p_2})[\dd \theta, F]$ can be compactly written as:
\begin{equation}
     \frac{\ii }{2\pi} \begin{bmatrix}
         \phi_1 & \phi_2 & \theta
     \end{bmatrix}
     \begin{bmatrix}
         N_1&0&1\\
         0&N_2&1\\
         p_1&p_2&0
     \end{bmatrix}
     \begin{bmatrix}
     c_1\\c_2\\F\end{bmatrix}.
\end{equation}

We can now perform the following redefinitions:
\begin{equation}
    \begin{bmatrix}
         \phi_1 & \phi_2 & \theta
     \end{bmatrix}=\begin{bmatrix}
         \phi_\beta & \phi_\alpha & \theta
     \end{bmatrix}\begin{bmatrix}
        -1&1&0\\
        yK_2&xK_1&0\\
        0&0&1
    \end{bmatrix}, \quad  \begin{bmatrix}
     c_1\\c_2\\F\end{bmatrix}=\begin{bmatrix}
         -x&k_2&0\\
         y&K_1&0\\
         0&0&1
     \end{bmatrix} \begin{bmatrix}
     c_\beta\\c_\alpha\\F\end{bmatrix},
\end{equation}
where $(x,y)\in\Z$ are Bézout coefficients for $K_1$ and $K_2$, i.e. $xK_1+yK_2=1$.

After the redefinition we obtain the coefficient matrix takes the form:
\begin{equation*}
    \begin{bmatrix}
        -1&1&0\\
        yK_2&xK_1&0\\
        0&0&1
    \end{bmatrix}
    \begin{bmatrix}
         N_1&0&1\\
         0&N_2&1\\
         p_1&p_2&0
     \end{bmatrix}
    \begin{bmatrix}
         -x&k_2&0\\
         y&K_1&0\\
         0&0&1
     \end{bmatrix}=\begin{bmatrix}
         M&0&0\\
         0&L&1\\
         -xp_1+yp_2&K_2p_1+K_1p_2&0
     \end{bmatrix}.
\end{equation*}
This is the coupling matrix of the desired $\mathcal{A}_2^{M}\otimes \mathcal{A}_2^{L,p_1K_2+p_2K_1}[\dd\theta, F]$ apart from the term:
\begin{equation}\label{eq: unwanted contribution}
  \frac{i(-xp_1+yp_2)}{2\pi}  c_\beta\dd \theta.
\end{equation}
However, the various parameters appearing in the definition of $(-xp_1+yp_2)$ are subject to ambiguities:
\begin{equation}
    p_i\sim p_i+N_i, \quad(x,y)\sim (x+K_2,y-K_1),
\end{equation}
and we can use that freedom to make $(-xp_1+yp_2)$ a multiple of $M$.
Let us say that we made a particular choice of parameters such that $(-xp_1+yp_2)=k\neq 0 \operatorname{mod} M$. After a shift $(x',y')\sim (x+tK_2,y-tK_1)$ we will have: $$(-x'p_1+y'p_2)=k+t(p_1K_2+p_2K_1)\operatorname{mod} M.$$ Since $\gcd((p_1K_2+p_2K_1),M)=M'=1$ by hypothesis, there exist $\overline{t}$ such that $\overline{t}(p_1K_2+p_2K_1)=1 \operatorname{mod} M$. Finally, we can satisfy the desired condition by imposing $t=-k\overline{t}$.
 With this choice of parameters, the term \ref{eq: unwanted contribution} does not contribute to the action, as the equation of motions fix the dynamical fields to have holonomies in $\frac{\Z}{M}$ and thus the whole term an integer (which exponentiate to $1$).
\end{remark}

\end{appendices}

\printbibliography
\end{document}